\documentclass[twocolumn]{aastex701}

\usepackage{graphicx}	
\usepackage{amssymb}    
\usepackage{bm}		
\usepackage{xcolor} 
\usepackage{ulem}
\usepackage[varvw]{newtxmath}  
\usepackage{comment}
\usepackage{tabularray}
\usepackage[utf8]{inputenc} 

\usepackage{amsmath}
\usepackage{mathtools} 

\DefTblrTemplate{firsthead, middlehead,lasthead}{default}{}

\newcommand{\BB}{{\bm B}}
\newcommand{\beq} {\begin{equation}}

\newcommand{\eeq} {\end{equation}}

\newcommand{\jR}{$j$-$R$}
\newcommand{\jM}{$j$-$M$}

\newcommand{\pcc}{{\rm cm}^{-3}}

\newcommand{\uu}{{\bm u}}

\definecolor{myblue}{RGB}{0, 90, 189}
\definecolor{mygreen}{RGB}{0, 80, 0}

\newcommand{\grn} {\color{mygreen} }

\defcitealias{Arroyo-Chavez.Vazquez-Semadeni2022}{ACVS22}

\usepackage[T1]{fontenc}
\usepackage{ae,aecompl}
\usepackage{newtxtext,newtxmath}

\shortauthors{Arroyo-Chávez et al.}

\begin{document}

\title{On the role of gravity, turbulence, and the magnetic field in angular momentum transfer within molecular clouds}

\author[orcid=0000-0002-7082-0587]{Griselda Arroyo-Ch\'avez}
\altaffiliation{Steward Observatory}
\affiliation{Steward Observatory, University of Arizona 933 North Cherry Avenue, Tucson AZ, 85721, USA}
\affiliation{Instituto de Radioastronom\'ia y Astrof\'isica, PO Box 3-72. 58090, Morelia, Michoac\'an, M\'exico}
\email[show]{arroyochavezg@arizona.edu}  

\author[orcid=0000-0002-1424-3543]{Enrique V\'azquez-Semadeni}
\altaffiliation{IRyA UNAM}
\affiliation{Instituto de Radioastronom\'ia y Astrof\'isica, PO Box 3-72. 58090, Morelia, Michoac\'an, M\'exico}
\email[]{e.vazquez@irya.unam.mx} 

\author[orcid=0000-0003-0688-5332]{James Wurster}
\altaffiliation{No longer affiliated with University of St. Andrews}
\affiliation{Scottish Universities Physics Alliance (SUPA), School of Physics and Astronomy, University of St. Andrews, North Haugh, St Andrews, Fife, KY16 9SS, UK}
\email[]{}

\begin{abstract}
Observations of molecular structures on scales of $\sim 0.1-50$ pc show that the specific angular momentum ($j$) scales with radius ($R$) as $j\sim R^{3/2}$. We study the effects of turbulence, gravity, and the magnetic field in shaping this scaling, by measuring clump size and specific angular momentum in three SPH simulations of the formation of giant molecular clouds, progressively adding these three ingredients. In each simulation, we define ``full'' and ``reduced'' clump samples, the latter restricted to aspect ratios $A<3$. We find that, in the non-magnetic runs, elongated clumps deviate the most from the \jR\ relation, which is best reproduced by the reduced sample in the gravity+turbulence run. In the purely hydrodynamic case, no dense elongated structures form, suggesting that turbulence alone is insufficient to generate dense filaments, although clumps have $j$ magnitudes consistent with observations. In the gravity+turbulence+magnetic field run, most of the clumps are filamentary, yet the full sample appears to follow the observed \jR\ relation. This result, rather than being a real trend, could be the combination of the increase in $j$ by the filamentary geometry, and its reduction by turbulence inhibition by the magnetic field. Finally, we measure the gravitational, magnetic, pressure-gradient, and hydrodynamic torques (which involve turbulent viscosity) in our clump samples. We find that, in magnitude, the hydrodynamic torques tend to be larger than the rest. This result is consistent with our previous work, where we proposed that gravity drives cloud formation and contraction, while turbulence redistributes angular momentum through fluid-parcel exchanges.
\end{abstract}

\keywords{\uat{Interstellar medium}{847},\uat{Giant molecular clouds}{653},\uat{Cloud collapse}{267}}



\section{Introduction}
\label{sec:introduction}

Although the star formation process is currently widely studied, many problems remain unsolved. One of them, the {\it angular momentum problem} (AMP), has persisted for several decades. \citet{Spitzer78} (and later \citet{Bodenheimer95}) described this problem using the example of a filamentary cloud that collapses to form a solar-type star. If angular momentum were conserved during the contraction, it would be impossible for the resulting star to remain as a single coherent object  due to the excessive centrifugal force. Therefore, most of the angular momentum must be removed by some mechanism during contraction.

This problem would later be reflected in measurements of velocity gradients for samples of molecular clouds in the Galaxy \citep{Fleck.Clark1981,Goldsmith.Arquilla85} and measurements of specific angular momentum (SAM, denoted $j$) for samples of objects on scales from tenths of parsecs up to a few \citep{Goodman+93,Caselli+2002,Pigorov+2003,Chen_X+2007,Tatematsu+16,Chen_Hope+2019b,Xu-Xuefang+2020a,Xu-Xuefang+2020b}. Together, these observations show that $j$ scales with the size ($R$) of the region as $j\sim R^{3/2}$ (see Fig. 1 of \citet[][\citetalias{Arroyo-Chavez.Vazquez-Semadeni2022} hereafter]{Arroyo-Chavez.Vazquez-Semadeni2022}). We will henceforth refer to this scaling as the \jR\ {\it relation}. 

The \jR\ scaling can also be observed in dense cores \citep{Pineda+2019}, while at smaller scales ($\sim 50$--a few $ \times 10^{3}$ au) $j$ tends to be independent of radius. The latter property is usually interpreted as a sign of angular momentum conservation during the formation of an accretion disk \citep{Gaudel+2020}, although other works suggest that the flat profile of $j$ in the inner region of a cloud is simply due to the elongation of this region by large radial velocity gradients in a runaway collapse, so that if originally $j$ varied with radius, this elongation would make the profile appear flat \citep{Takahashi+2016}. 

Different factors need to be taken into account, depending on the scale for which the solution to the angular momentum problem is intended. It is well known that the magnetic field plays an important role in removing angular momentum on protoplanetary disk scales \citep{Balbus_Hawley91,Balbus.Hawley1998,Balbus2003,Joos.Hennebelle.Ciardi2012,Hennebelle+2016,Marshall+2018}. However, the net effect of the magnetic field on angular momentum transport at molecular cloud scales is not entirely clear. Some magnetic field effects such as ambipolar diffusion or magnetic braking have also been invoked as a solution to the AMP on molecular cloud scales (\citet[Section 2.3.3]{Shu+87}, \citet[Section 2.1]{Mouschovias91}. However, it is now known that the \jR\ relation is also recovered in simulations that do not involve the magnetic field \citep[] [hereafter ACVS22] {Jappsen.Klessen04,Arroyo-Chavez.Vazquez-Semadeni2022}. Alternatively, gravitational torques have been proposed as the mechanism responsible for transporting angular momentum out of molecular clouds \citep{Larson84,Jappsen.Klessen04,Kuznetsova+2019}. Other works \citep[e.g.,] [] {Fleck.Clark1981, Burkert.Bodenheimer2000, Offner2008, Chen_Che.Ostrikers2018, Misugi+2019} suggest that the cores acquire rotation from the properties of the turbulent medium in which they are immersed. Continuing this line, \citet{Fleck1982} proposed that the \jR\ relation (sometimes also expressed as \jM, with $M$ being the mass of the clump) at all scales depends on the parameters of the turbulence associated with specific systems. Recently, \citet{Misugi+2023} have studied the evolution of angular momentum in simulated molecular cores formed from filament fragmentation, and measured the evolution of gravitational and pressure torques, concluding that the angular momentum of the cores is transferred mainly by pressure-gradient torques. 

In \citetalias{Arroyo-Chavez.Vazquez-Semadeni2022} we studied the evolution of the specific angular momentum in an SPH simulation carried out with the SPH \textsc{Gadget-2} code \citep{Springel+01} of the formation and collapse of giant molecular clouds in the turbulent warm neutral medium, without including stellar feedback nor the magnetic field. Furthermore, from an inspection of the angular momentum evolution equation,\footnote{Obtained by taking the cross product of the position vector with the momentum equation including self-gravity and the magnetic field.} we identified the torques that come into play in the transport of angular momentum, namely pressure-gradient, magnetic, gravitational, and hydrodynamic torques. The latter arise from the advective term of the momentum equation, and are related to the turbulent (or eddy) viscosity. We proposed that the angular momentum transfer mechanism at molecular cloud scales is mainly carried out by hydrodynamic torques since the loss of angular momentum occurred more efficiently in regions containing a larger number of external (``intruder'') particles with which the SPH particles initially belonging to the clump can interact. However, we did not perform any direct measurements of the different types of torques. 

Since the nature of the mechanism that transfers and distributes angular momentum in molecular clouds remains uncertain, the goal of this work is to clarify the relative importance of the torques involved in this process through SPH numerical simulations that include turbulence, gravity, and the magnetic field. The structure of this paper is as follows. In Section \ref{Sec:Numerical Data} we describe the main characteristics of the simulations used to define numerical samples of clumps, to continue with Section \ref{sec:results}, in which we present the \jR\ scaling for each sample as well as the magnitude of the four different torques measured on them. In Section \ref{sec:discussion} we discuss the implications of these results, and the conclusions in Section \ref{sec:conclusions}.


\section[]{Numerical data}
\label{Sec:Numerical Data}


\subsection{The simulations}
\label{subsec: Simulation}

We use the smoothed-particle hydrodynamics (SPH) code \textsc{Phantom} \citep{Price+2018}, to perform three simulations of decaying turbulence in the warm neutral medium (see Table \ref{tab:simulations}): two purely hydrodynamic runs, one with gravity (labeled HDG3) and one without it (labeled HD3), and another one including both gravity and an initially uniform magnetic field of $3\, \mu$G (labeled MHDG3) along the $x$ direction. For these simulations we use the cluster \textsc{Phantom} default setup, which emulates the setup used by \cite{Bate_Bonnell_Bromm2003} on the formation of a stellar cluster from a collapsing, turbulent, massive core, combined with the turbulent driving module based on the works of \citet{Price.Federrath2010} and \citet{Tricco.Price.Federrath2016,Tricco.Price.Laibe2017}.

We use a resolution of $145^{3}$ particles, in a numerical box of $256$ pc per side. The initial density and temperature were set at $n_{0} \equiv n(t=0) = 3\, \pcc$ and $T_{0} \equiv T(t=0) = 730$ K, respectively, the latter being the thermal-equilibrium (heating equals cooling) temperature at $n_0$. This corresponds to a mass resolution of $0.5$ M$_{\odot}$. Stellar (or ``sink'') particles are formed once an SPH particle passes all the checks \citep[][Section 2.8.4]{Price+2018} once it reaches a density of $3.3 \times 10^{6}\, \pcc$. The critical radius (no new sinks are created within this radius around an existing sink), the accretion radius (particles within this radius are accreted if they pass a series of tests) and the merge distance condition (sinks will unconditionally merge within this separation) were all set to the same value of $0.04$ pc\footnote{This choice is made in order to avoid the timestep from becoming excessively small when failure to produce a sink particle (due to nearness to another one) causes an SPH particle to unrestrictedly increase its density.}. The cooling and heating processes are included implicitly via adjusted functions from \cite{Koyama.Inutsuka2000}, with the typographical correction given by \citet{Vazquez-Semadeni+07}. The boundaries are periodic for the hydrodynamics, but isolated for gravity, as \textsc{Phantom} does not include a provision for periodic gravity.

We apply Fourier-space forcing for the first $0.65$ Myr for wave numbers in the range $1<kL/2\pi<4$, reaching a maximum velocity dispersion of $10.7$ km/s. The rms sonic Mach number reached after the forcing time in runs HD3 and HDG3 is $\mathcal{M}_{\rm s}\sim 4.3$, although the maximum of $\mathcal{M}_{\rm s}\sim 9.1$ is reached later at $t\sim 3.4$ Myr. This late increase of the rms Mach number is due to the rapid cooling in the flow, which increases the Mach number by quickly reducing the sound speed in the cooled regions. On the other hand, the Alfvénic Mach number at the same time in run MHDG3 is $\mathcal{M}_{\rm A}\sim 1$. Given the same forcing parameters applied to the HD3 and HDG3 simulations, the velocity dispersion reaches a maximum value of $\sim 3.5$ km/s for run MHDG3. Unlike \citetalias{Arroyo-Chavez.Vazquez-Semadeni2022}, where turbulence driving was purely solenoidal, in this case we use a mixture of compressible and solenoidal modes such that the {\it solenoidal weight} $w (\in [0,1]) = 0.4$ for all the simulations. This implies a value for the ratio of compressive to total forcing power $\approx 0.5$ for 3D numerical data, as defined in \citet{Federrath+2010a}. The main motivation for adding compressible modes is to compensate to some extent for the lack of stellar feedback and supernovae in our simulation, which naturally add a compressible component to the interstellar medium. On the other hand, since our simulation does not include any form of feedback, regions where local feedback effects would be important due to their high sink content are excluded from our analysis with the selection criteria presented in the following section.

\begin{deluxetable*}{ccccccc ccc}
\tablecaption{Simulations\label{tab:simulations}}

\tablehead{
\colhead{Simulation} &
\colhead{$T_{0}$ (K)} &
\colhead{$n_{0}$ (cm$^{-3}$)} &
\colhead{Resolution} &
\colhead{$L$ (pc)} &
\colhead{$M/M_{\rm J}$\tablenotemark{a}} &
\colhead{$L/L_{\rm J}$\tablenotemark{b}} &
\colhead{Turbulence} &
\colhead{Gravity} &
\colhead{Uniform magnetic field} \\
&
&
&
&
&
&
&
\colhead{($\sigma_{0}\!\approx\!10.7~{\rm km\,s^{-1}}$)} &
&
\colhead{($3~\mu{\rm G}$)}
}

\startdata
HD3   &     &   &           &     &      &      & x &   &   \\
HDG3  & 730 & 3 & $145^{3}$ & 256 & 2.09 & 1.03 & x & x &   \\
MHDG3 &     &   &           &     &      &      & x & x & x \\
\enddata

\tablenotetext{a}{$M_{\rm J}$: Jeans mass.}
\tablenotetext{b}{$L_{\rm J}$: Jeans length.}

\end{deluxetable*}


\subsection{The numerical clump sample}
\label{subsec:sample}

The clump definition was carried out using the clump-finding algorithm presented in \cite{Camacho+16}, which is based on the search for ``connected'' structures (i.e., particles within a smoothing length of a test particle with density above a certain threshold). For lack of a more specific nomenclature, here we will refer to any connected structures defined in the simulations using the clump-finding algorithm as ``clumps'', regardless of their size. Therefore, our use of the term should not be confused with its standard interpretation as structures of size ($\sim 1$ pc) intermediate between that of cores ($\sim 0.1$ pc) and that of clouds ($\sim 10$ pc). We apply the clump-finding algorithm at times $14$, $15$ and $16$ Myr for runs HD3 and MHDG3, and times $31$, $32$ and $33$ Myr for run HDG3\footnote{Owing to the well-known physical delay in the dynamical evolution induced by magnetic fields, later snapshots in run MHDG3 were selected to ensure a meaningful comparison between simulations.}, for six density threshold values: $n_{\rm th} = 3 \times 10^{2}$, $10^{3}$, $3 \times 10^{3}$, $10^{4}$, $3 \times 10^{4}$ and $10^{5}$ cm${}^{-3}$. Currently, there is a wide variety of methods for searching for and defining structures in both simulations and observations. Some of these provide not only information about the structure itself but also its position within a hierarchical network in relation to its environment and substructure \citep[Dendrograms,][for example]{Rosolowsky+2008}. However, when comparing structures positioned at the same level in this network, despite sharing topological characteristics, they do not necessarily share the same physical properties, and therefore, the categories can be confusing. For this reason, we consider a density thresholding algorithm appropriate here to subsequently compare the behavior of the various groups of clumps.

It is important to mention that not all the thresholds were used for the three simulations, since these reach different maximum densities at the chosen times. We will discuss these differences later in Section \ref{subsec:diff_in_sim}. Also, we allow for the possibility that some of the denser clumps are found within more diffuse structures, so the sample may include nested objects defined at different density thresholds. In order to guarantee that our clumps are resolved, we restrict our sample to clumps with at least 58 SPH particles (the number of particles within a smoothing length). Similarly we restrict the mass in sinks in our clumps such that
\begin{equation}
\centering
\frac{M_{*}}{M_{\text{tot}}} \equiv \frac{M_{*}}{M_{*}+M_{\text{gas}}} < 30 \%,
\label{eq:efficiency}
\end{equation}
where $M_{*}$ is the mass in stars (sinks). We do this because we do not include feedback in our simulations. Since real clumps with a stellar mass fraction larger than this value are most likely strongly affected by feedback, they are not modeled correctly in our simulation, and thus we discard them. 

We compute the angular momentum per particle with respect to the clump's center of mass (denoted by the subindex CM) as 
\begin{equation}
{{\bm J}_{p, CM}} \equiv {{\bm r}_{\rm p,CM}} \times {{\bm p}_{{\rm p, CM}}} = {{\bm r}_{{\rm p, CM}}} \times m_{{\rm p}} {{\bm v}_{{\rm p, CM}}},
\label{eq:ang_mom}
\end{equation}
such that ${{\bm r}_{{\rm p, CM}}} = {{\bm r}_p} - {\bm X_{\rm CM}}$ and ${{\bm v}_{{\rm p, CM}}} = {{\bm v}_p} - {\bm V_{\rm CM}}$, where the capital letters represent the quantities of the center of mass, and the subscript $p$ refers to each particle. Then the specific angular momentum of the clump is $(\Sigma_p {\bm J_{p, CM}})/M_{{\rm gas}}$. The position and velocity of the center of mass are computed as 
\begin{equation}
{\bm X_{\rm CM}} = \frac{\Sigma_p m_p {\bm r}_{\rm p}} {\Sigma_p m_p}, \; {\bm V_{\rm CM}} = \frac{\Sigma_p m_p {\bm v}_{\rm p}} {\Sigma_p m_p}
\label{eq:X_CM}
\end{equation}

The clump radius (or ``size''), $R$, is calculated as that of a sphere with the same volume as the clump's; i.e., as $R =(3V/4 \pi)^{1/3}$ \citep{Camacho+16}, where the clump's volume is:
\begin{equation}
V = \sum_{p} V_{p} = \sum_{p} \frac{m_{p}}{\rho_{p}}=m_{p} \sum_{p} \rho_{p}^{-1}.
\label{eq:total vol}
\end{equation}

\subsection{The residual angular momentum of clumps}
\label{subsec:residual AM}

It is important to clarify that the calculation of angular momentum in observational samples tends to rely on the identification of velocity gradients associated with certain rotation patterns \citep{Goodman+93,Caselli+2002,Pigorov+2003,Chen_X+2007,Tatematsu+16,Pandhi+2023}. In this work, no rotation pattern is assumed in advance, and the angular momentum of a clump is defined simply as the resultant vector of the contribution from all SPH particles in the clump with respect to its center of mass. Therefore, this resultant vector should be considered as a {\it residual} angular momentum, not necessarily indicative of a {\it coherent} rotation. By performing the calculation with respect to the center of mass, we also ensure that it is not affected by the translational motion of the clumps  (see Section \ref{sec: CM discussion} and Appendix \ref{Appe:AM for system of particles}).

In this work we will consider only the magnitude of the angular momentum vector. However, the study of its direction is also relevant to compare its orientation with respect to other vector fields of interest, such as the magnetic and velocity fields. We defer such a study to a future contribution.


\section{Results}
\label{sec:results}


\subsection{General features of the simulations}
\label{subsec:diff_in_sim}

\begin{figure*}
\centering \offinterlineskip
\includegraphics[width=\textwidth]{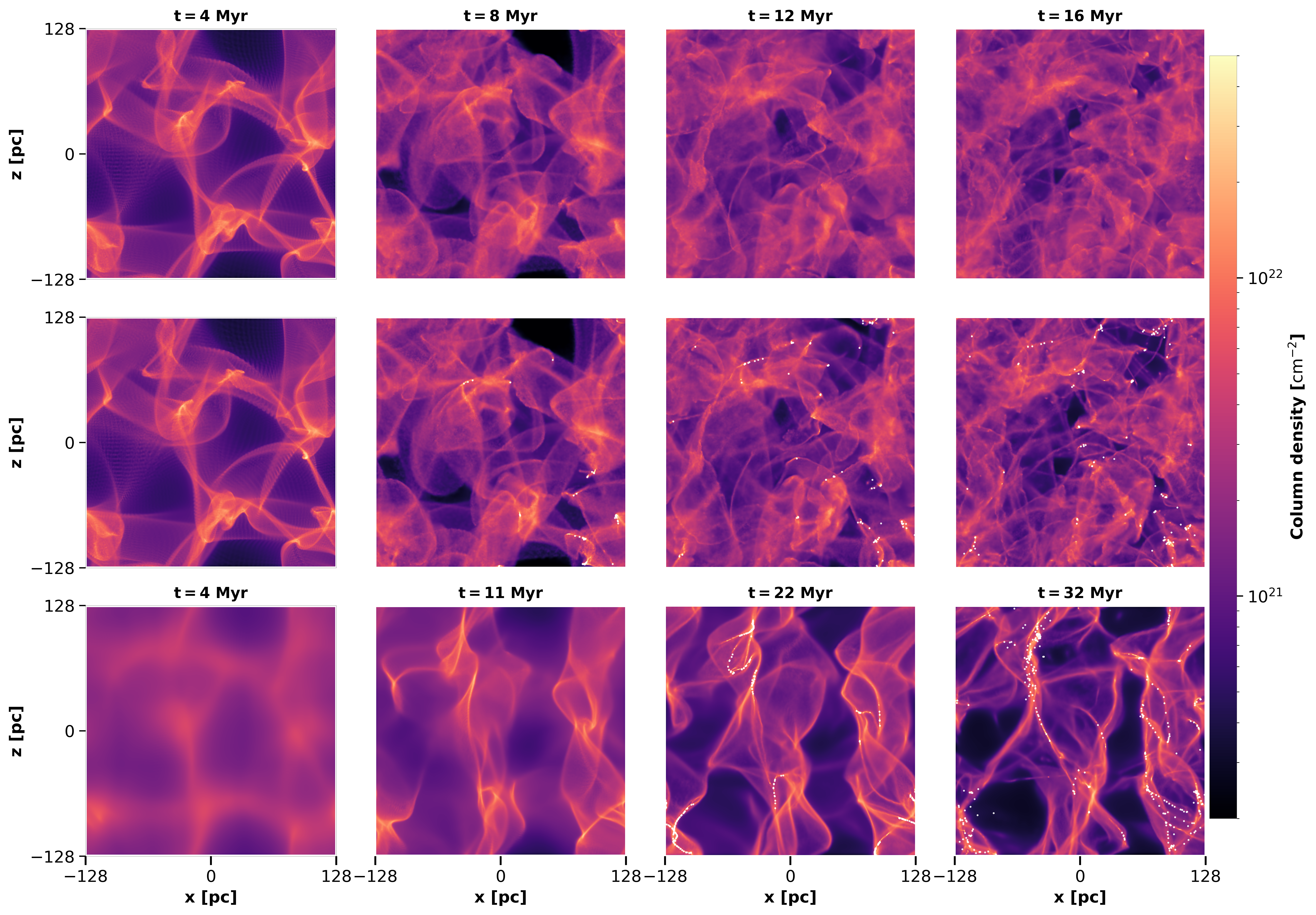}
 \caption{Evolution of runs HD3  ({\it top panels}), HDG3 ({\it middle panels}) and MHDG3 ({\it bottom panels}). The columns show increasing times from left to right, as indicated by the labels at the top. The color bar shows the column density along the z-axis of the entire numerical box, with the same range of values for the three simulations. Sinks are represented by white dots. Sink formation is noticeably delayed in the magnetic simulation and completely nonexistent in the simulation without gravity. The HD3 and HDG3 simulations appear significantly more fragmented than MHDG3, which looks smoother and more filamentary.}
 \label{fig:snaps_sims}
\end{figure*}

In Figure \ref{fig:snaps_sims} we present a sequence of column density images from the three simulations studied here. The columns correspond, from left to right, to times $t= 4,8,12$ and $16$ Myr for runs HD3 (firs row) and HDG3 (second row), and $t= 4,11,22$ and $32$ Myr for run MHDG3 (third row). Sinks are represented by white dots. The formation of the sink particles is considerably delayed in the MHDG3 simulation by the presence of the magnetic field. To illustrate the typical values of the magnetic field strength in run MHDG3, the relation between the magnetic field strength, $B$, and the number density, $n$, for the SPH particles in the MHDG3 run is shown in Appendix \ref{Appe:B vs n}. Specifically, the formation of the first sink in run HDG3 occurs at $t= 5.4$ Myr, while in MHDG3 it occurs at $t = 14.7$ Myr. On the other hand, no sinks are formed in the HD3 simulation since the fragments formed by turbulence cannot continue to contract due to the absence of gravity, although relatively high densities of up to $n \sim$ a few times $10^3 \pcc$ are reached just by the combined action of the cooling and the turbulent compressions. 

\begin{figure}
\centering
\includegraphics[width=\linewidth]{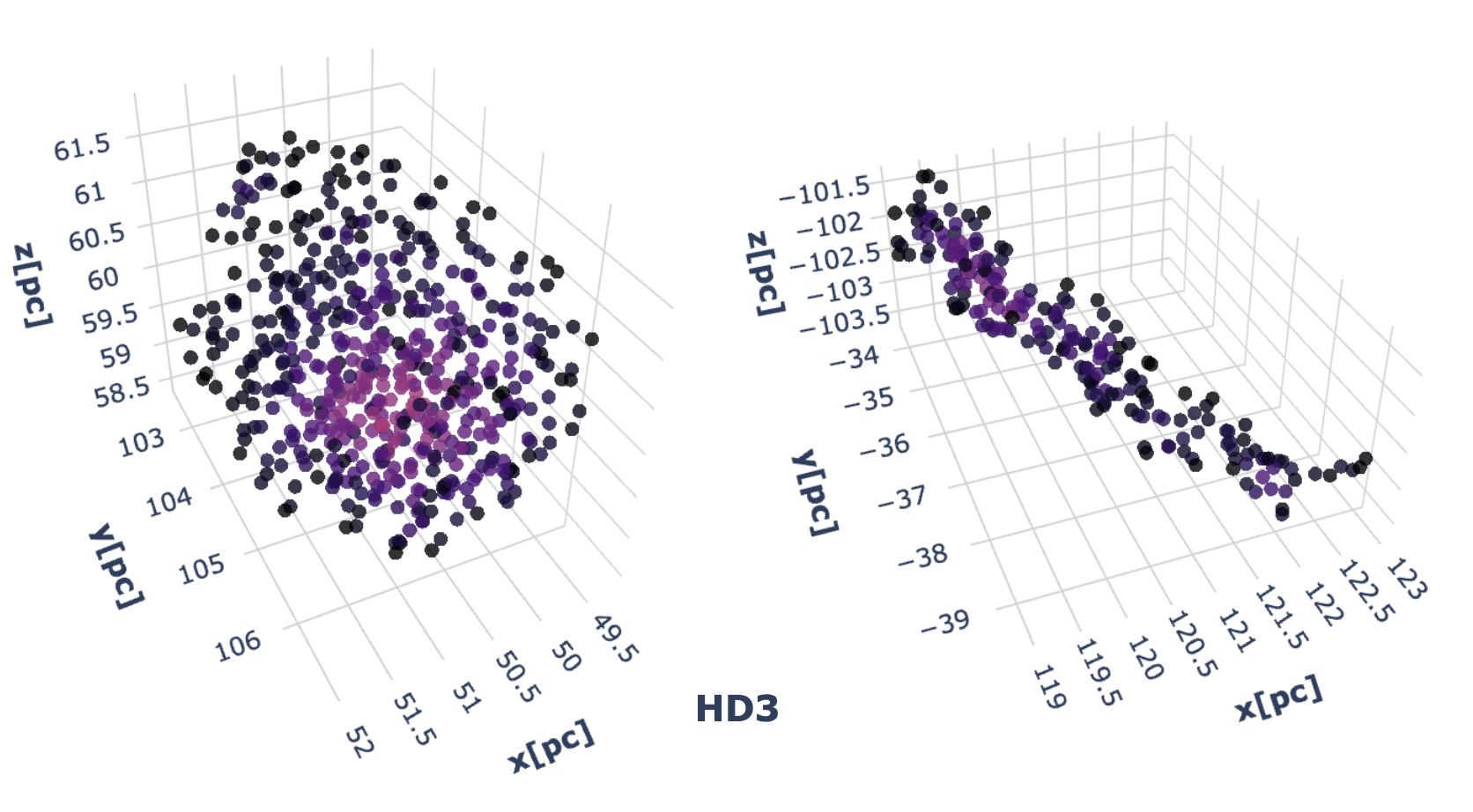}
\includegraphics[width=\linewidth]{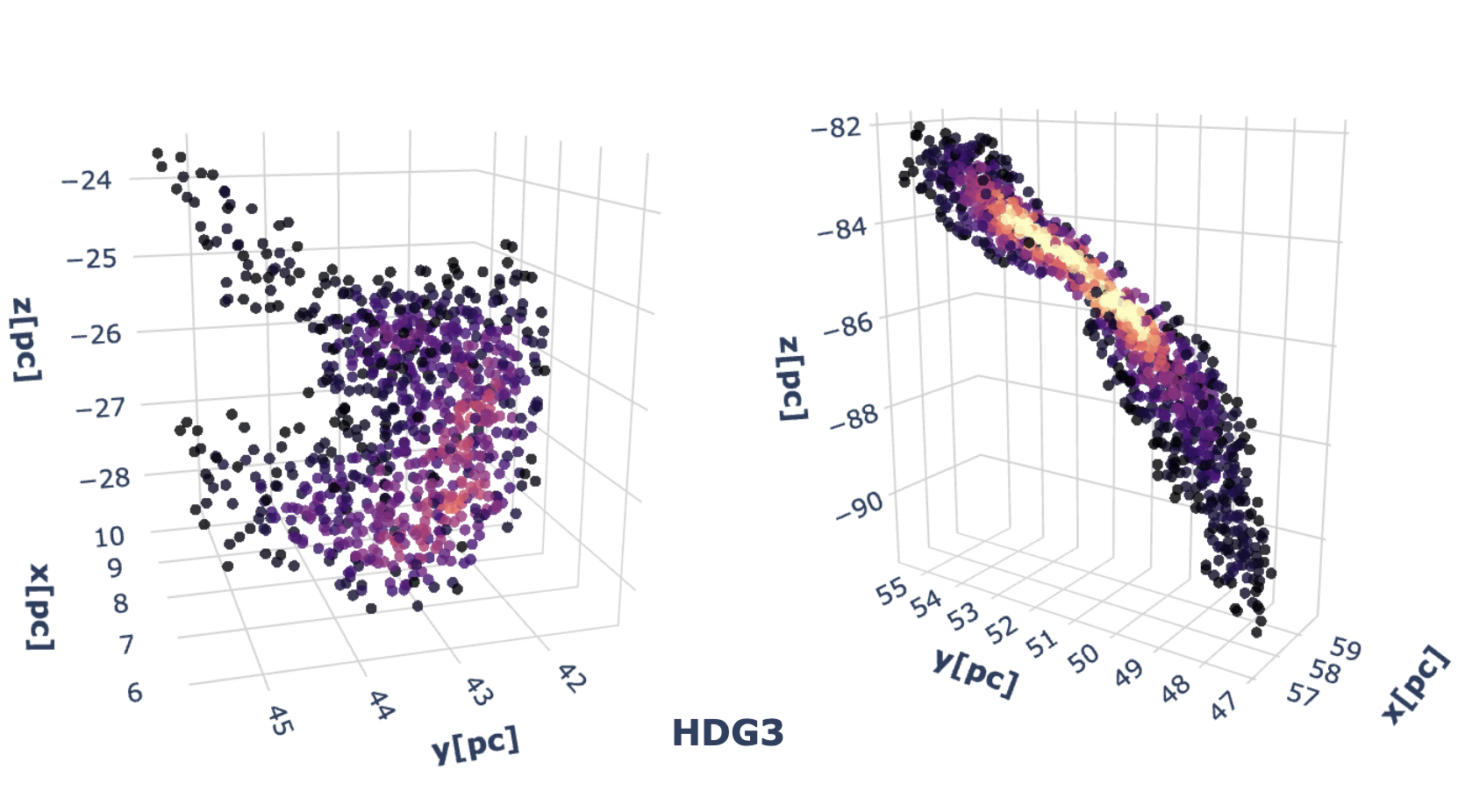}
\includegraphics[width=\linewidth]{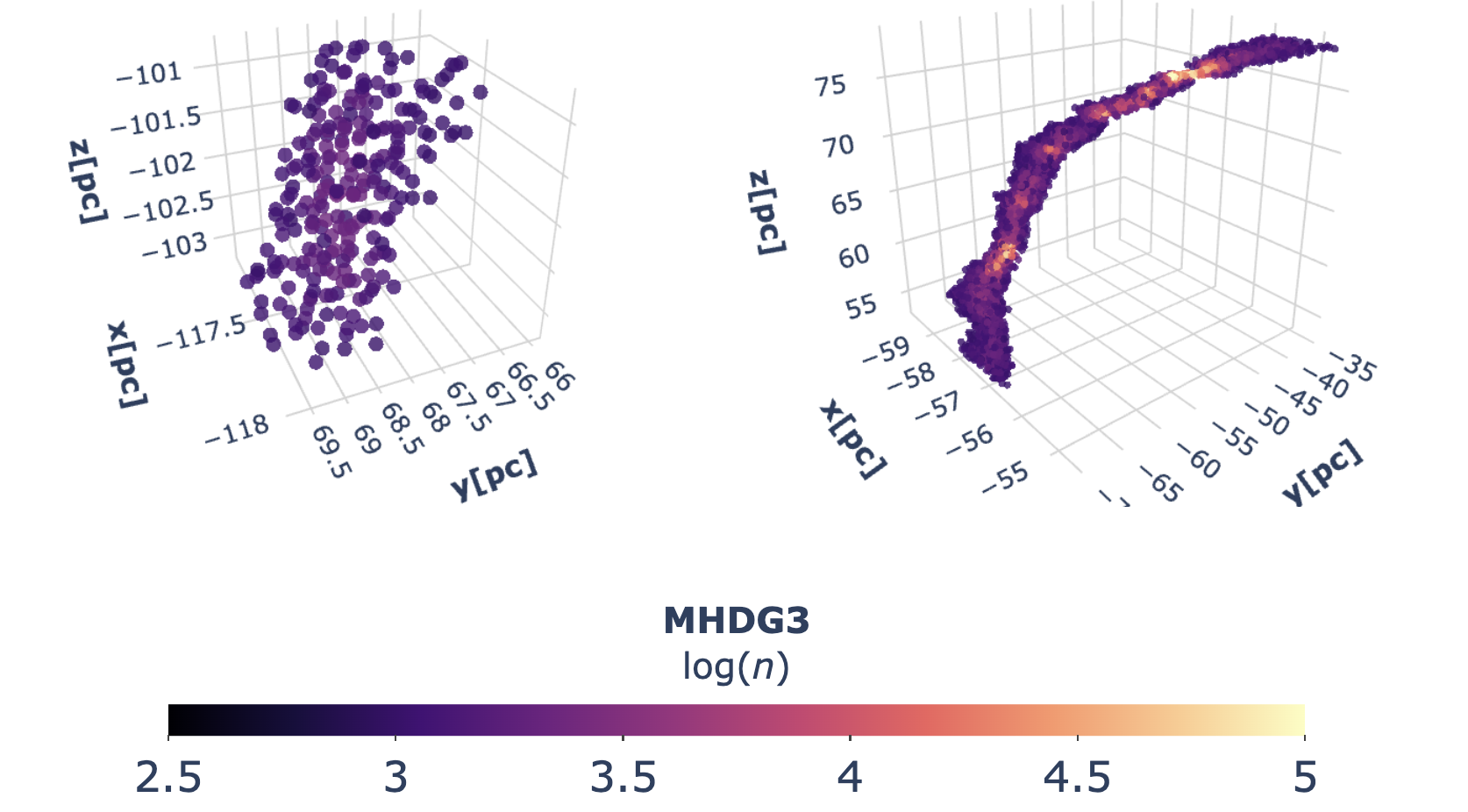}
 \caption{Representative round (left column) and elongated (right column) clumps of each numerical sample in the HD3 (top row), HDG3 (middle row) and MHDG3 (bottom row) simulations. The color code represents the density in units of cm${}^{-3}$. The most elongated clumps are recovered in the HDG3 and MHDG3 simulations, the latter having the most defined and narrow filaments. Densities around $10^5\, \pcc$ are only reached in the HDG3 and MHDG3 simulations.}%
 \label{fig:com_clumps}
\end{figure}

Regarding the morphology of the structures, overdensities in run MHDG3 tend to be more elongated compared to those generated in simulations without a magnetic field, in agreement with earlier results \citep{HennebelleP2013,Inoue.Inutsuka2016,Inoue+2018}. The filaments are also thinner in this run. On the other hand, filamentary structures also appear in the runs HD3 and HDG3, although in HD3 the filaments tend to be wider and more diffuse due to the absence of gravity. We have verified that these are real filaments, and not just projections of curved sheets seen at grazing angles by rotating the projected images. Figure \ref{fig:com_clumps} shows two representative clumps from each of the simulations. In the first column we show clumps with roundish shapes, while in the second column we show clumps with filamentary morphologies. For the purpose of distinguishing between these morphologies, following the definition adopted by \citet{Arzoumanian+2019}, from now on we will consider as ``elongated structures'' (i.e., filaments) those clumps with aspect ratios, $A > 3$. The aspect ratio is calculated as the ratio of the longest to the shortest principal axis, with the principal axes being those obtained from the eigenvalues of the moment of inertia tensor calculated for each clump. 


\subsection{The \jR\ relation in the set of simulations}
\label{subsec:all jR}

\subsubsection{General features for the complete samples}
\label{subsibsec:General features for the complete samples}

\begin{figure*}
\centering

\includegraphics[width=0.48\linewidth]{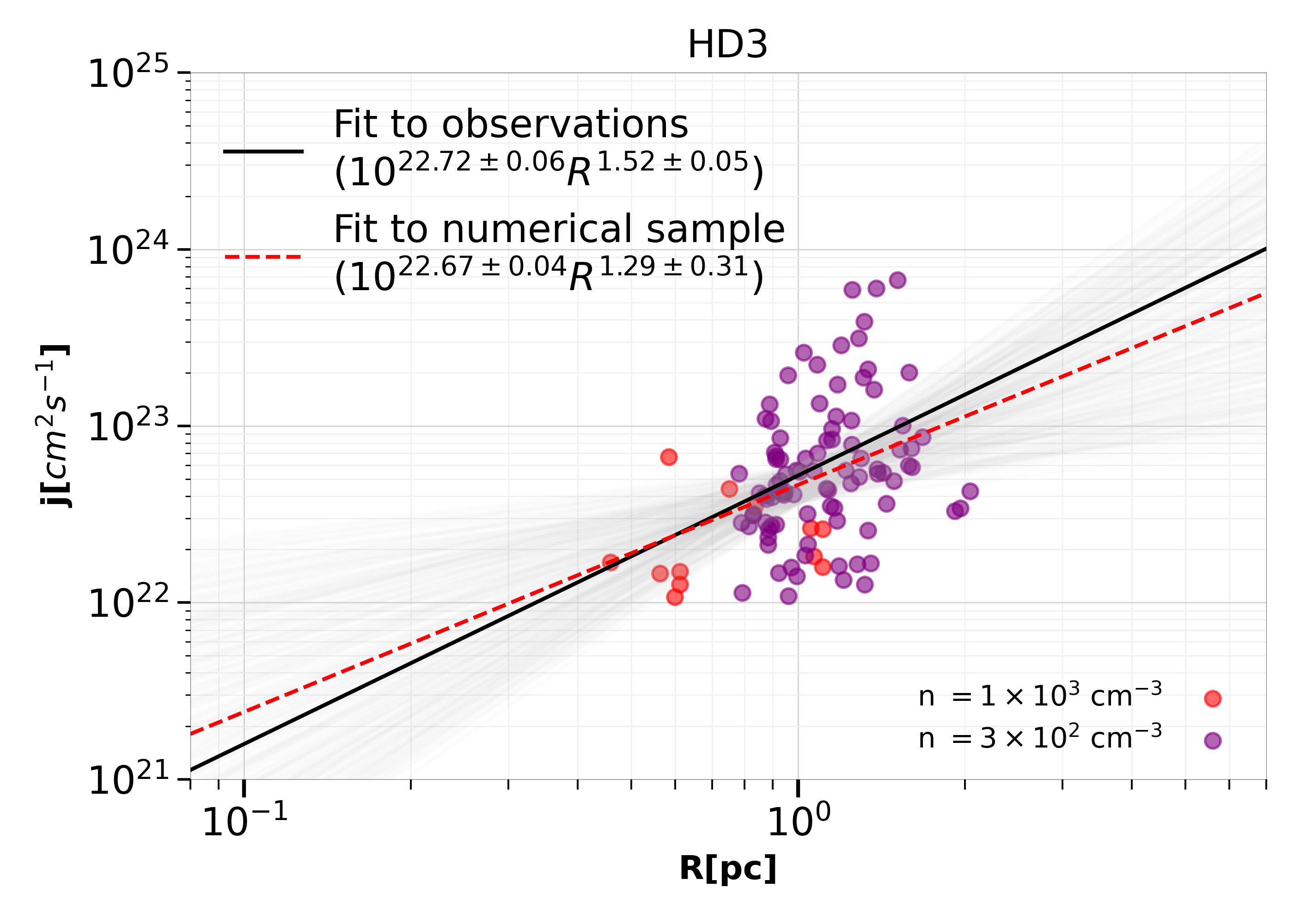}
\includegraphics[width=0.48\linewidth]
{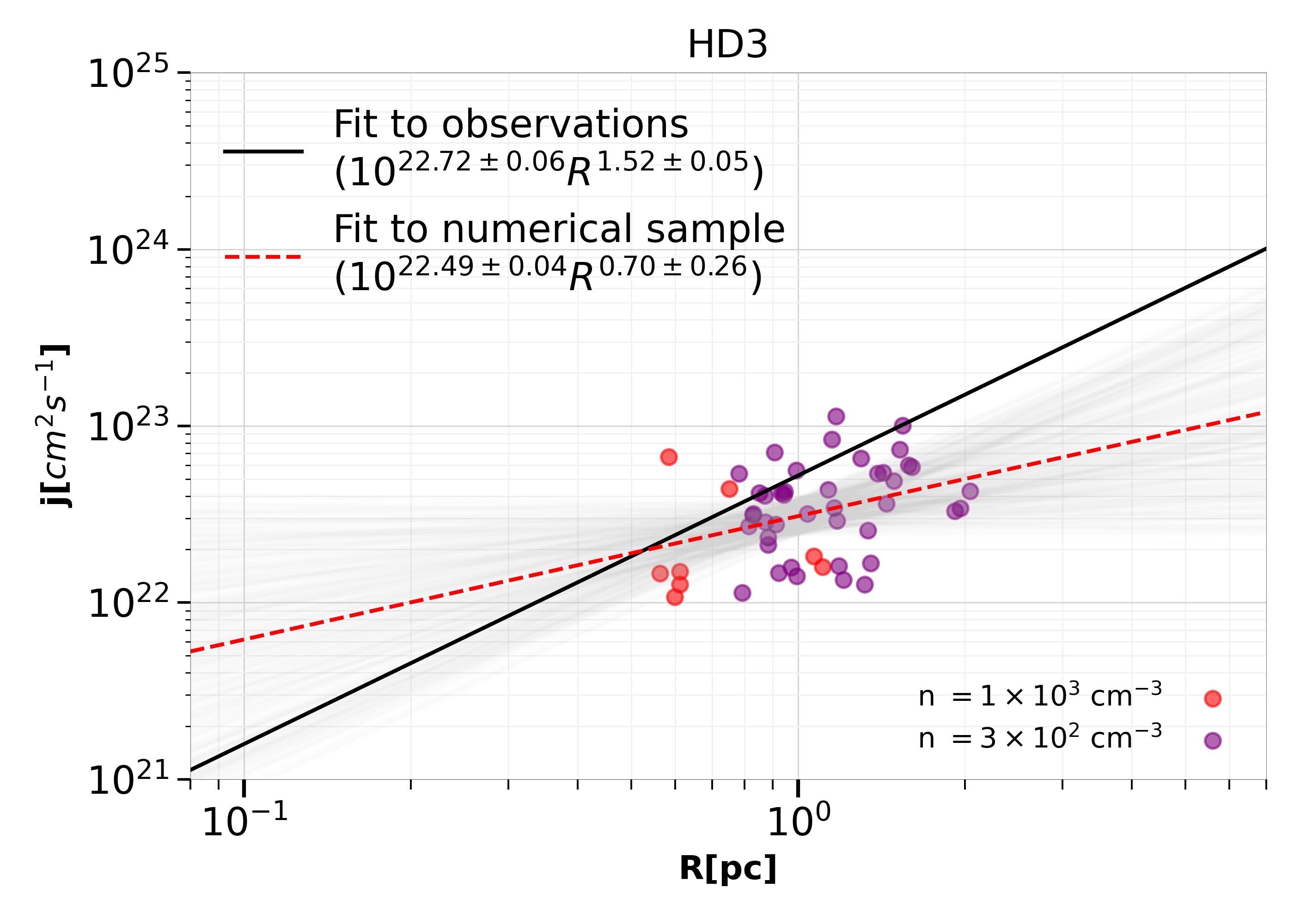}

\includegraphics[width=0.48\linewidth]{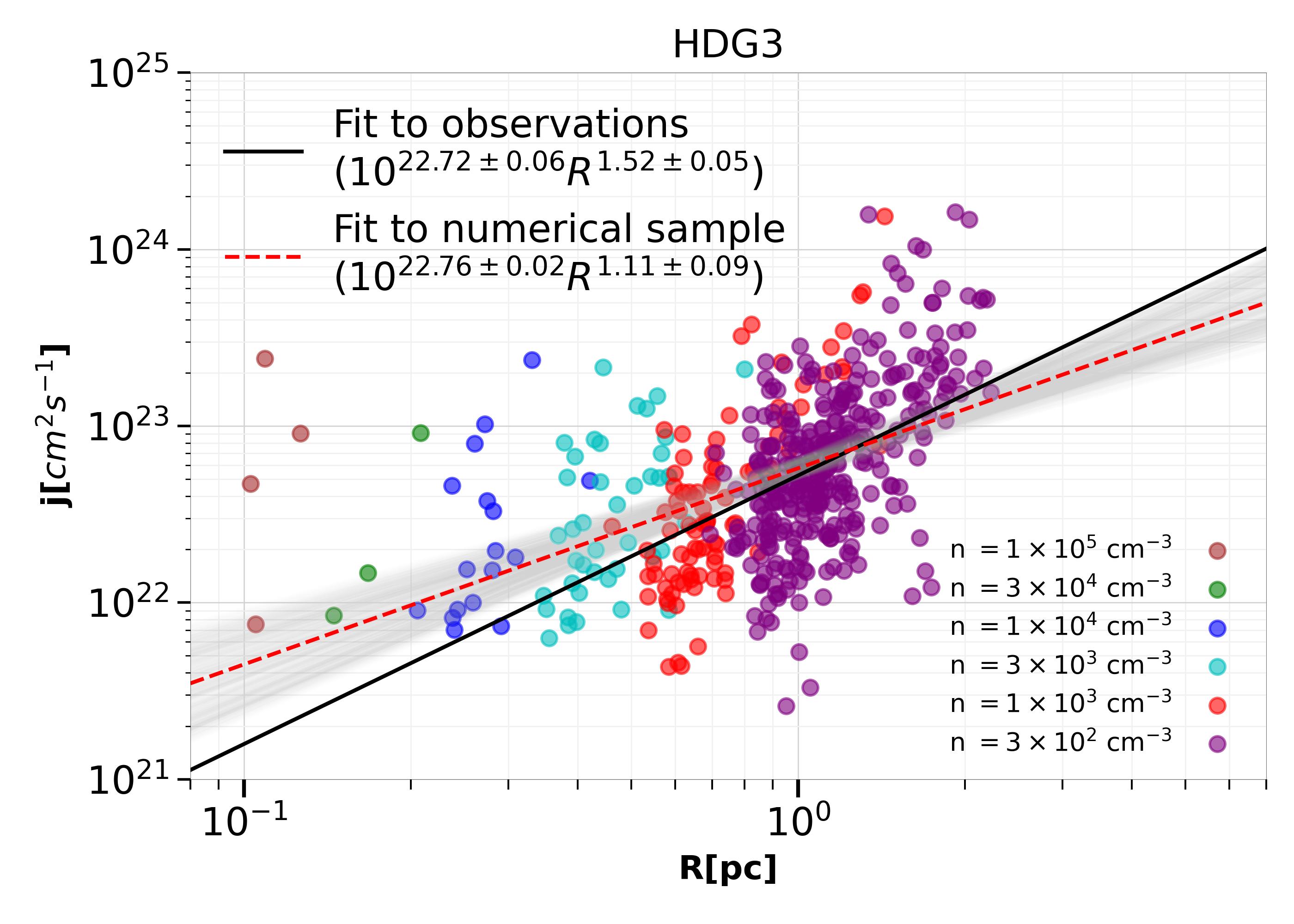}
\includegraphics[width=0.48\linewidth]{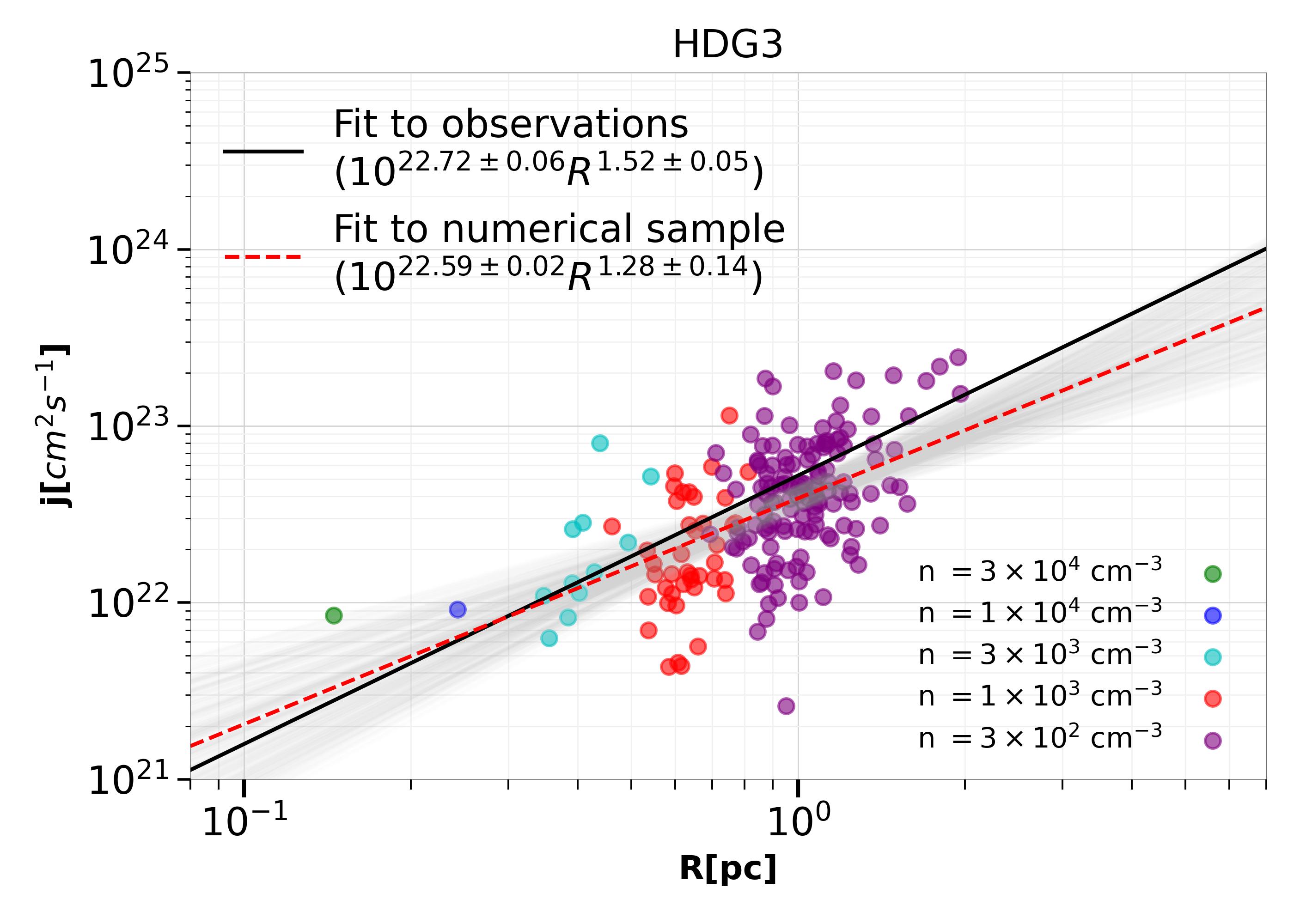}

\includegraphics[width=0.48\linewidth]{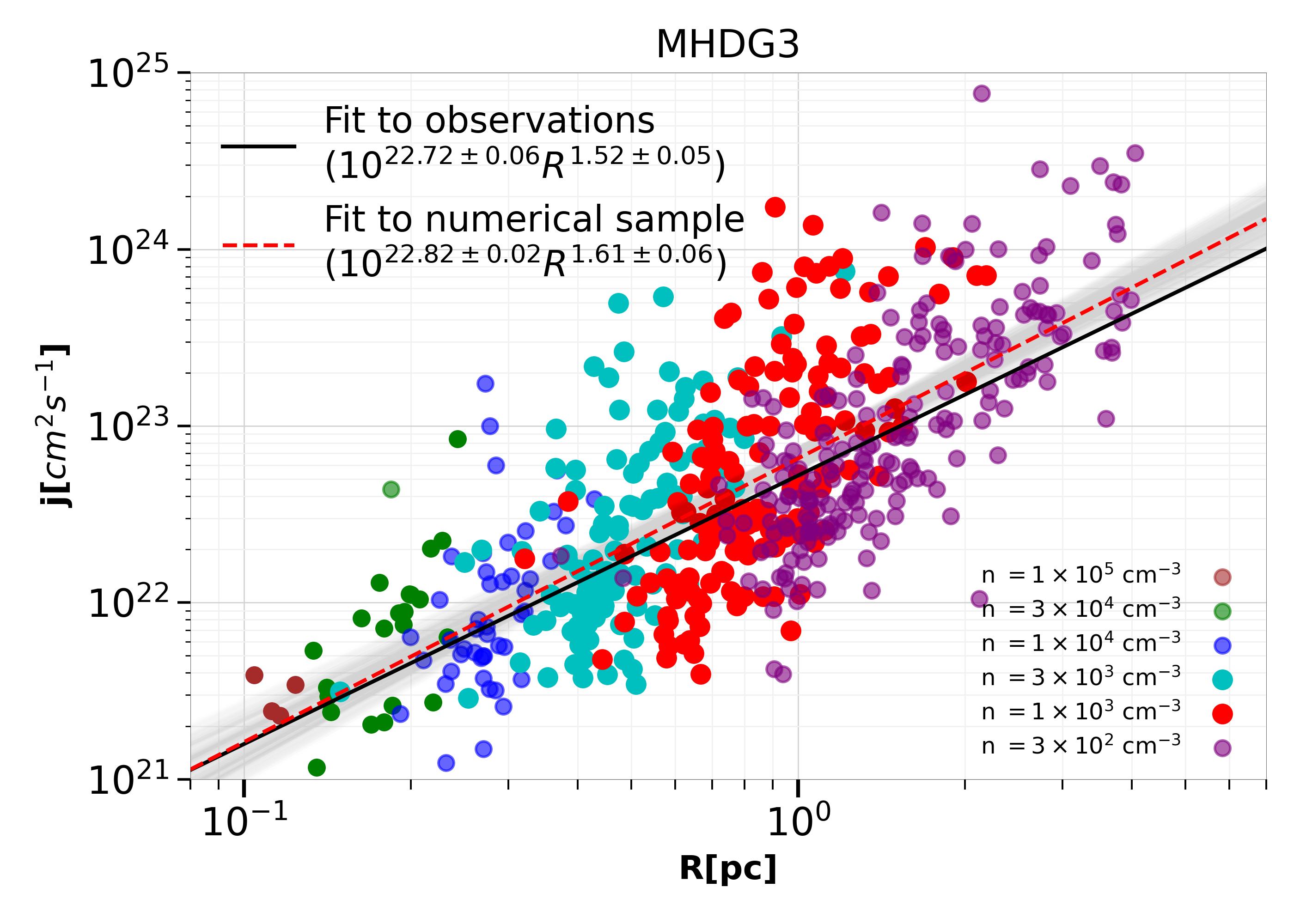}
\includegraphics[width=0.48\linewidth]{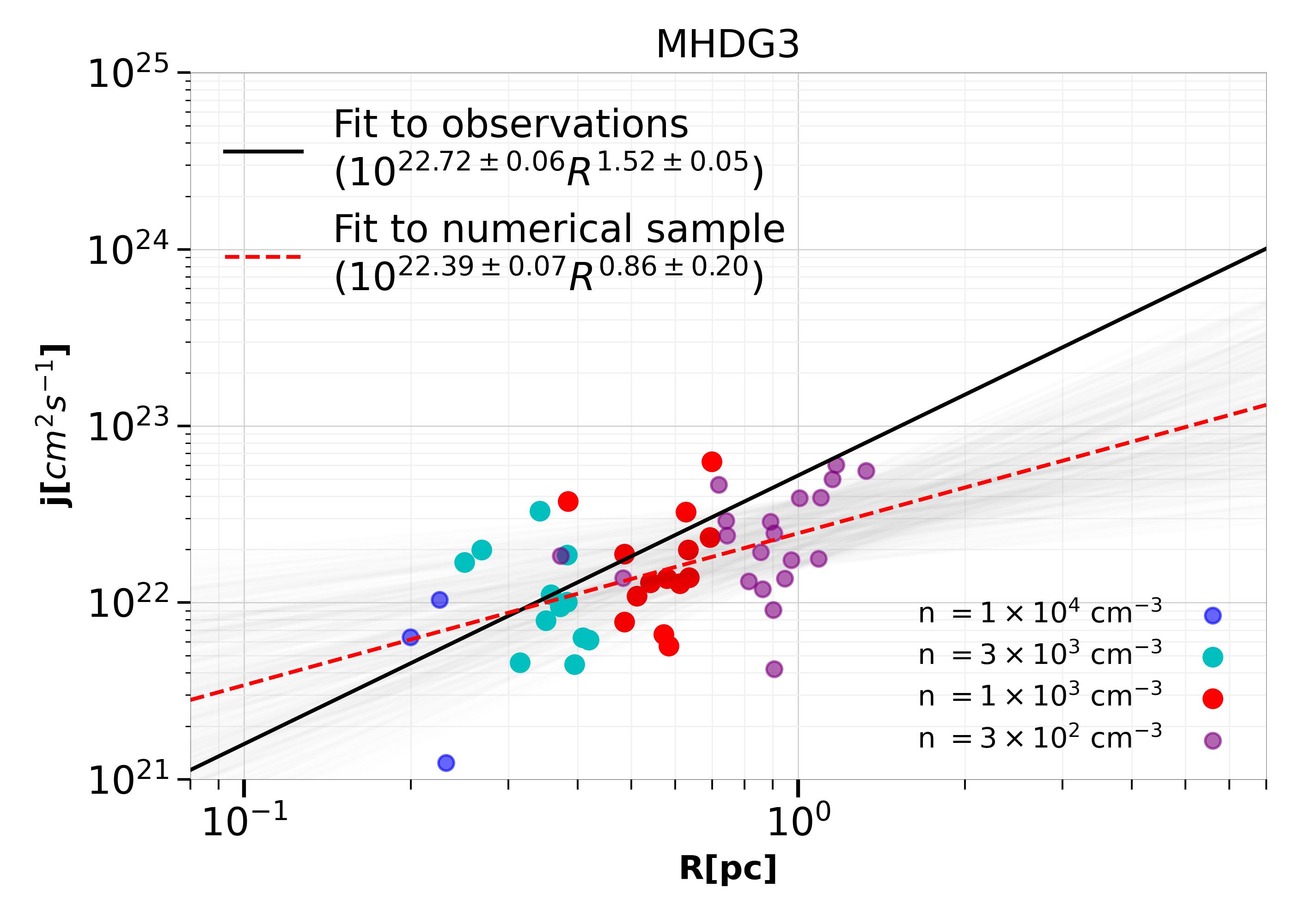}

 \caption{{\it Left column}: plots of the \jR\ relation for the three full numerical clump samples corresponding to each simulation. {\it Right column}: plots of the \jR\ relation for the three reduced numerical clump samples, i.e., after removing structures with aspect ratios $>3$. The density thresholds is represented by the color code. The solid black line represents the fitting to the observational sample compiled in Figure 1 of \citet{Arroyo-Chavez.Vazquez-Semadeni2022}, while the dashed red line represents the fitting to the numerical clump samples for the three simulations. The shaded gray region represents a variation in fitting parameters of 1$\sigma$. The \jR\ relation is most closely reproduced for the reduced sample of the HDG3 simulation.}
\label{fig:com_jR}
\end{figure*}

With the SAM and radius calculated as described in Section \ref{subsec:sample} for clumps in the three numerical samples, we show the \jR\ relation for the full clump sample for each simulation in the left column of Figure \ref{fig:com_jR}. The color code represents the density threshold used to define the clumps. The solid black line represents the fit to the observational sample compiled in \citetalias{Arroyo-Chavez.Vazquez-Semadeni2022}, while the dashed red line represents the fit to the numerical clump samples in this paper.

The number of clumps in the numerical sample from run HD3 is considerably smaller than that from runs HDG3 and MHDG3 due to the absence of gravity. On the other hand, although it is possible to identify clumps in the HD3 simulation, structures as small as the smallest ones in run HDG3 (with radii $\sim 0.1-0.5$ pc) are not found. \citet{VS+08} similarly found that in simulations of non-magnetic isothermal turbulence in a $9$ pc box, turbulence alone cannot generate structures of the same density and size as those expected from the \citet{Larson81} scaling relations. 

As gravity and magnetic fields are progressively included from run HD3 to run MHDG3, the range of clump sizes increases. In the case of the HD3 simulation, due to its small dynamic range of sizes, a power-law fit is not representative of the sample, and therefore neither is the value of its slope. However, in general the points are located around the observational trend.

It is noteworthy that the clumps in the full sample of the HDG3 simulation (Figure \ref{fig:com_jR}, middle panel in left column) exhibit a behavior not previously observed in \citetalias{Arroyo-Chavez.Vazquez-Semadeni2022}, namely that some clumps defined with the highest density thresholds (and thus with the smallest sizes) have values of the specific angular momentum comparable to that of low-density clumps, deviating from the observed relation, thus seemingly conserving angular momentum during their contraction. In turn, this causes that, although a linear fit is shown in the plot, it is not representative of the sample. On the other hand, clumps in the full sample of MHDG3 simulation appear to follow a clear trend, although they present a steeper slope than that observed in the HD3 and HDG3 simulations. 

\subsubsection{Effect of filamentary structures and their removal}
\label{subsibsec:Effect and removal of filamentary structures}

It is important to mention that clumps in the three simulations generally tend to maintain their filamentary structure over time, as can be seen in Figure \ref{fig:snaps_sims}. Therefore, it is possible that we are considering clumps for which the calculation of the angular momentum with respect to a single point (the center of mass in this case) rather than with respect to an axis may not be the most appropriate.  

In order to reduce the effect of elongated structures, we discard those structures with aspect ratios $A > 3$ for the three simulations. We will refer to the samples of remaining clumps as the {\it reduced samples}. The new set of \jR\ plots for the reduced samples is presented in the right column of Figure \ref{fig:com_jR}.

\begin{figure}
\centering
\includegraphics[width=\linewidth]{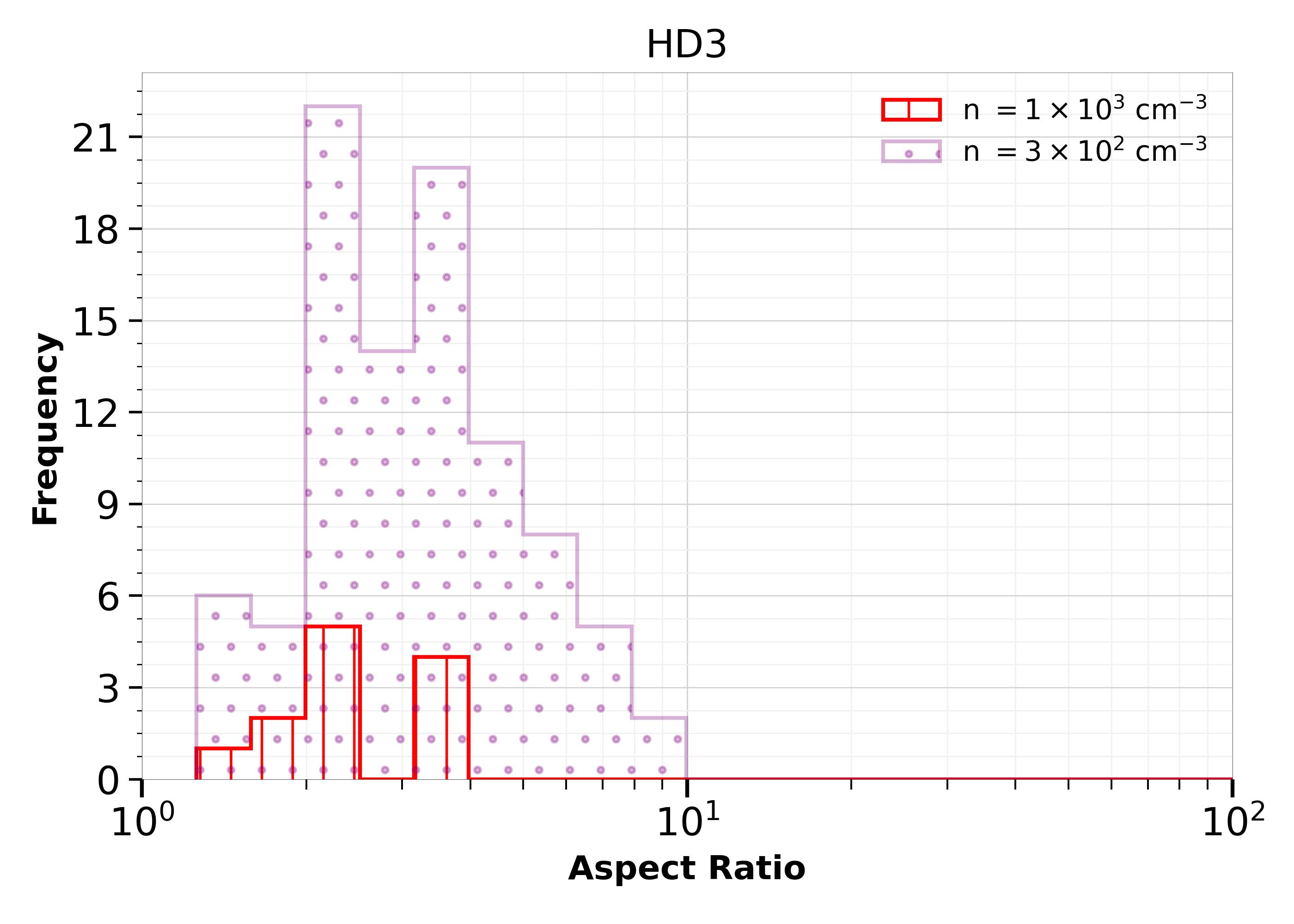}
\includegraphics[width=\linewidth]{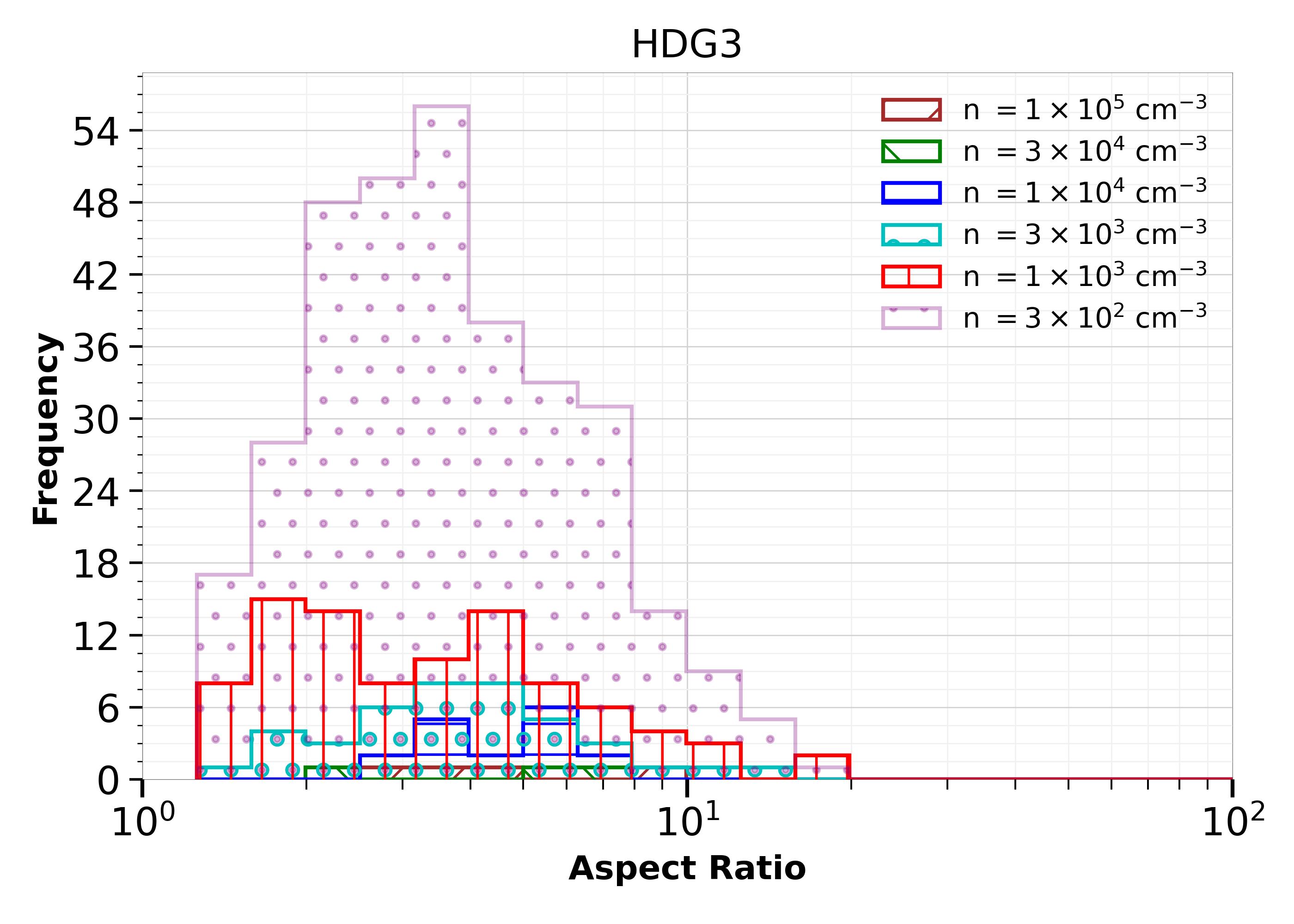}
\includegraphics[width=\linewidth]{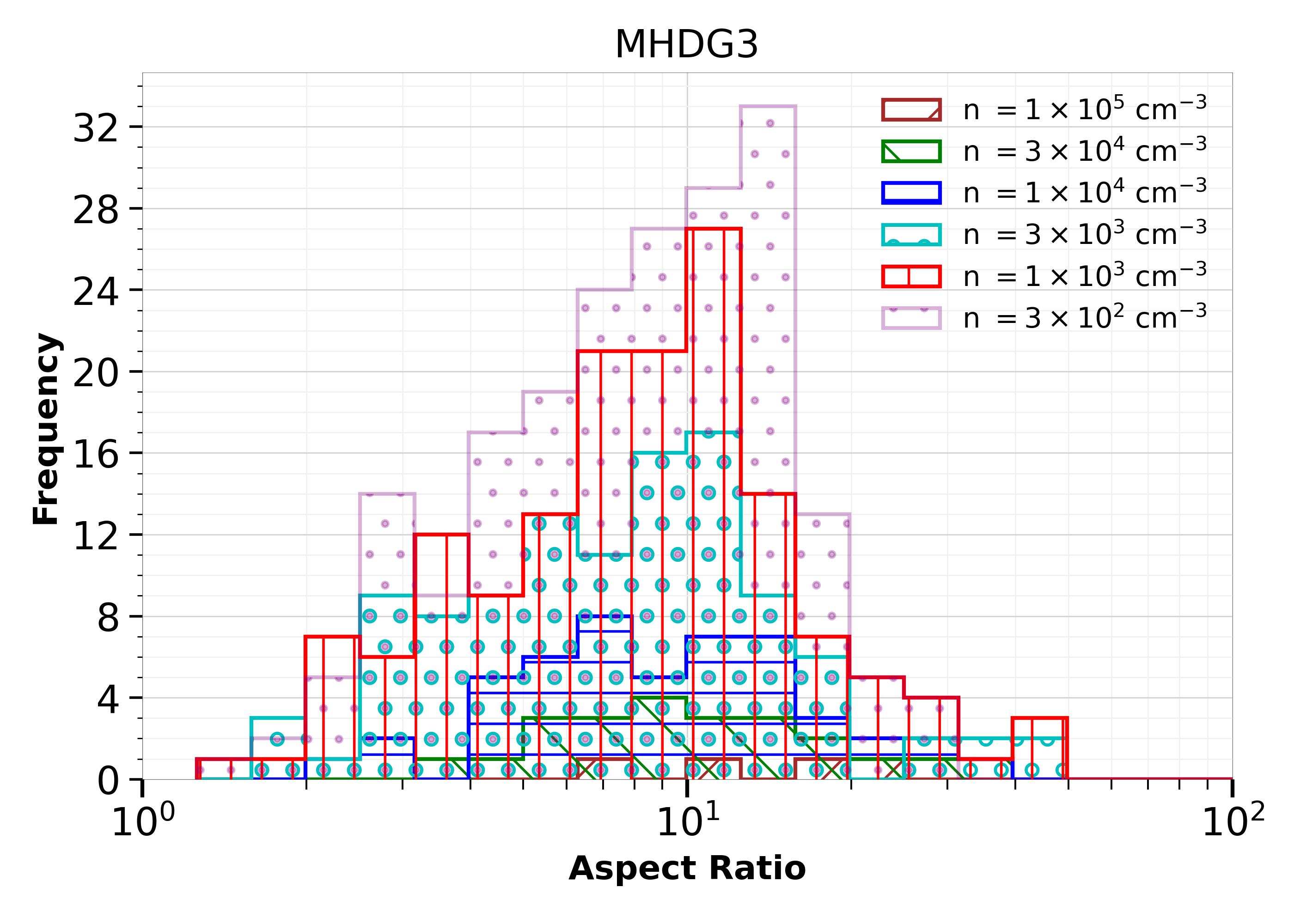}
 \caption{Histogram of aspect ratios obtained as the ratio between the largest and shorted principal axis of inertia for the full sample of clumps in simulations HD3, HDG3, and MHDG3. Color represents the density threshold used to define the clumps samples, following a similar color patter as in Figure \ref{fig:com_jR}.}
 \label{fig:aspect hist}
\end{figure}

The number of discarded clumps in the reduced samples appears to increase progressively as we add gravity and magnetic field. This can also be seen in Figure \ref{fig:aspect hist}, where we present the aspect ratio histograms for the clumps defined at each density threshold in the three simulations. The color represents the density threshold, following the same color pattern as in Figure \ref{fig:com_jR}. For every density threshold, the aspect ratios increase with progressive addition of gravity and magnetic field. Furthermore, although filamentary structures can be clearly seen in the HD3 simulation (Figure \ref{fig:snaps_sims}), these are not reflected in the aspect ratio histograms given the thresholds selected here, indicating that they consist of more diffuse gas. 

As can be seen from the \jR\ diagrams for the reduced samples, by removing elongated structures, we are mainly removing clumps with large specific angular momentum values in the HD3 and HDG3 simulations. The clumps in the reduced sample for the HD3 simulation are distributed around the black line that represents the observed relation, while the clumps with excessively large values of $j$ for each density threshold in the HD3 and HDG3 simulations have been removed. In other words, in both simulations, those structures that deviate the most from the observed relation tend to be mostly filaments. At the same time, some of the clumps defined at $n_{\rm th} = 3 \times 10^{2}\, \pcc$ with the smallest $j$ values are also removed. Therefore, the removal of the elongated structures in general decreases the scatter. 

Our results are consistent with \citetalias{Arroyo-Chavez.Vazquez-Semadeni2022} in the sense that the observed \jR\ relation is better recovered in the HDG3 simulation with the reduced sample, i.e., including only turbulence and gravity, and without considering elongated structures. However, it is important to note that the clumps in the HD3 simulation have a magnitude of $j$ that is not far from the expected values, even in the absence of gravity or magnetic field, suggesting that the angular momentum transport mechanism, which in this case would involve only turbulence and the pressure gradient, is capable of recovering the expected magnitude of $j$. 

Self-gravity, however, seems to be fundamental for expanding the dynamic range of sizes necessary to recover the observed relation which, otherwise, turbulence by itself cannot achieve.

\subsection{The \jR\ relation and the turbulent velocity dispersion}
\label{subsec:Deviations}

In the previous section we showed that most of the structures with atypically large values of $j$ in the HD3 and HDG3 simulations are mainly filaments. One possible explanation may be based on the geometry of a filament itself. Filaments with large aspect ratios will have regions farther from their center of mass compared to a round clump. These regions might have a larger contribution to the angular momentum simply by having a larger position vector despite the velocities being comparable to those of a round clump. On the other hand, there is also the possibility that the velocity of the gas inside the filaments is intrinsically larger. Both effects could come into play to different extents given the conditions in each simulation.

\begin{figure*}
\centering
\includegraphics[width=0.33\linewidth]
{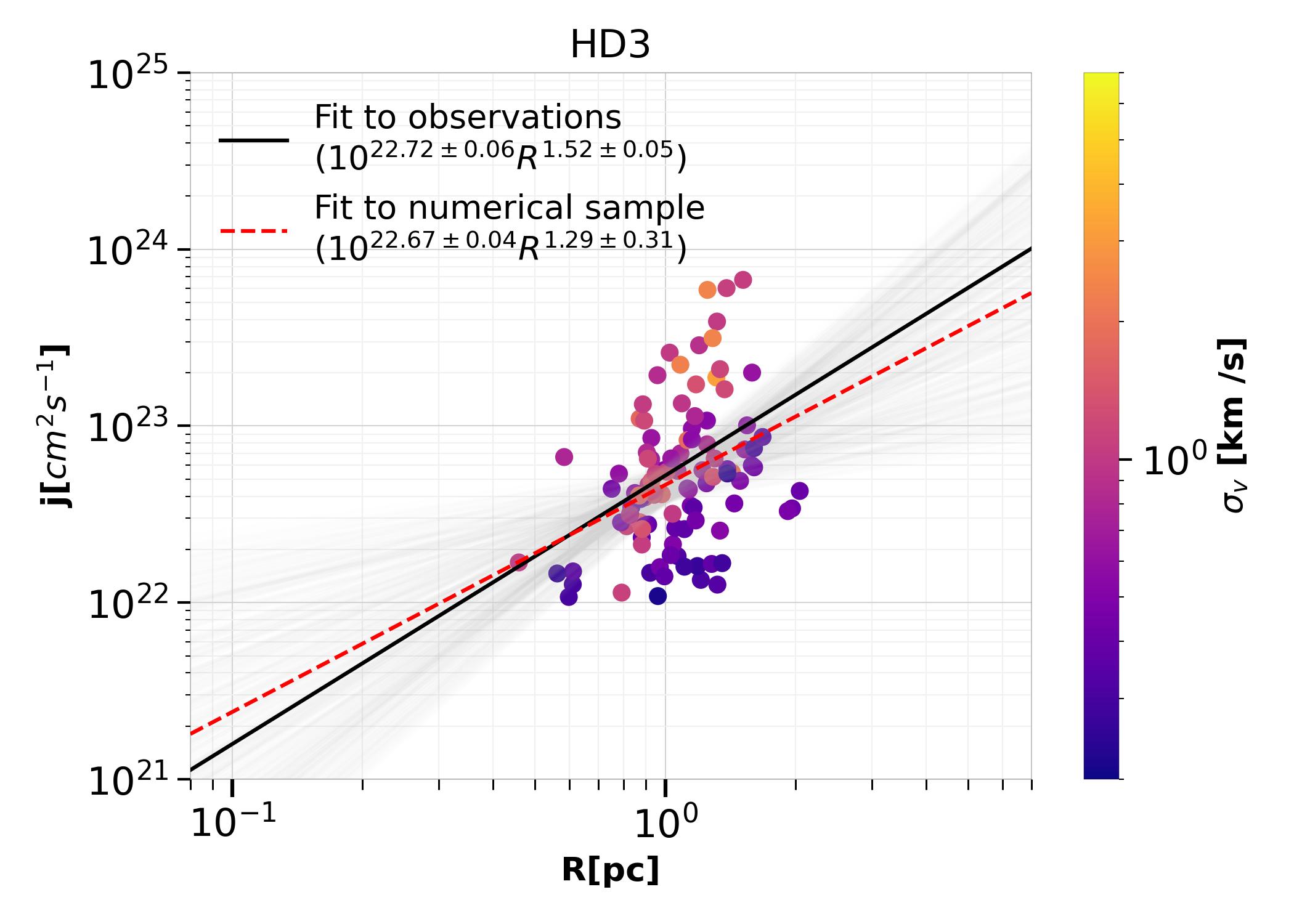}
\includegraphics[width=0.33\linewidth]{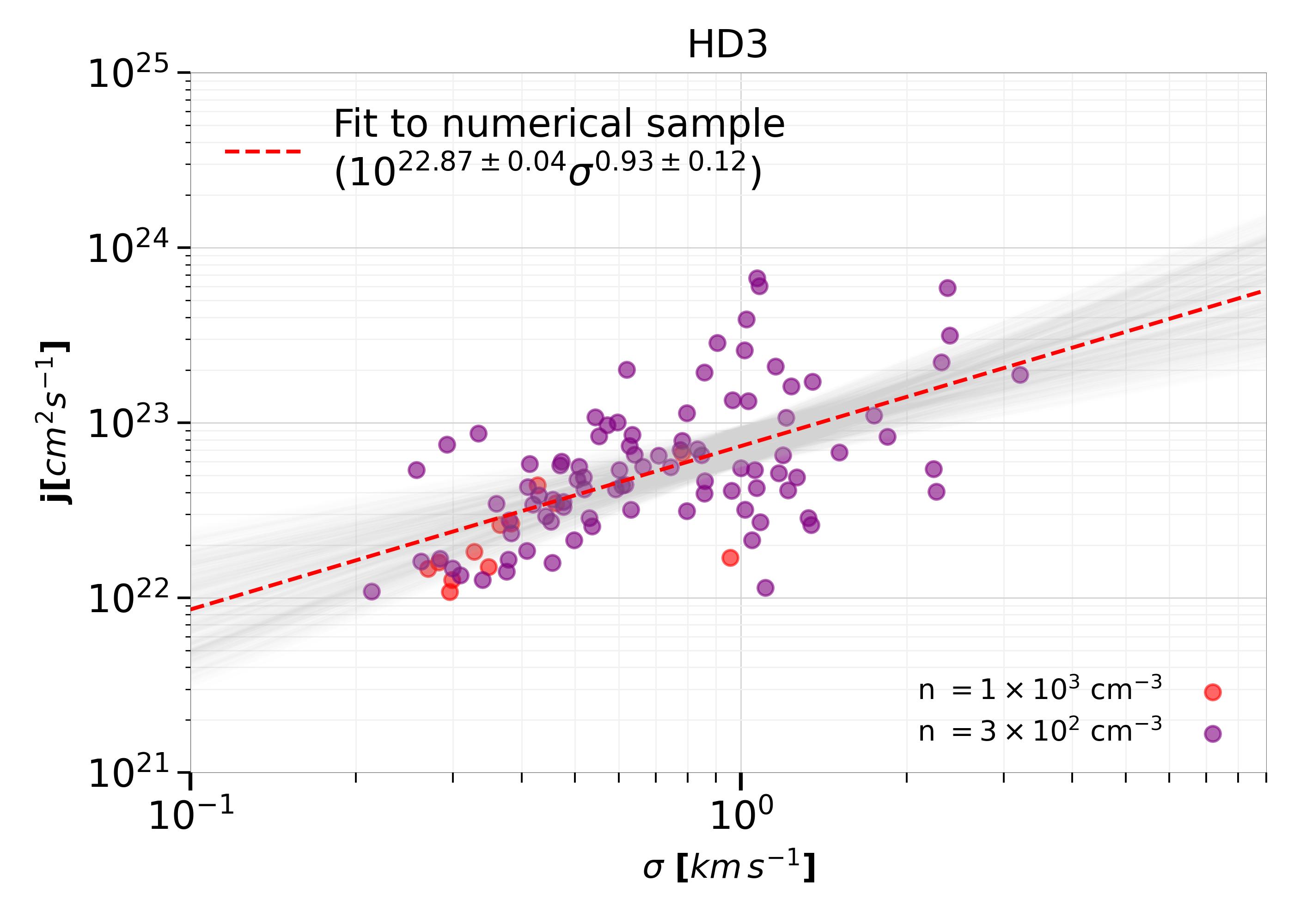}
\includegraphics[width=0.33\linewidth]
{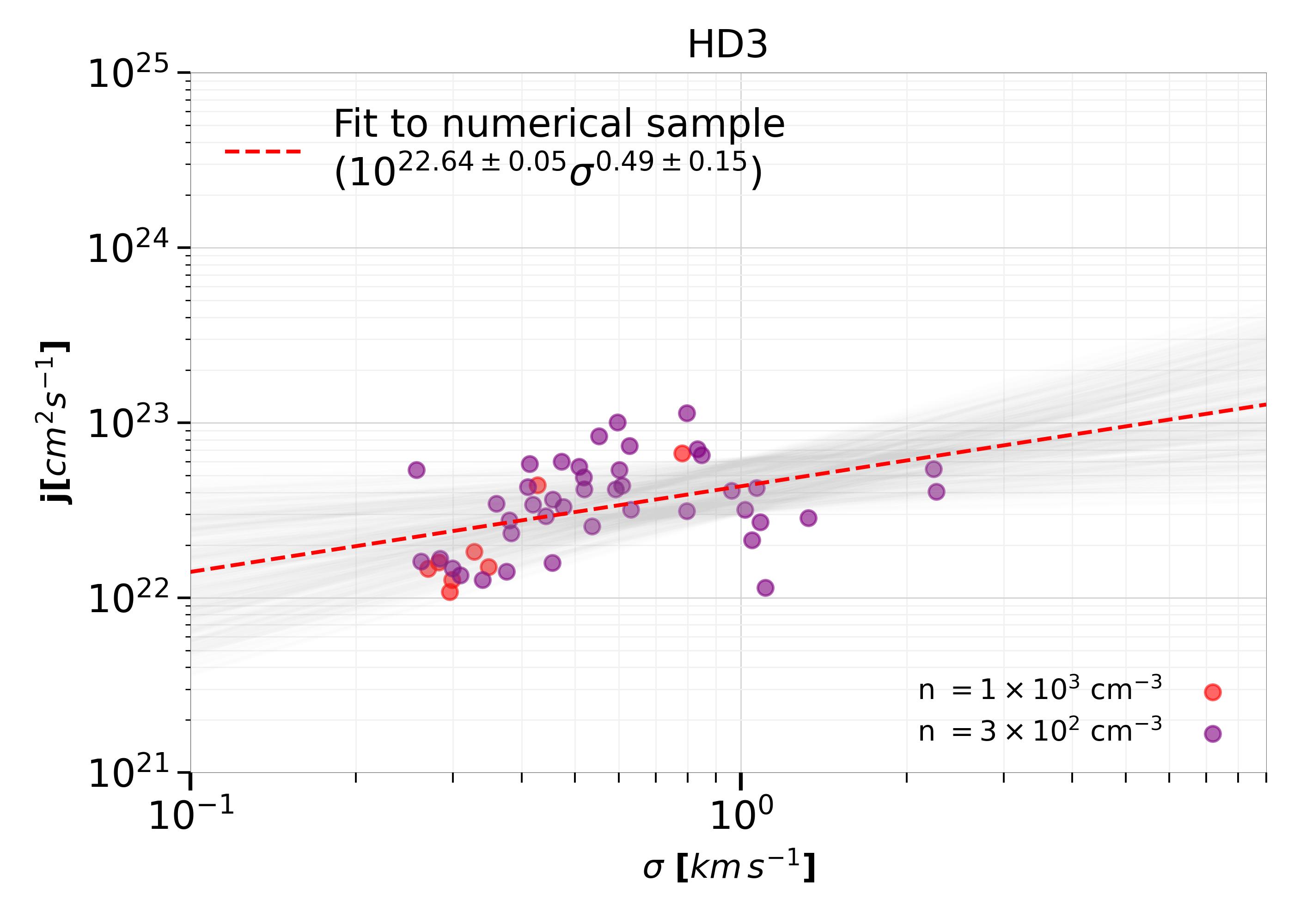}

\includegraphics[width=0.33\linewidth]
{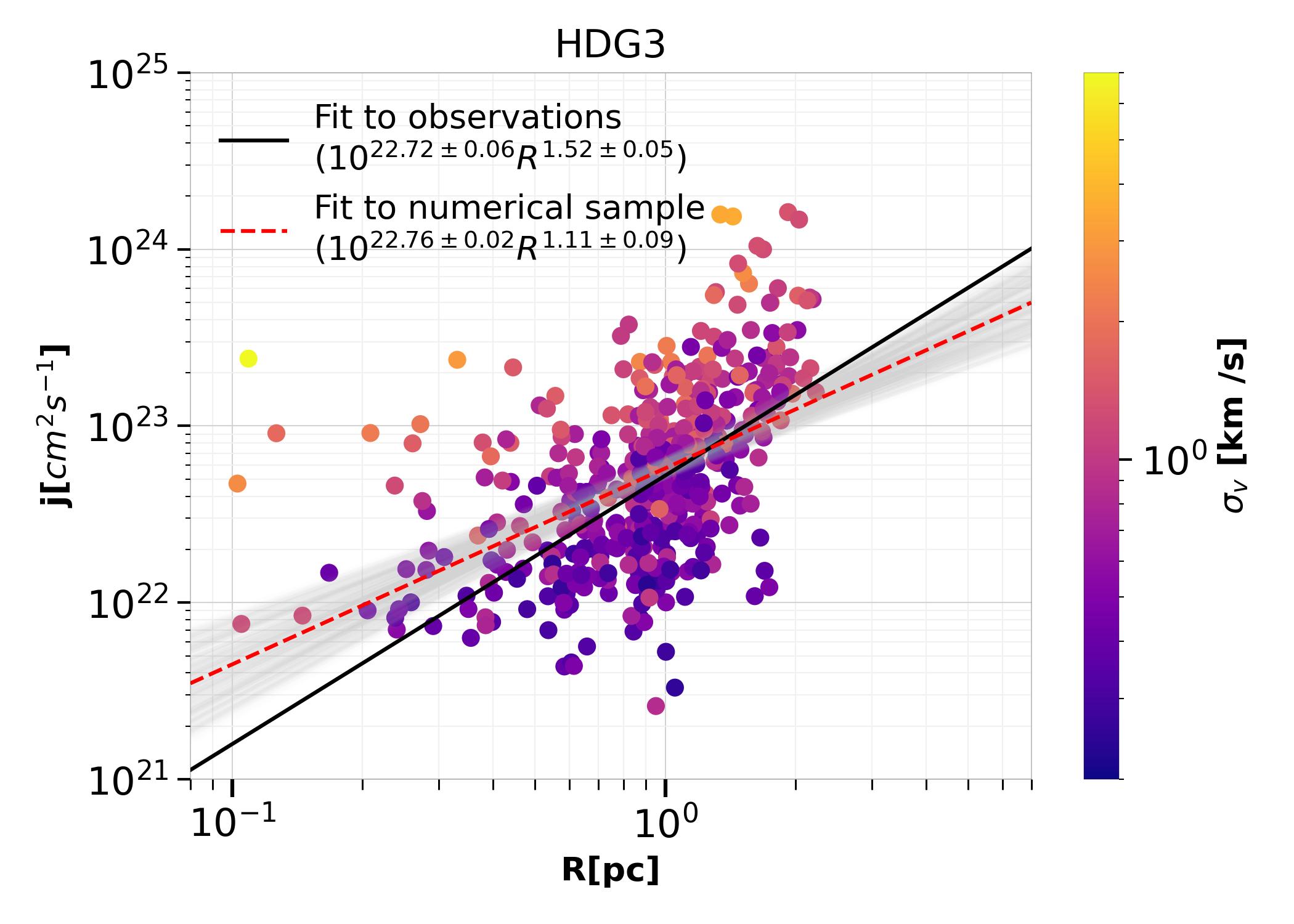}
\includegraphics[width=0.33\linewidth]{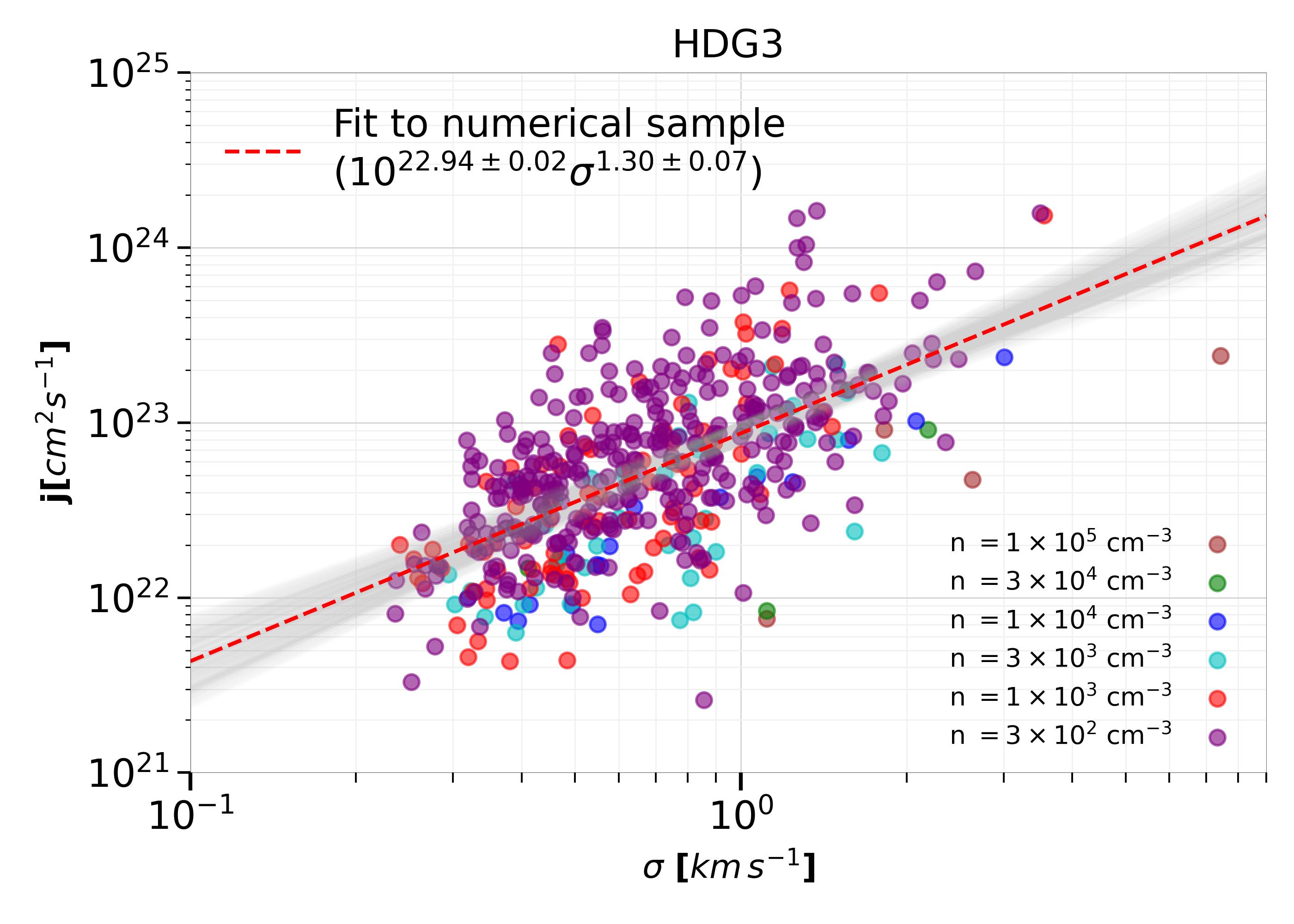}
\includegraphics[width=0.33\linewidth]{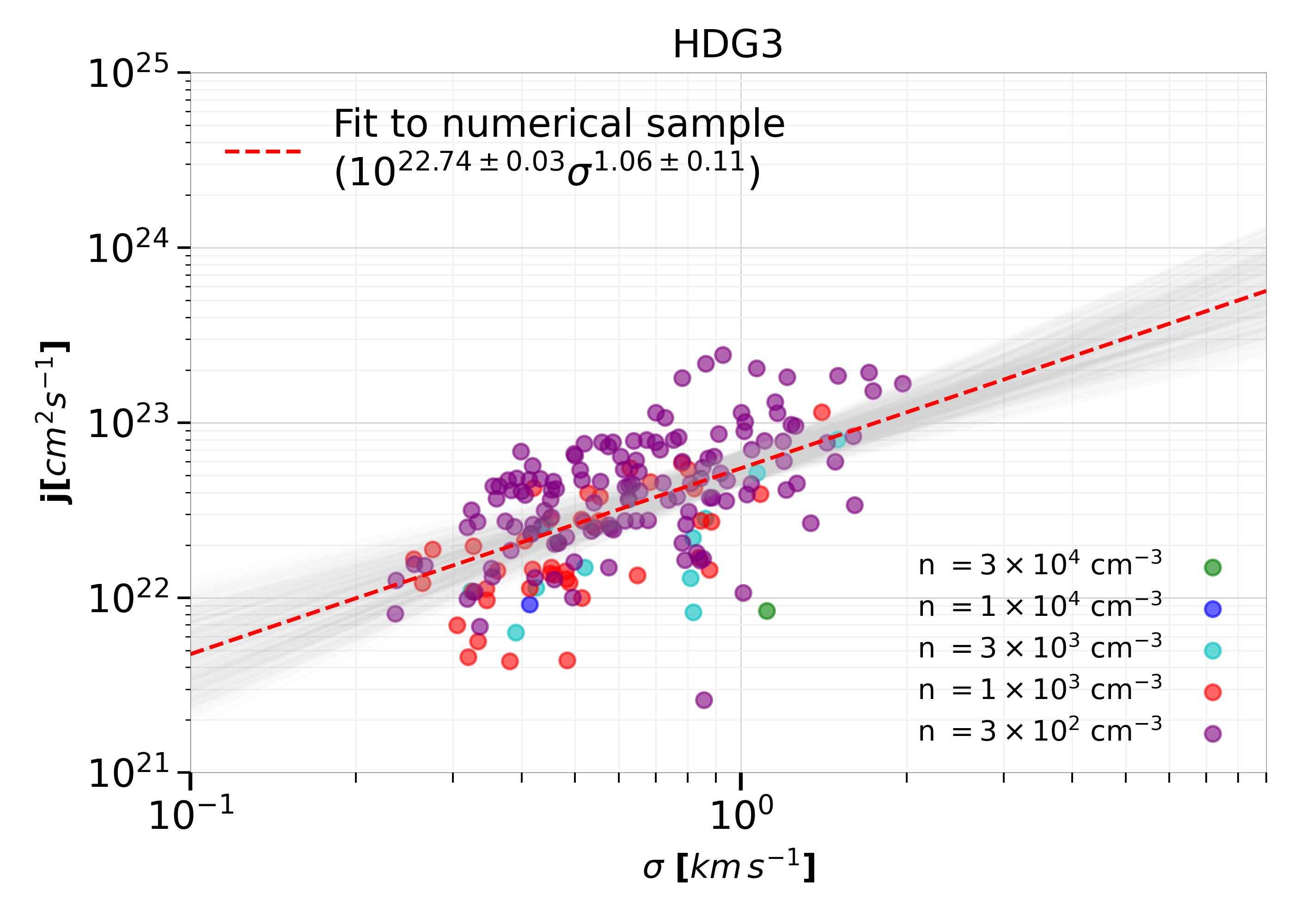}

\includegraphics[width=0.33\linewidth]
{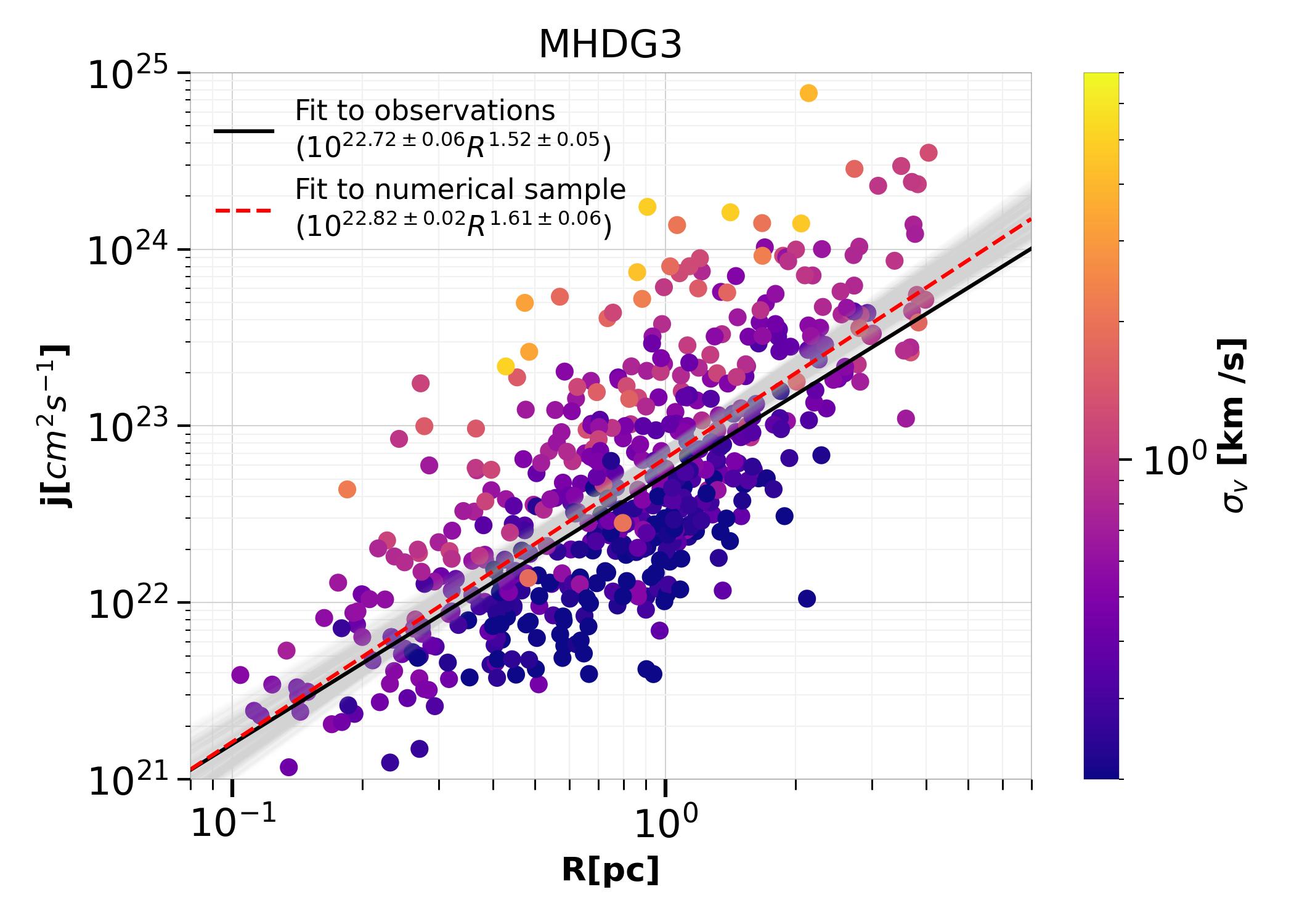}
\includegraphics[width=0.33\linewidth]{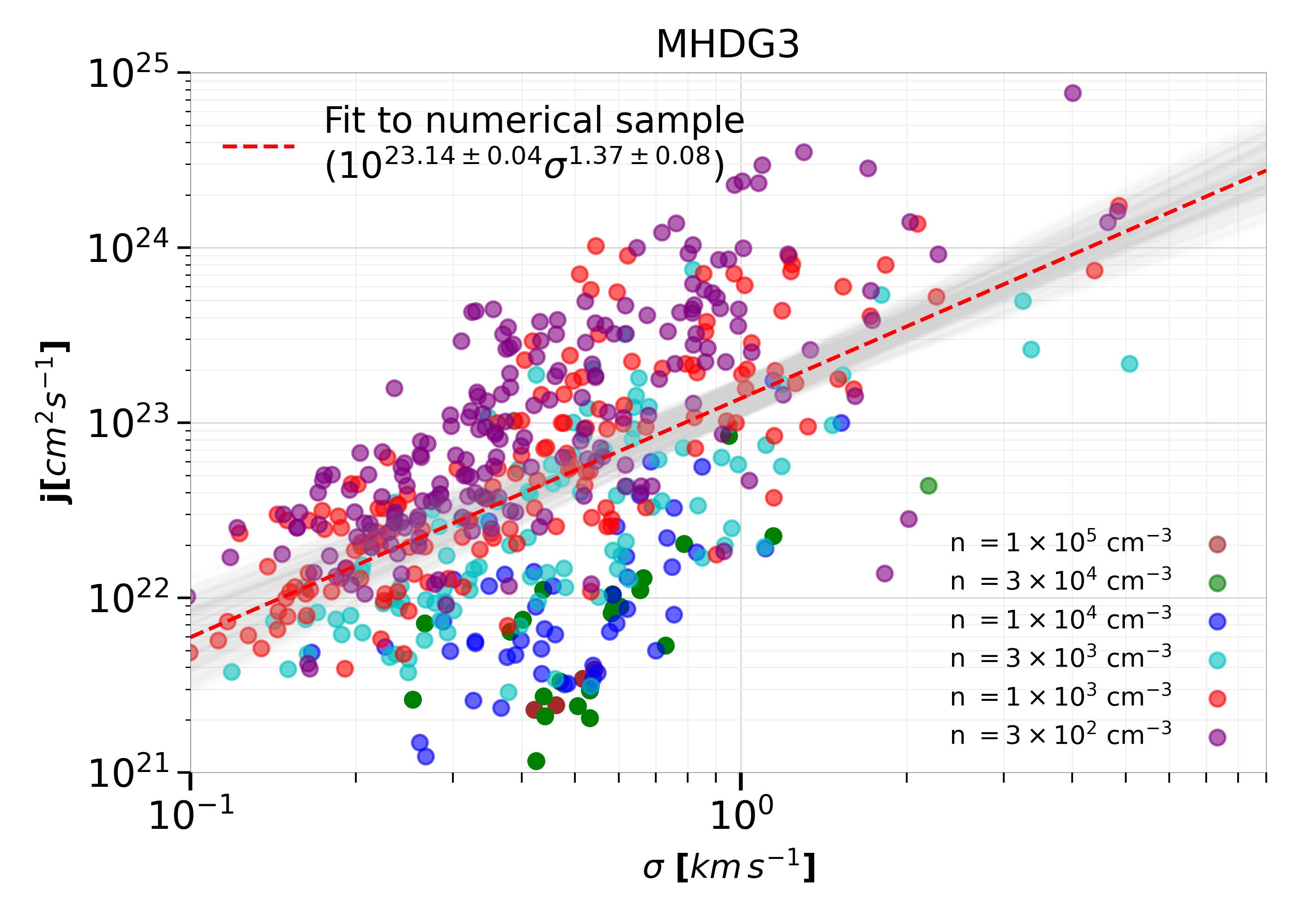}
\includegraphics[width=0.33\linewidth]{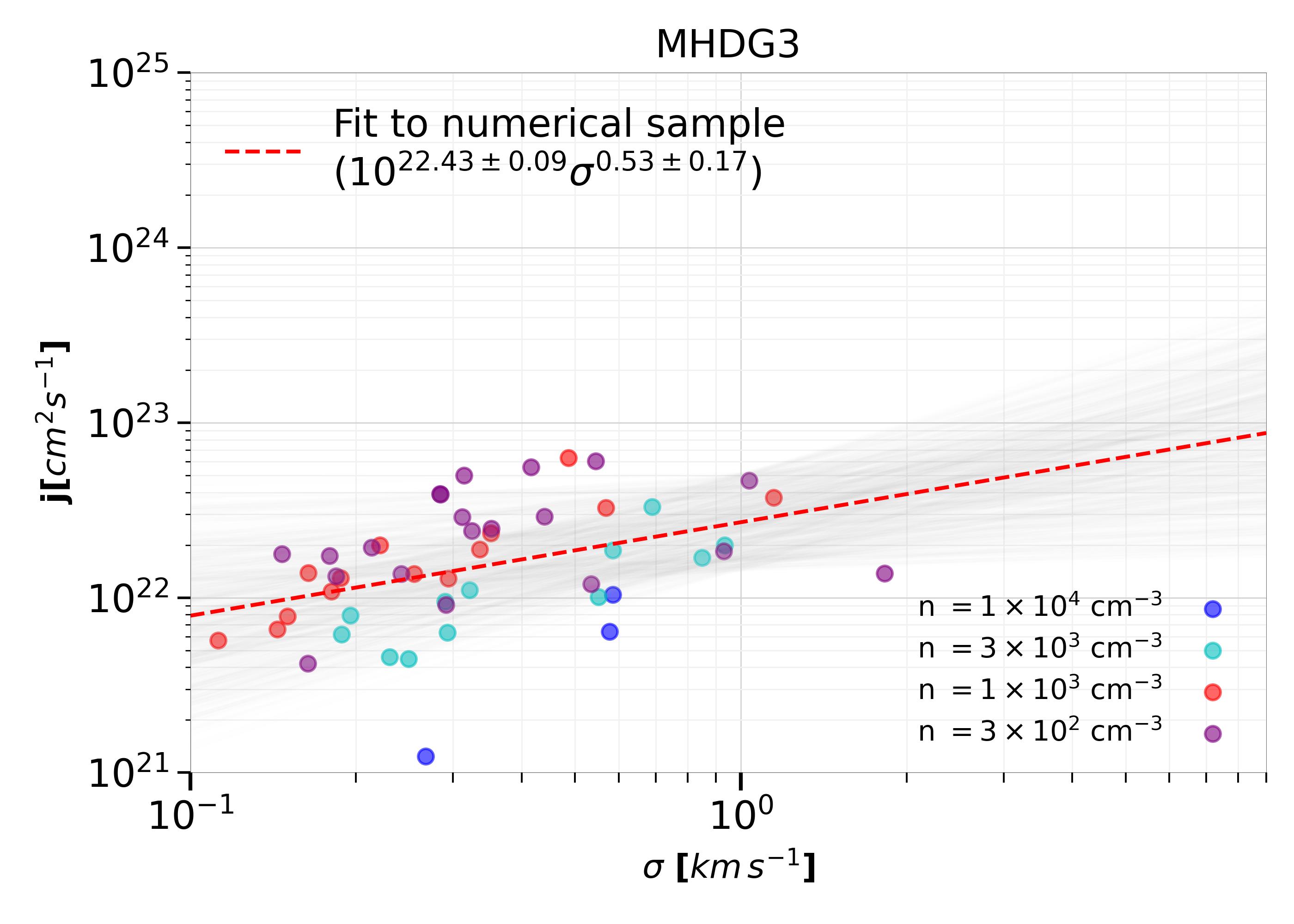}

 \caption{{\it Left column}: \jR\ relation for the full samples in each simulation (as in the left column of Figure \ref{fig:com_jR}), where the color now represents the value of the 3D velocity dispersion, $\sigma_{v}$. {\it Middle column}: $j$ as a function of the velocity dispersion, $\sigma_{v}$, for the full samples. {\it Right column}: same as in the middle column for the reduced samples. Clumps with larger $j$ values that deviate further from the observational trend in the HD3 and HDG3 simulations, are those with a higher velocity dispersion, and at the same time, are the structures with larger aspect ratios. A clear correlation in the $j$-$\sigma_{v}$ plots can be observed for all three simulations, which also seems to have a dependence on the density threshold, and the slope becomes steeper as gravity and magnetic field are added.}
\label{fig:j-sigma}
\end{figure*}

To identify the possible origin of the deviations in the \jR\ plot for our clump samples, in the left column of Figure \ref{fig:j-sigma} we show the \jR\ relation for the full samples in each simulation (as in the left column of Figure \ref{fig:com_jR}), where the color now represents the value of the 3D velocity dispersion. It is clearly seen that the clumps with larger $j$ values that deviate further from the observational trend in the HD3 and HDG3 simulations, are those with a higher velocity dispersion, and at the same time, are also the structures with larger aspect ratios, i.e., filaments (as can be seen in right column of Figure \ref{fig:com_jR}, where these clumps do not appear in the reduced sample). For these two simulations, both geometry and velocity dispersion could contribute to filamentary structures deviating from the observational relation. However, this correlation is not as clear for the MHDG3 simulation. As expected due to the inhibition of turbulence by the magnetic field, the velocity dispersion in MHDG3 does not reach values as large as in the HD3 and HDG3 simulations. In this case, the large values of the aspect ratio can compensate for the decrease in residual angular momentum caused by the low velocity dispersion in the clumps, such that both effects counteract each other, causing the overall sample not to deviate considerably from the observational trend. 

However, a third effect needs to be considered when taking the magnetic field into account. It is known that the magnetic field contributes to transferring more energy into rotational modes compared to a non-magnetic case \citep[see, e.g., Fig. 6 of][]{VS+96}. Thus, the inclusion of the magnetic field may increase the vorticity in general in the flow, and, in particular, it may organize the motions, causing the structures to have a larger angular momentum at a given value of the velocity dispersion. 

To further examine the velocity dispersion as the source of the angular momentum, in the middle column of Figure \ref{fig:j-sigma} we show $j$ as a function of the velocity dispersion, $\sigma_{v}$, for the full samples of the three simulations, while in the right column we show the same plots for the reduced samples, i.e., after removing structures with aspect ratios $A >3$. A clear correlation can be observed for all three simulations, which also seems to have a dependence on the density threshold, such that clumps defined at higher thresholds tend to have lower $j$ than those defined at lower thresholds. This dependence can be seen more clearly in the $j$-$\sigma$ plot for the full sample in the MHDG3 simulation.

It can be seen that the slope of the $j$-$\sigma$ plot increases as we progressively include gravity and magnetic field, with the exception of the reduced sample from run MHDG3, whose slope is closer to that of the HD3 run and does not continue to increase. However, for the full samples, given the same velocity dispersion, there is more specific angular momentum in the clumps of the MHDG3 simulation than in the HD3. This is consistent with the fact that the MHDG3 simulation has more filamentary structures, which are thus subject to the effect discussed in Sec.\ \ref{subsibsec:Effect and removal of filamentary structures} that calculation of the angular momentum with respect to the center of mass may overestimate it. This separation by density threshold may imply an extra dependence on this relation that we are not considering. We address this discussion below with the goal of reducing the scatter in the $j$-$\sigma$ plot.

\subsubsection{On the origin of the $j$-$\sigma$ relation.}
\label{jsigma origin}

In \citetalias{Arroyo-Chavez.Vazquez-Semadeni2022} we proposed a derivation of the dependence of $j$ with $R$ for a clump ensemble similar to that of \citet{Goodman+93}, based on assuming that the ratio of the rotational to the gravitational energy $\beta$ may approach a constant, finding
\begin{equation}
 j \approx (2\pi \beta G\Sigma)^{1/2} R^{3/2},
 \label{eq:j-R deriv}
\end{equation}
where $\Sigma$ is a clump's column density and $R$ is its radius. Introducing the Keto-Heyer relation\footnote{This relation can be considered the generalization of Larson's linewidth-size relation for non-constant-column density objects \citep[e.g.,] [] {BP+11}.} \citep{Keto_Myers86, Heyer+09},
\begin{equation}
\frac{\sigma_{v}}{R^{1/2}} \approx \left( \frac{\pi G\Sigma}{5} \right)^{1/2},
 \label{eq:Larson-Heyer}
\end{equation}
 in eq. \eqref{eq:j-R deriv}, we obtain that
\begin{equation}
j\Sigma \sim \sigma_{v}^{3}.
 \label{eq:j-Sigma-sigma}
\end{equation}

\begin{figure*}
\centering
\includegraphics[width=0.47\linewidth]{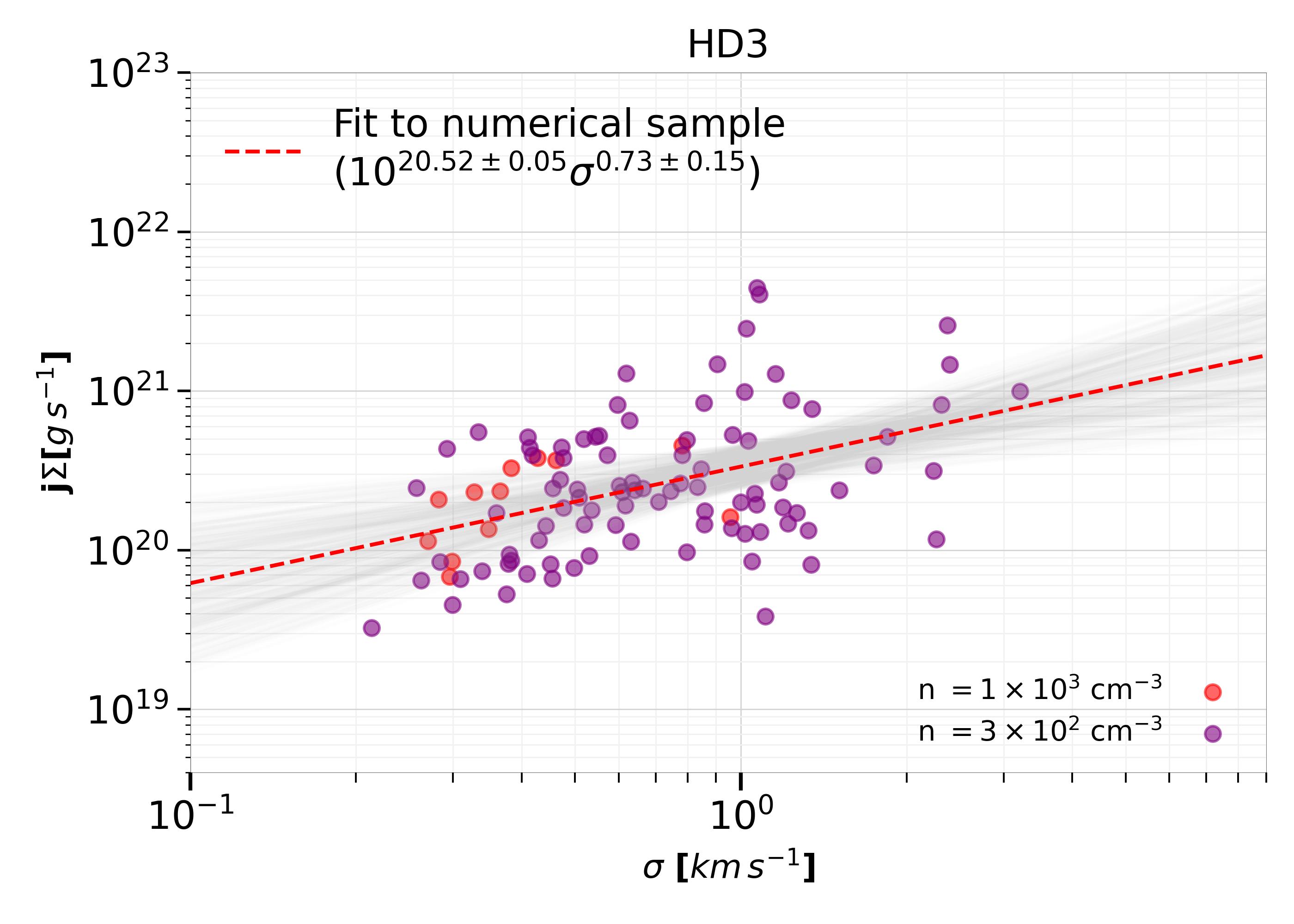}
\includegraphics[width=0.47\linewidth]
{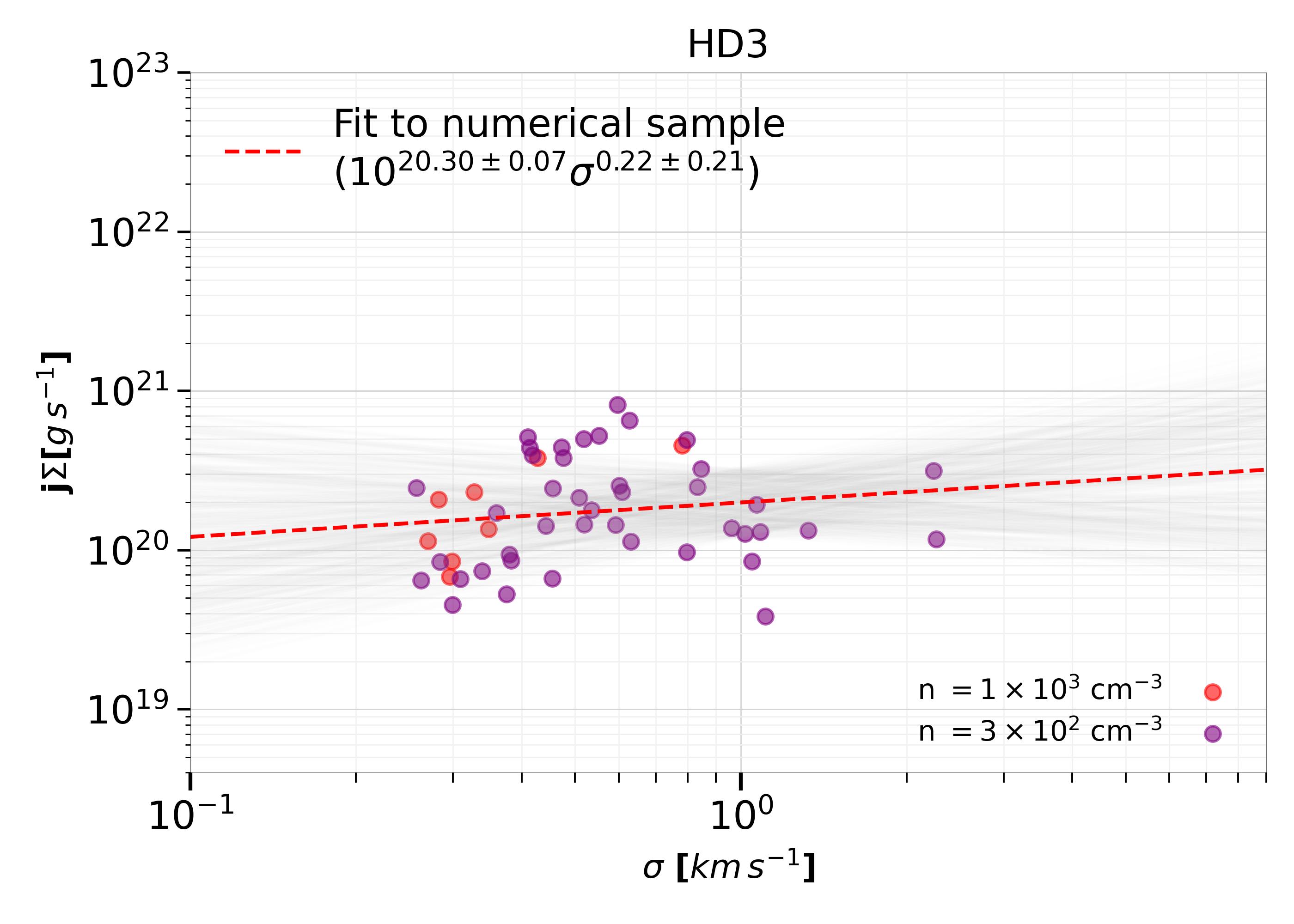}
\includegraphics[width=0.47\linewidth]{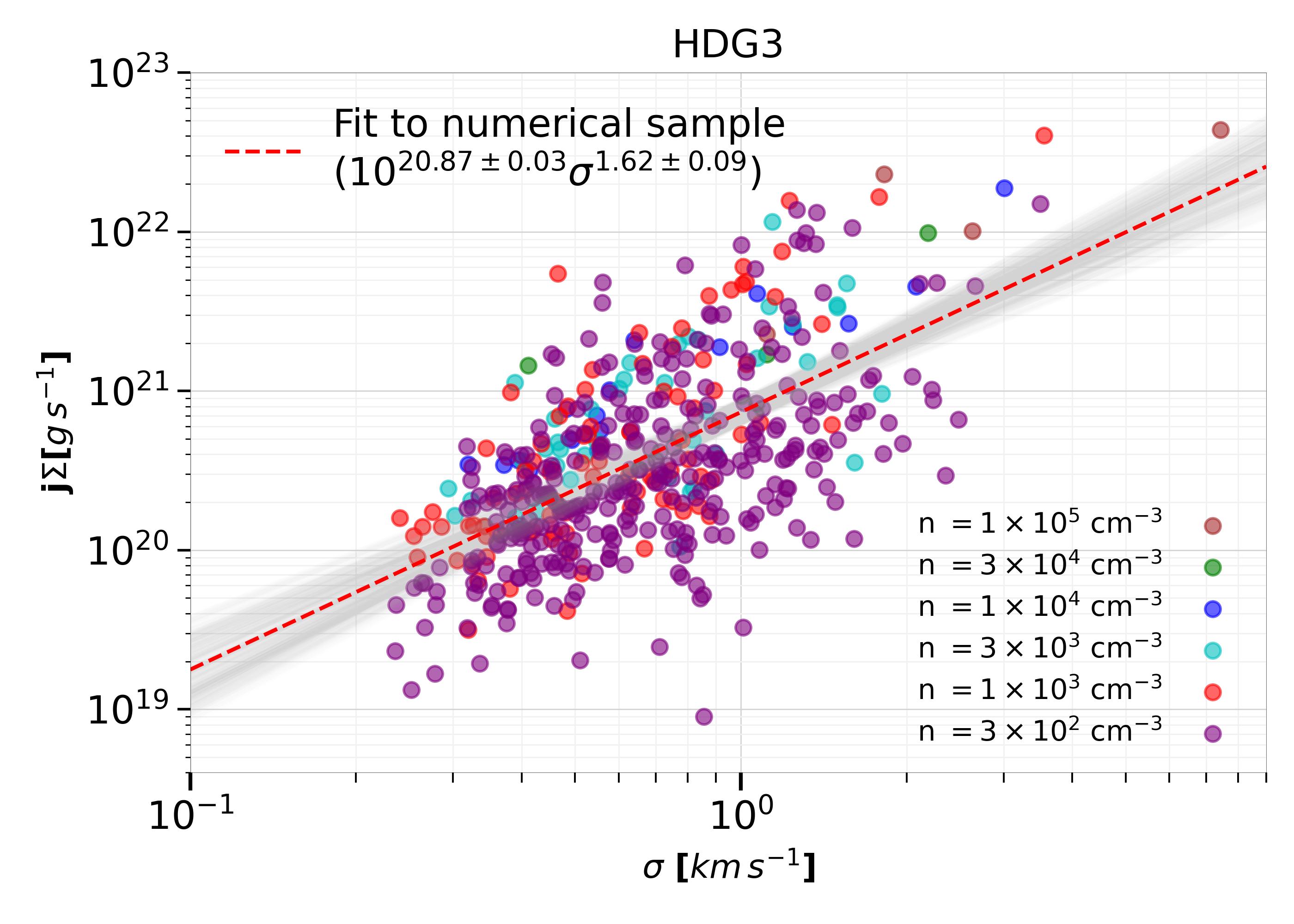}
\includegraphics[width=0.47\linewidth]{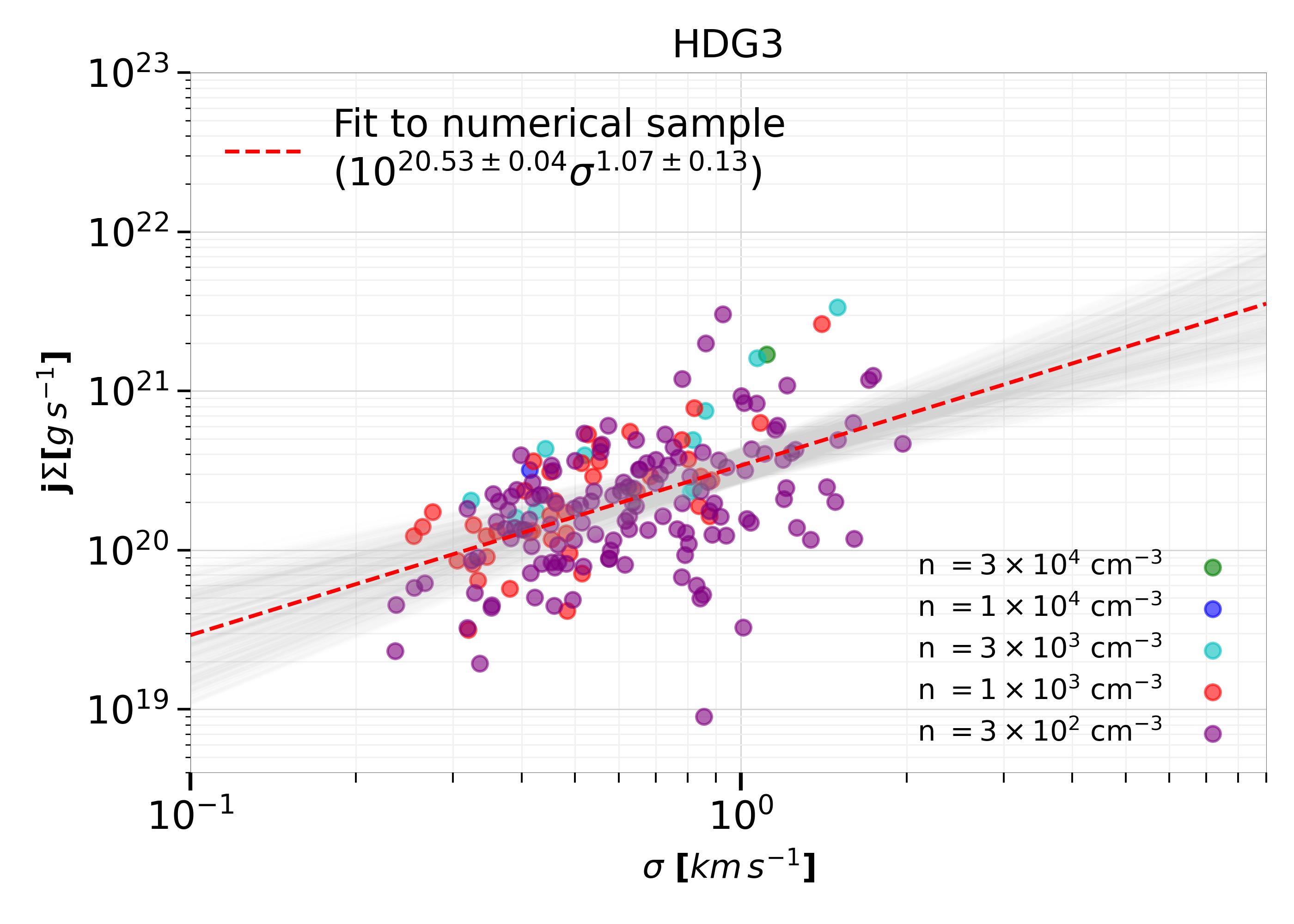}
\includegraphics[width=0.47\linewidth]{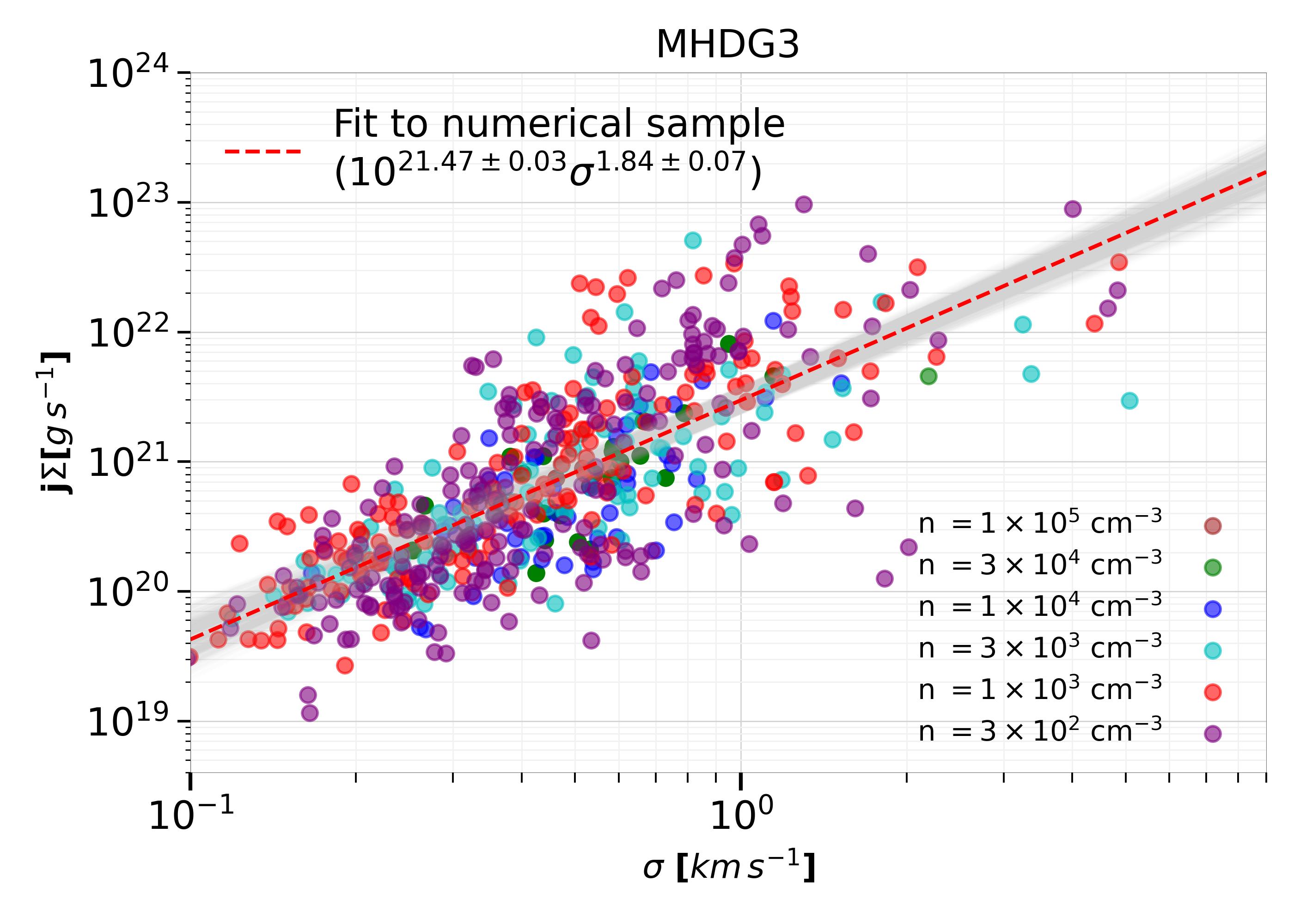}
\includegraphics[width=0.47\linewidth]{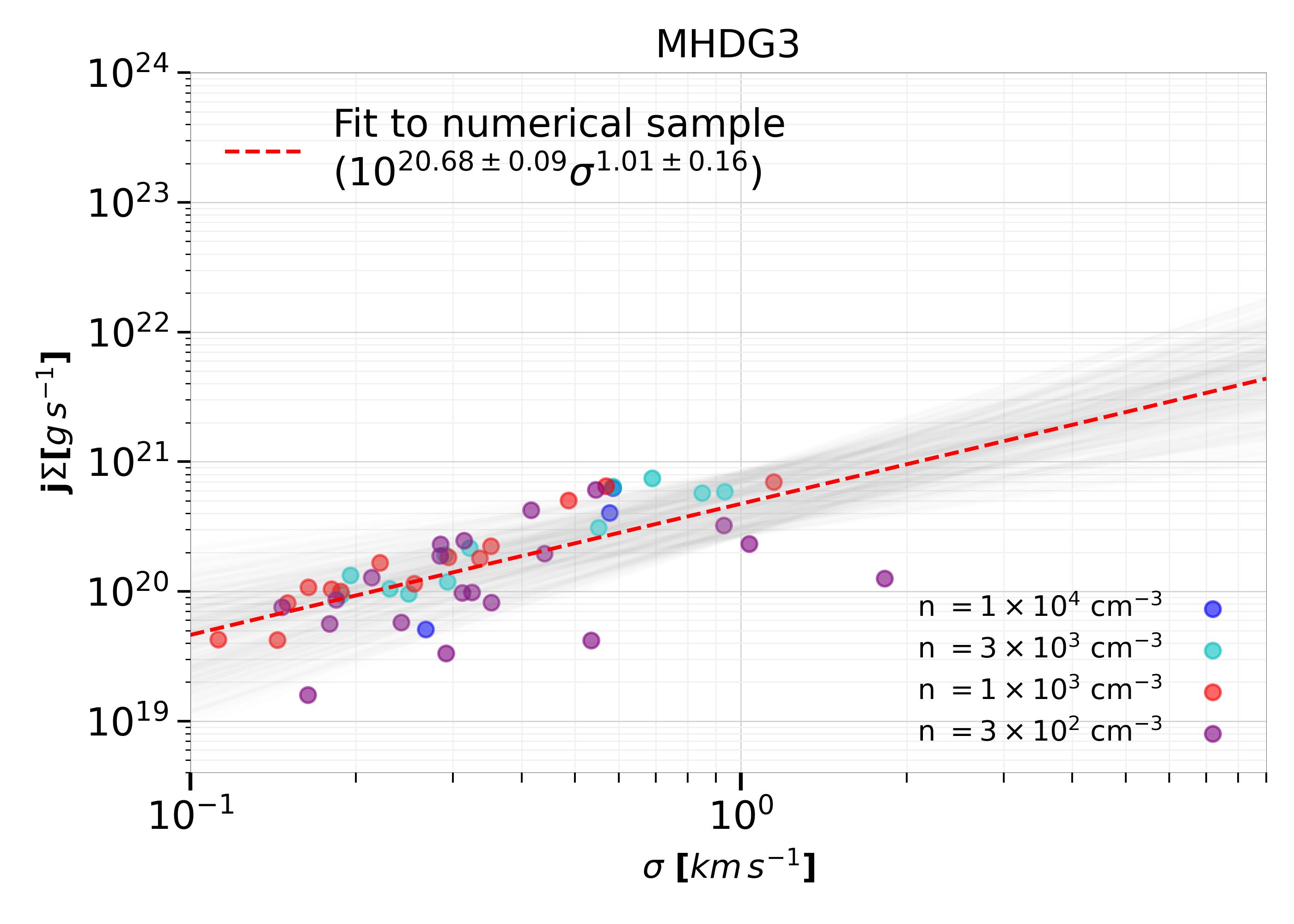}
 \caption{$j\Sigma$ as function of $\sigma_{v}$ for the full (left column) and reduced (right column) samples in HD3, HDG3 and MHDG3 simulations. The density thresholds is represented by the color code. Dashed red line represents the fitting to the numerical clump samples. The shaded gray region represents a variation in fitting parameters of 1$\sigma$. The reduction in scatter can be seen more clearly for the full sample case in the MHDG3 simulation, with this simulation being also the closest to the expected slope value of $3$ according to eq. \eqref{eq:j-Sigma-sigma}. }
\label{fig:j-Sigma-sigma}
\end{figure*}

In Figure \ref{fig:j-Sigma-sigma} we plot the quantity $j\Sigma$ as a function of $\sigma_{v}$ for the full (left column) and reduced (right column) samples in HD3, HDG3 and MHDG3 simulations. A reduction in scatter with respect to the $j$-$\sigma_v$ plot can be seen more clearly for the full sample case in the MHDG3 simulation, with all clumps defined by different thresholds seeming to follow a single trend. Furthermore, the slope of the $j\Sigma$-$\sigma_v$ plot for run MHDG3 seems to be the steepest, but still below the expected value of $3$ according to eq.\ \eqref{eq:j-Sigma-sigma}. The effect for runs HD3 and HDG3 is less evident.

\begin{figure*}
\centering
\includegraphics[width=0.47\linewidth]{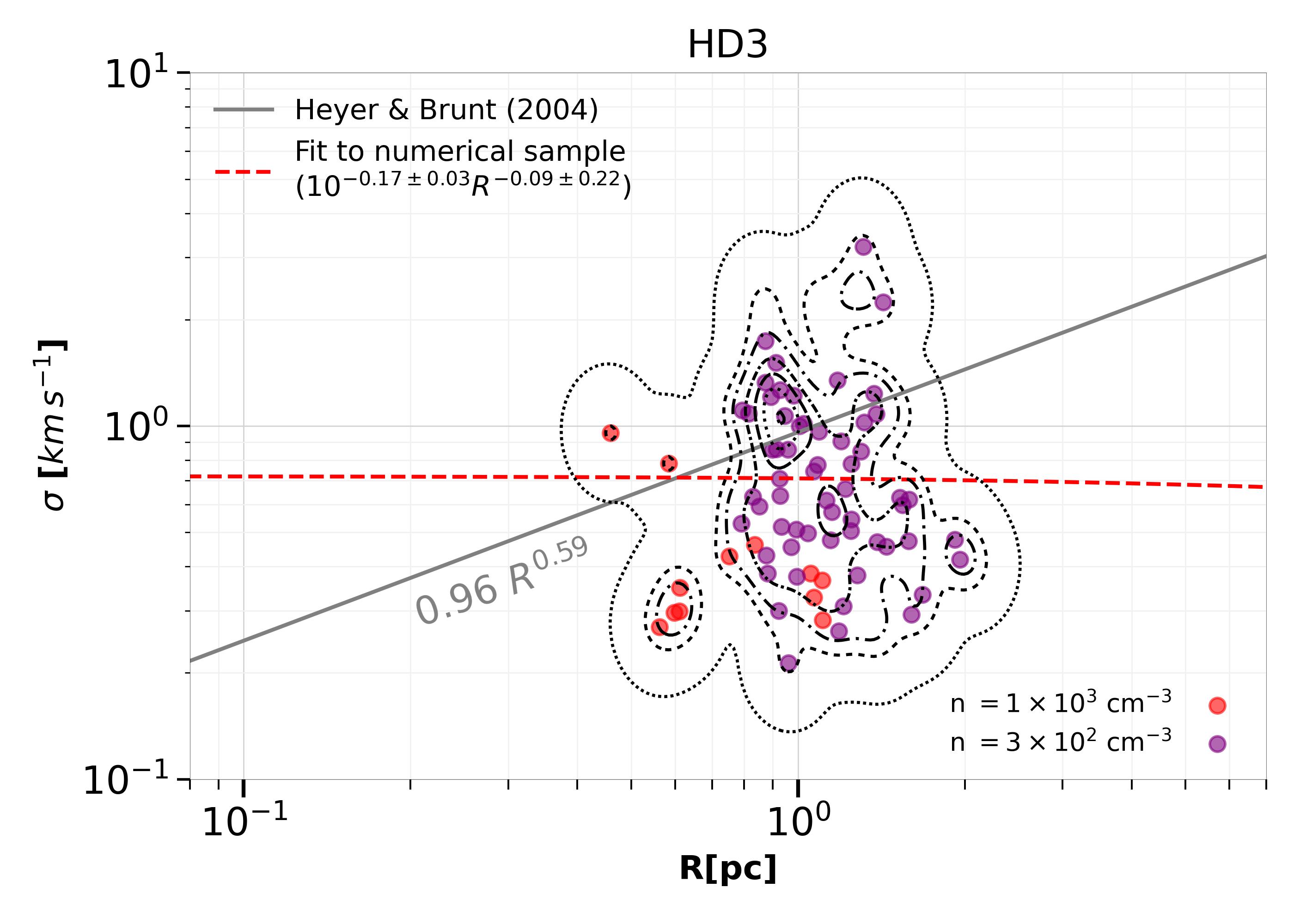}
\includegraphics[width=0.47\linewidth]
{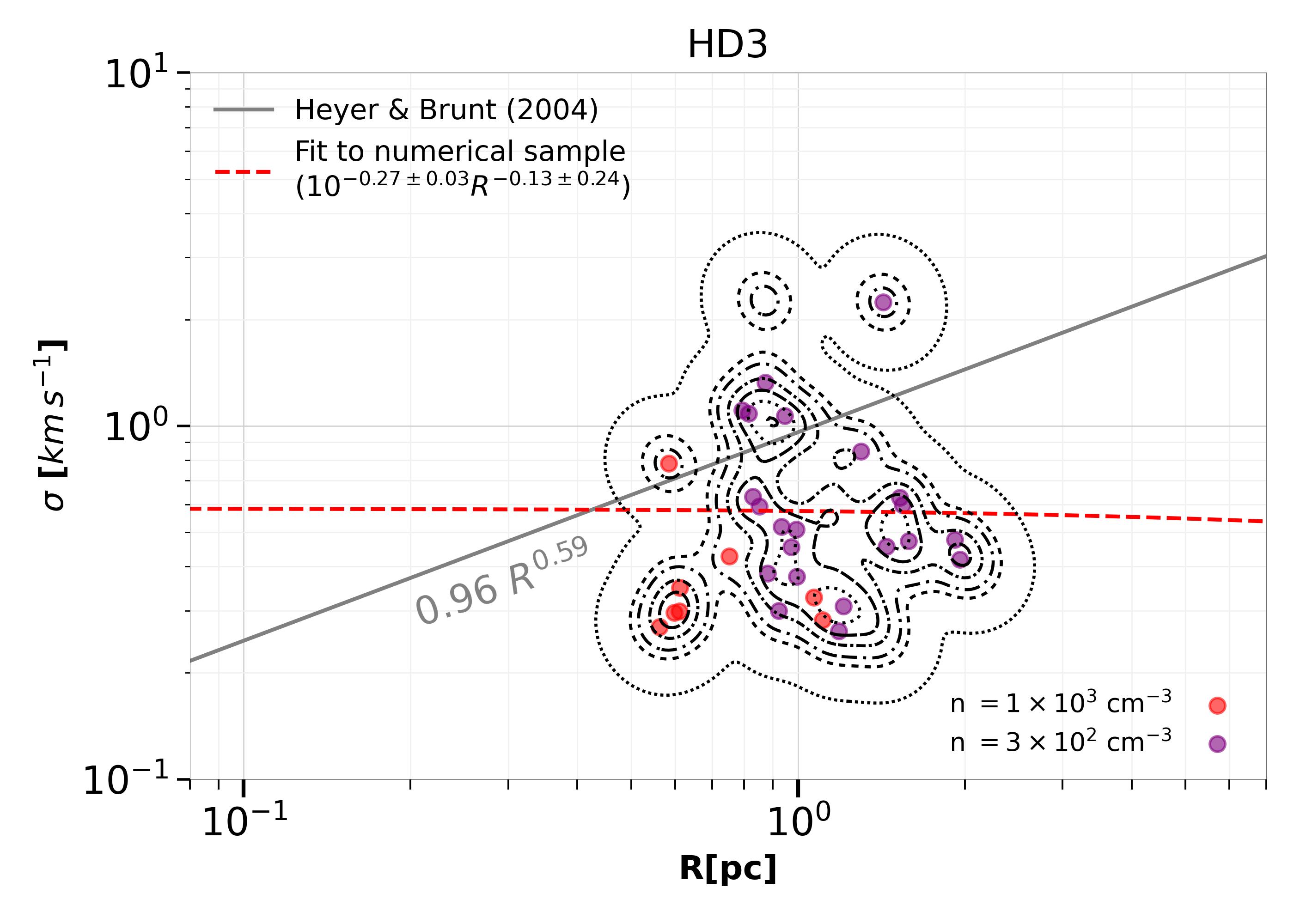}
\includegraphics[width=0.47\linewidth]{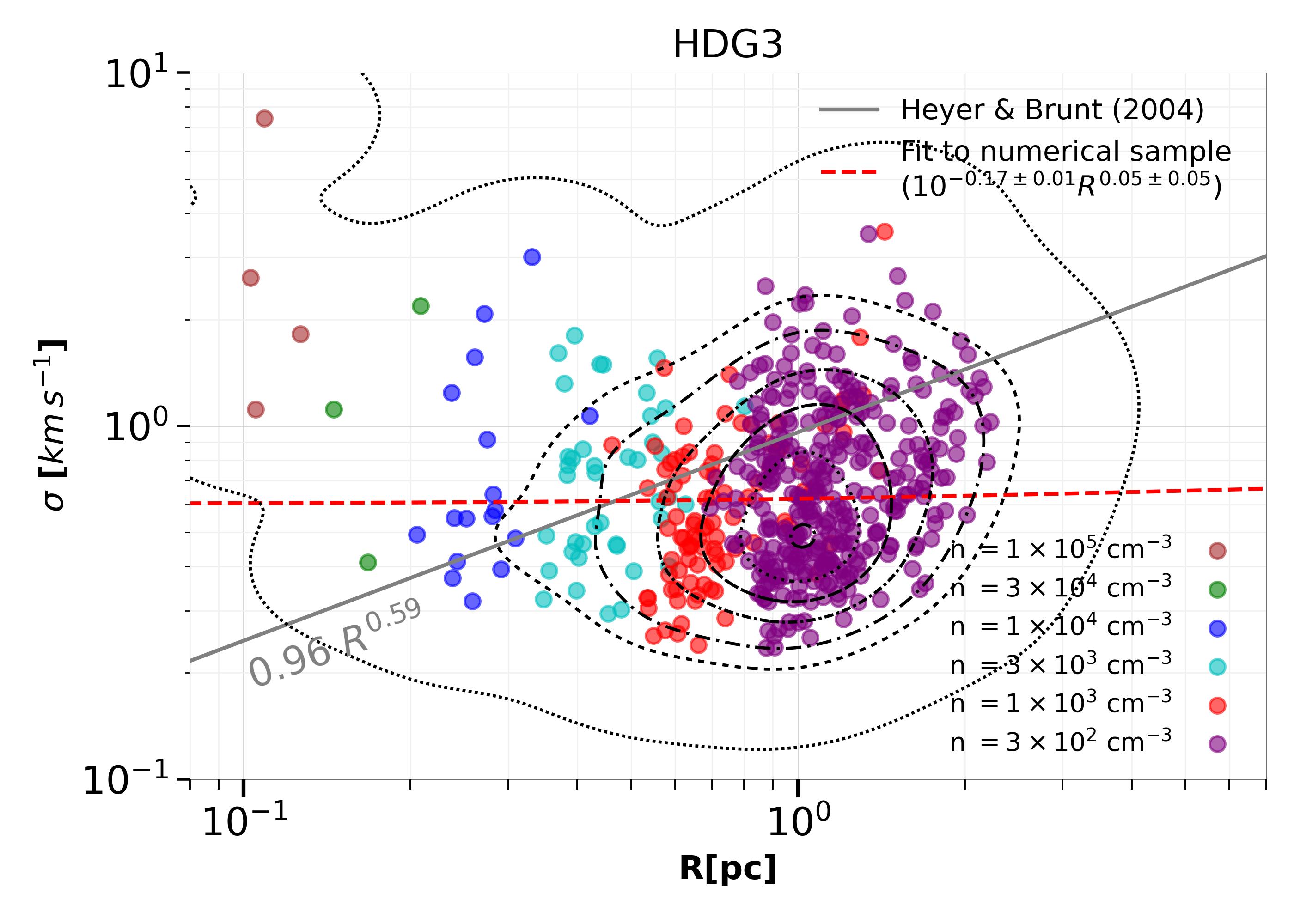}
\includegraphics[width=0.47\linewidth]
{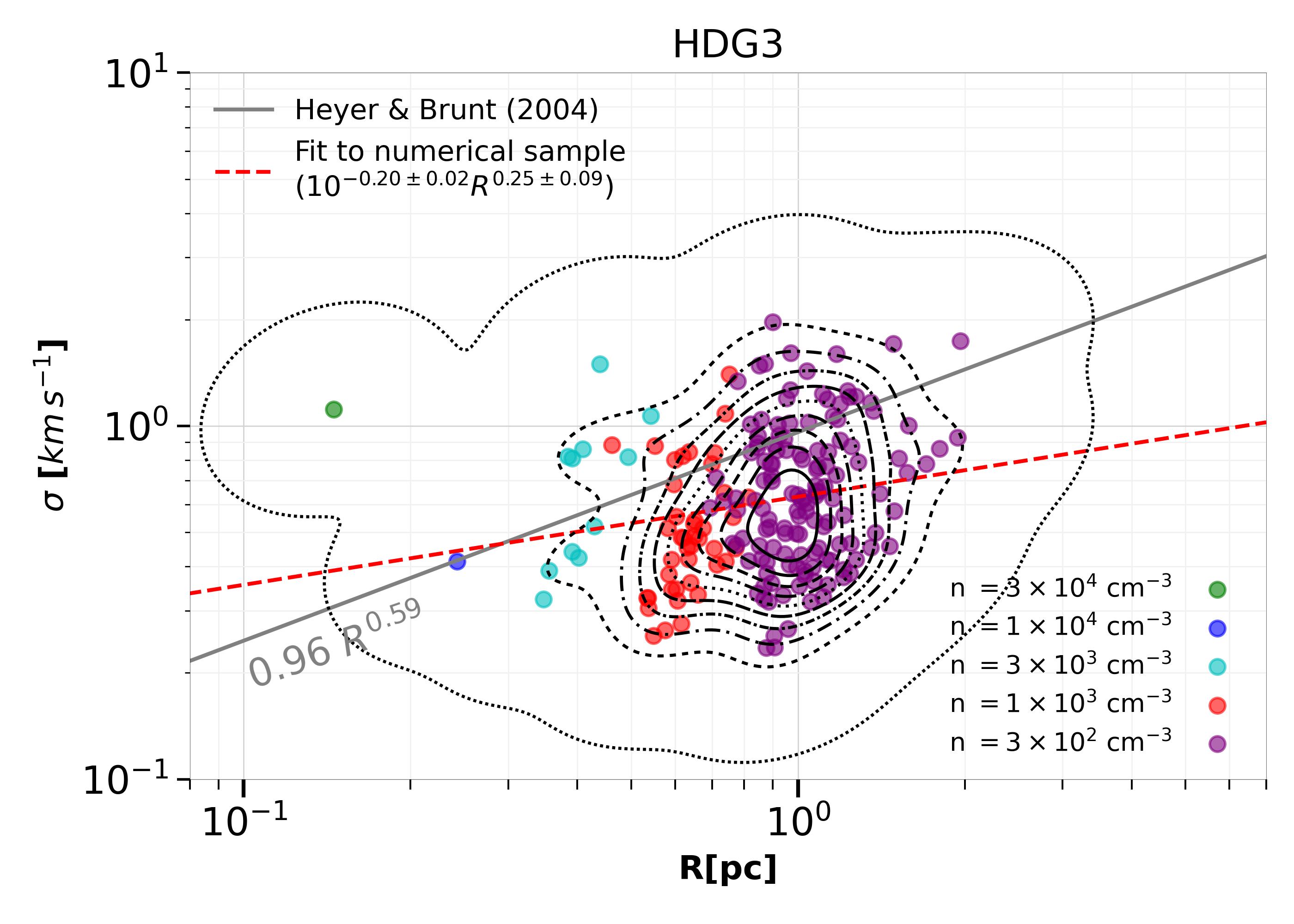}
\includegraphics[width=0.47\linewidth]{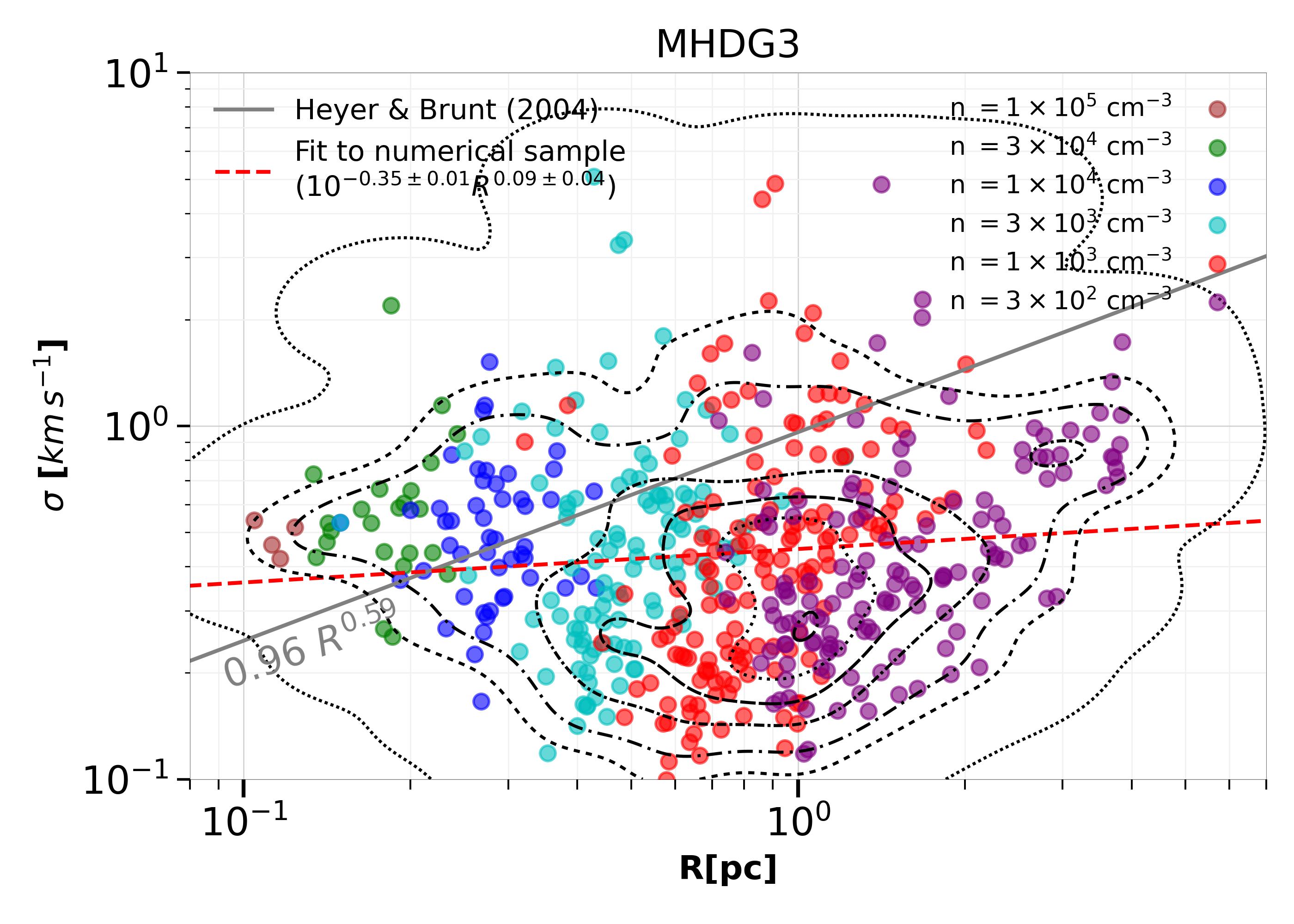}
\includegraphics[width=0.47\linewidth]
{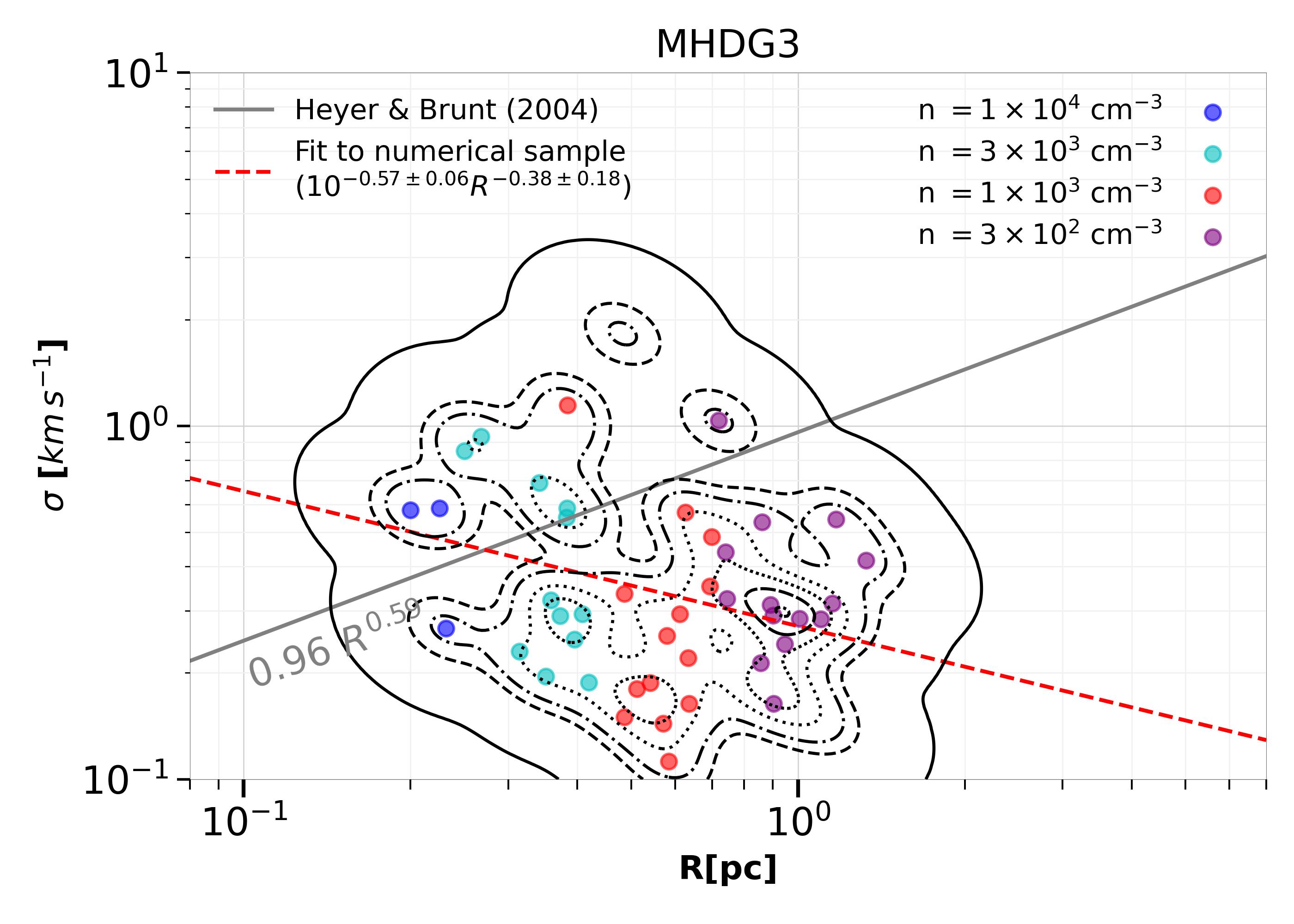}
 \caption{Larson's $\sigma$-$R$ scaling relation for the full (left) and reduced (right) samples in HD3 (first row), HDG3 (middle row), and MHDG3 (bottom row) simulations. The black lines show isocontours of the 2D density distribution of the points, with levels normalized to the peak in point density and set at $0.001$, $0.1$, $0.2$, $0.4$, $0.6$, $0.8$, and $0.99$ shown with different line styles. The relation is marginally recovered only in the reduced sample in run HDG3. Elongated structures deviate from this relation.}
\label{fig:Sigma-R}
\end{figure*}

To determine whether the dependence of $j$ and $j\Sigma$ on $\sigma$ is consistent with the Larson's $\sigma$-$R$ scaling relation, in Figure \ref{fig:Sigma-R} we show the latter relation for the full (left) and reduced (right) samples in runs HD3 (first row), HDG3 (middle row), and MHDG3 (bottom row).
The solid gray line represents the fit by \citet{Heyer.Brunt2004}. The black lines show isocontours of the 2D density distribution of the points. It can be seen that the relation is marginally recovered only in the reduced sample in run HDG3, and that, in general, our clumps are below the fitting line. Note that it is actually not surprising that our samples do not reproduce this relation, since, as previous studies have pointed out \citep{VS+08,Mejia-Ibanes+2016}, this relation holds only for samples of self-gravitating clumps with similar column densities. In our case, clumps are identified using multiple volume density thresholds, yielding structures with a range of sizes and morphologies and, consequently, a natural spread in column densities. 

On the other hand, although it is not possible to recover this relation in the HD3 simulation, since it has no gravity, some clumps in this simulation reach $\sigma$ values comparable to those observed. Given the adopted definition of angular momentum for clumps, computed as the sum of the individual SPH particle contributions, velocity dispersions comparable to those shown in the Larson relation may be responsible for the observed \jR relation. 

Regarding run MHDG3, both the full and the reduced samples appear to differ even more from the Larson scaling, exhibiting behavior that seems to be better described by a flatter slope. Deviations from Larson's velocity dispersion-size relation have been previously reported in filament observations, particularly within filaments at parsec-scale that exhibit a flattened velocity dispersion profile \citep{Arzoumanian+2013,Hacar+2016}. 

In general, if the velocity dispersion follows a scaling of the form $\sigma \propto R^{n}$, and given that in our approach $j \sim \sigma R$, we expect $j \propto R^{n+1}$. In particular, for a Larson-like relation with $\sigma \propto R^{1/2}$, this leads to $j \propto R^{3/2}$, consistent with observations \citep[] [] {Goodman+93}. However, for the filamentary structures in run MHDG3, the $\sigma$–$R$ relation appears significantly flatter (i.e., $n \sim 0$), albeit at magnitudes greater than those of the non-elongated ones, which would naturally imply $j \propto R$, in contrast with the behavior of the full MHDG3 sample, which instead shows a slope closer to the observed value of $3/2$. This apparent discrepancy can be explained if there is a selective increase of $j$ in elongated structures. I.e., if the largest clumps tend to be the most filamentary, and angular momentum is computed with respect to the center of mass, particles located near the filament ends contribute disproportionately to $j$ due to their larger position vectors, introducing an excess that steepens the slope and increases the intercept in the full sample. By contrast, the reduced sample, composed mainly of roundish clumps, does not include these strong end-dominated contributions, resulting in a shallower slope, close to unity, as expected from our simple argument, and lower intercept at $R = 1$ pc. This interpretation is supported by Figure \ref{fig:AR} in Appendix \ref{Appe:A vs R}, in which the aspect ratio, $A$, shows a mild positive correlation with size, $R$, in run MHDG3, and where clumps reach systematically larger values of both $R$ and $A$ than in the non-magnetic simulations, reinforcing the role of filamentary geometry in shaping the \jR\ relation.

\begin{figure*}
\centering
\includegraphics[width=0.47\linewidth]{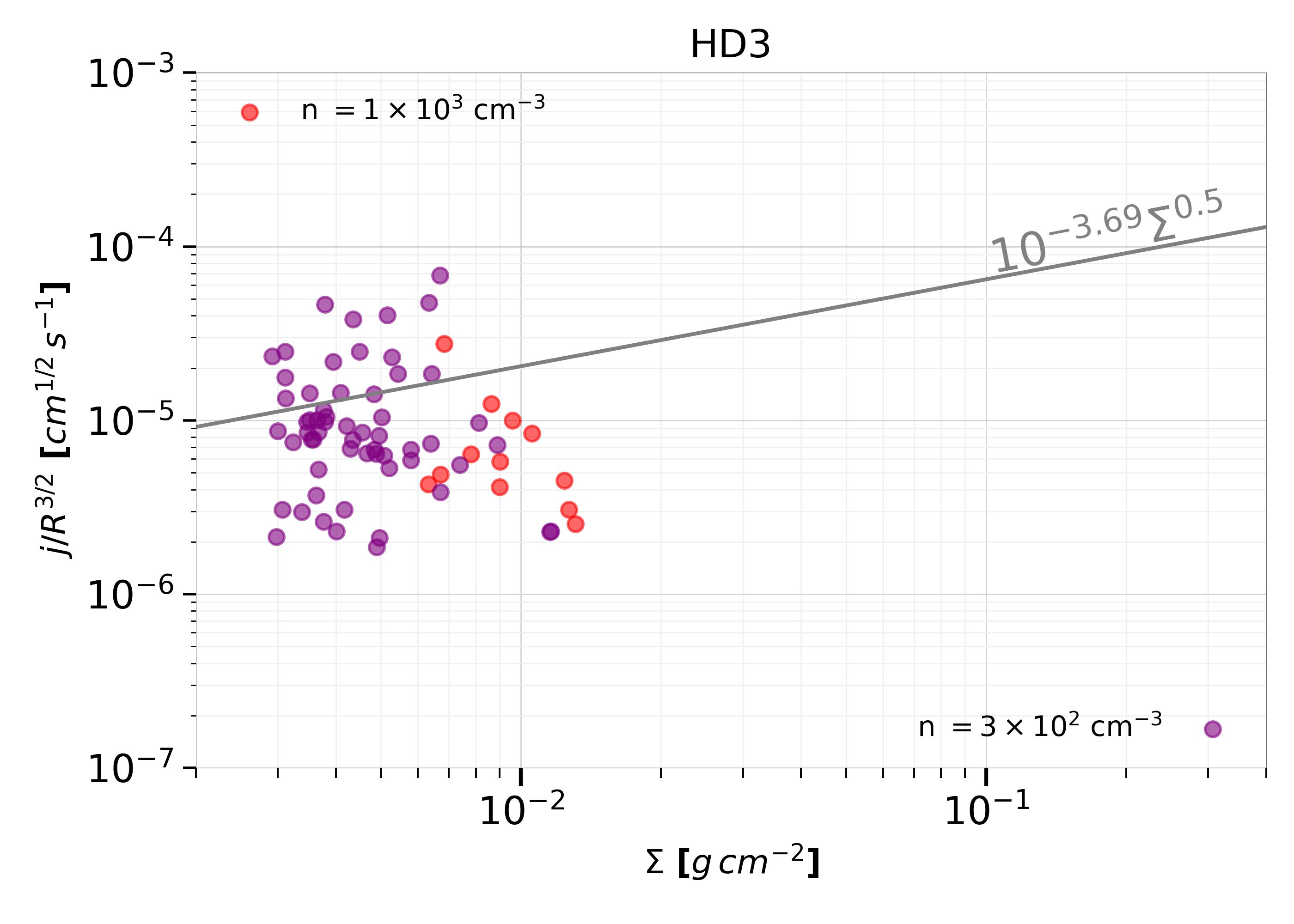}
\includegraphics[width=0.47\linewidth]
{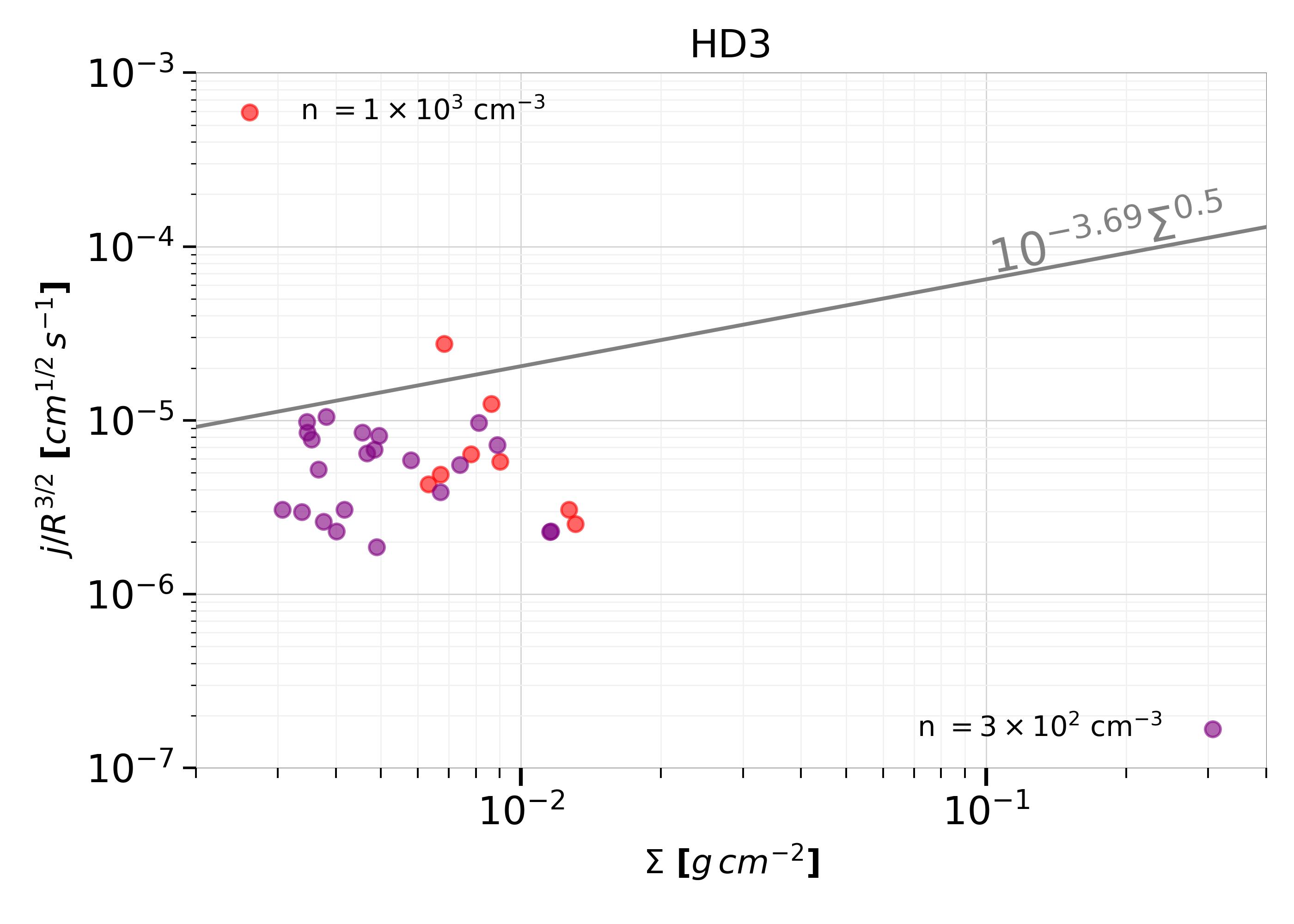}
\includegraphics[width=0.47\linewidth]{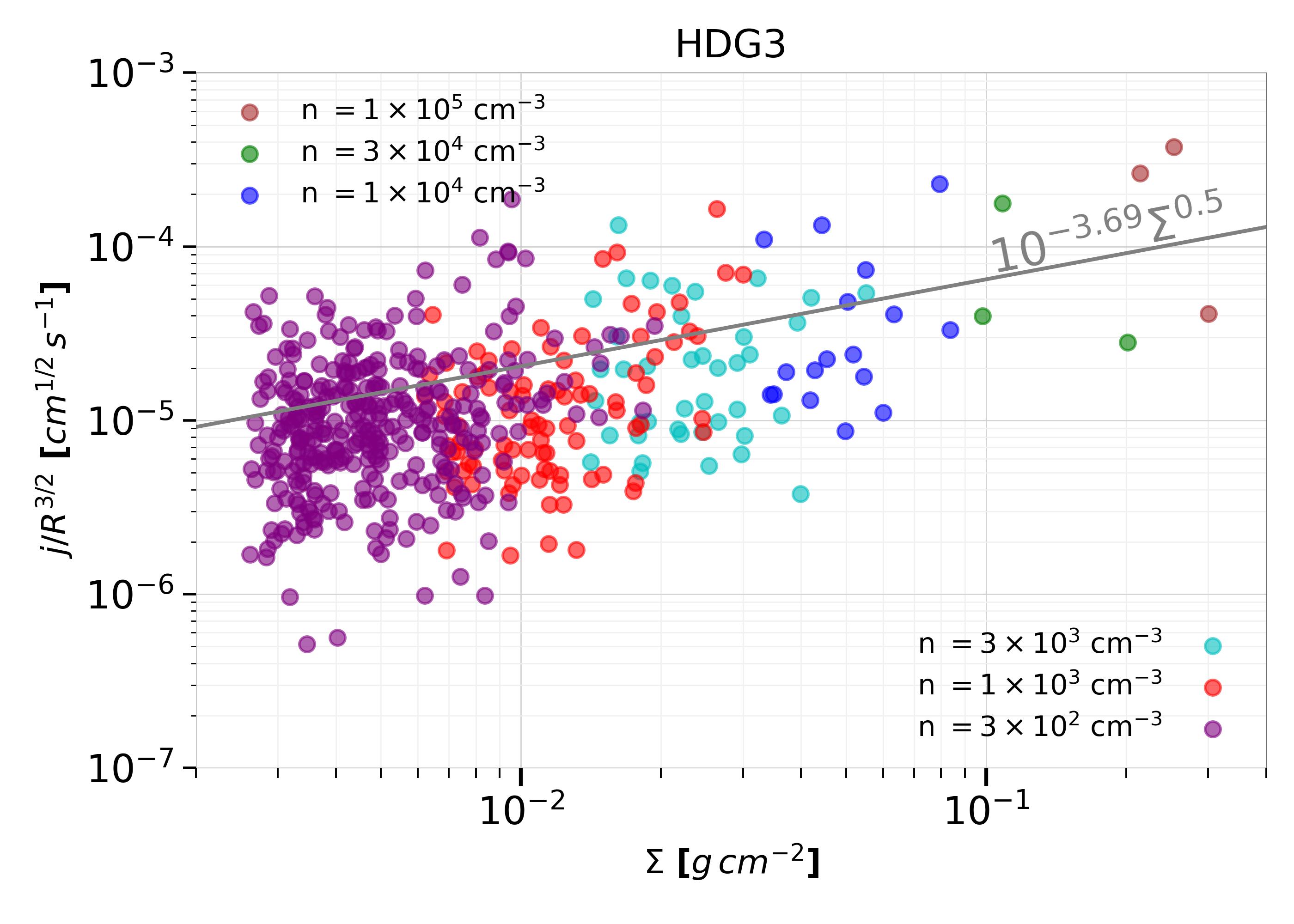}
\includegraphics[width=0.47\linewidth]
{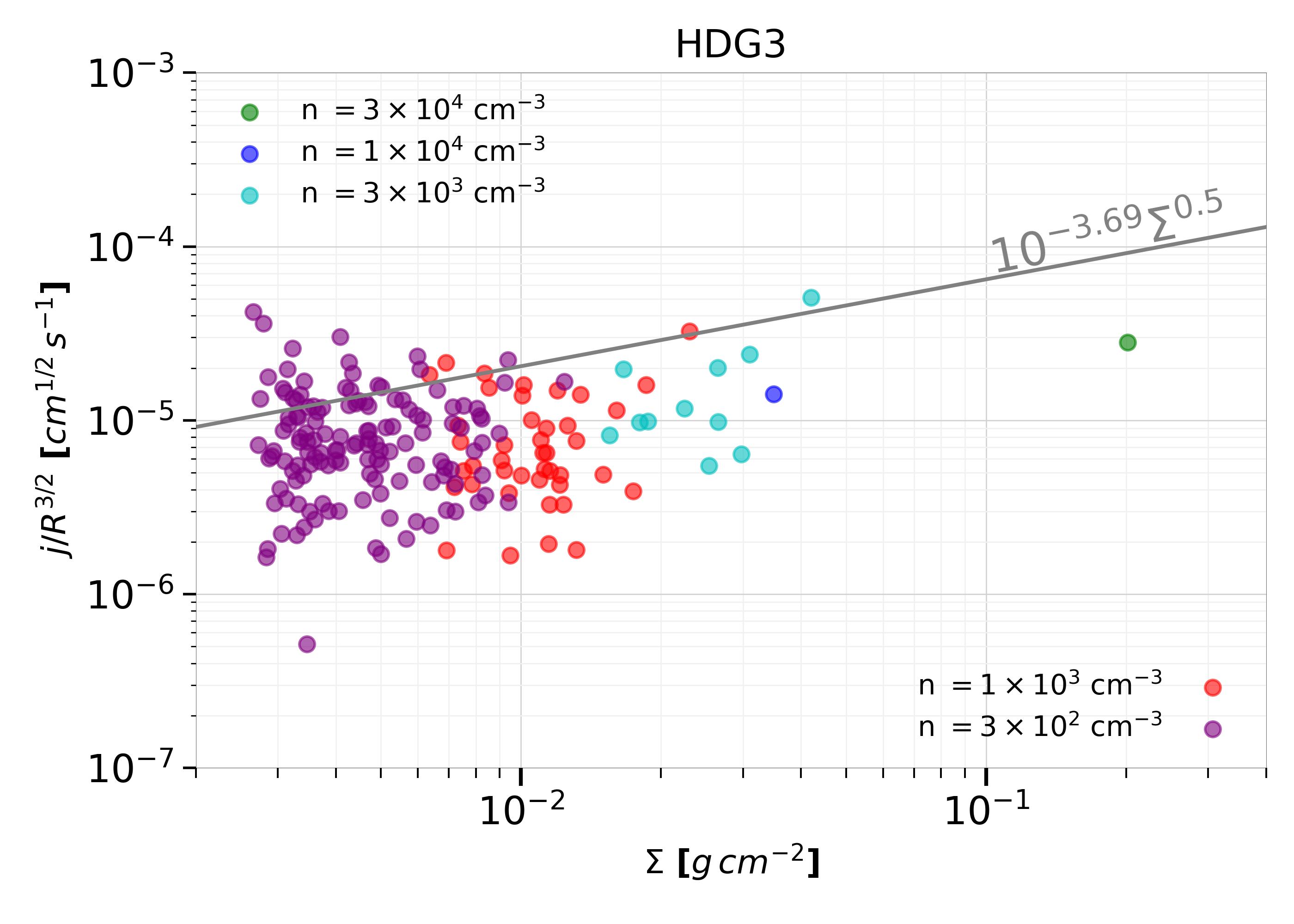}
\includegraphics[width=0.47\linewidth]{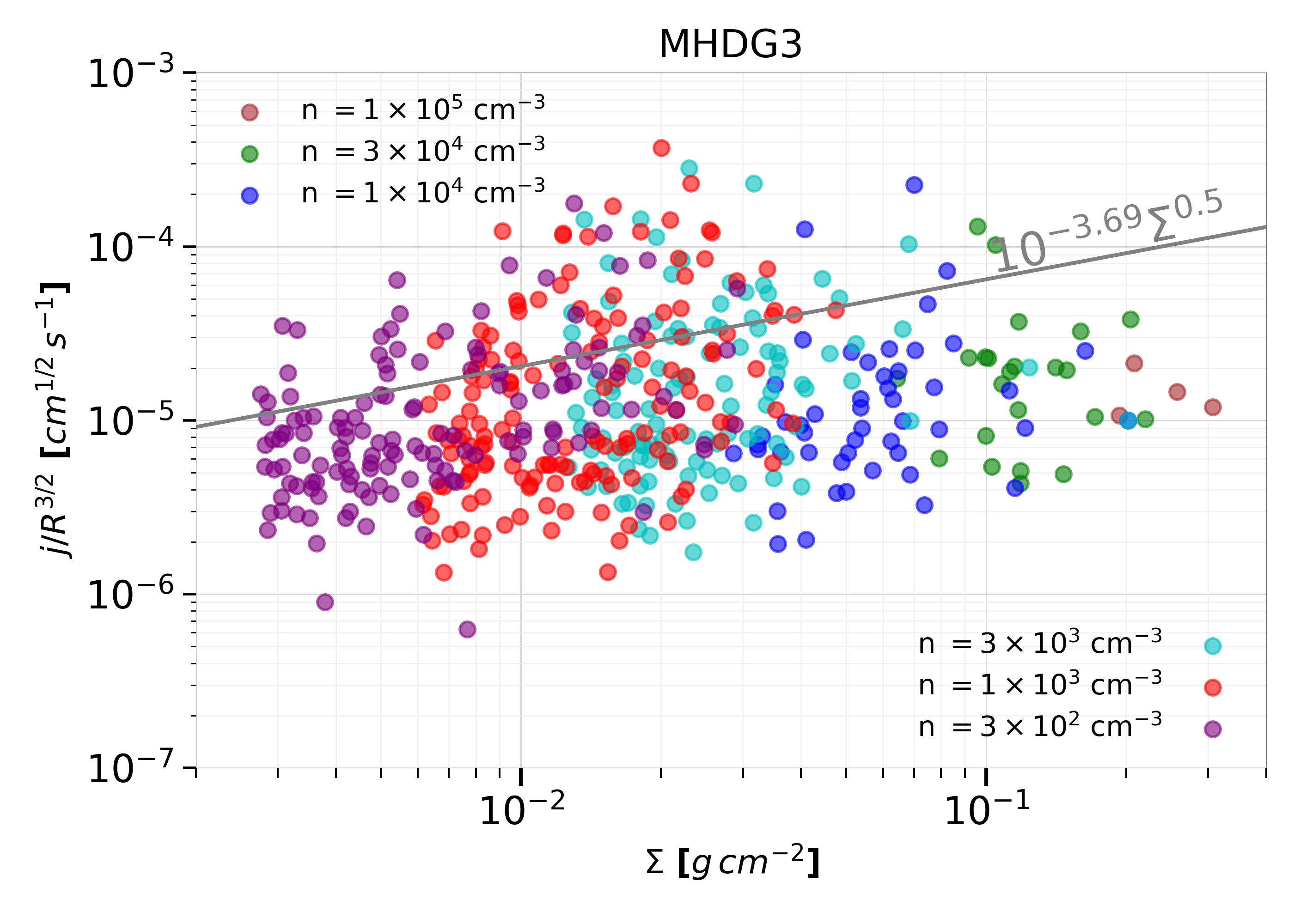}
\includegraphics[width=0.47\linewidth]{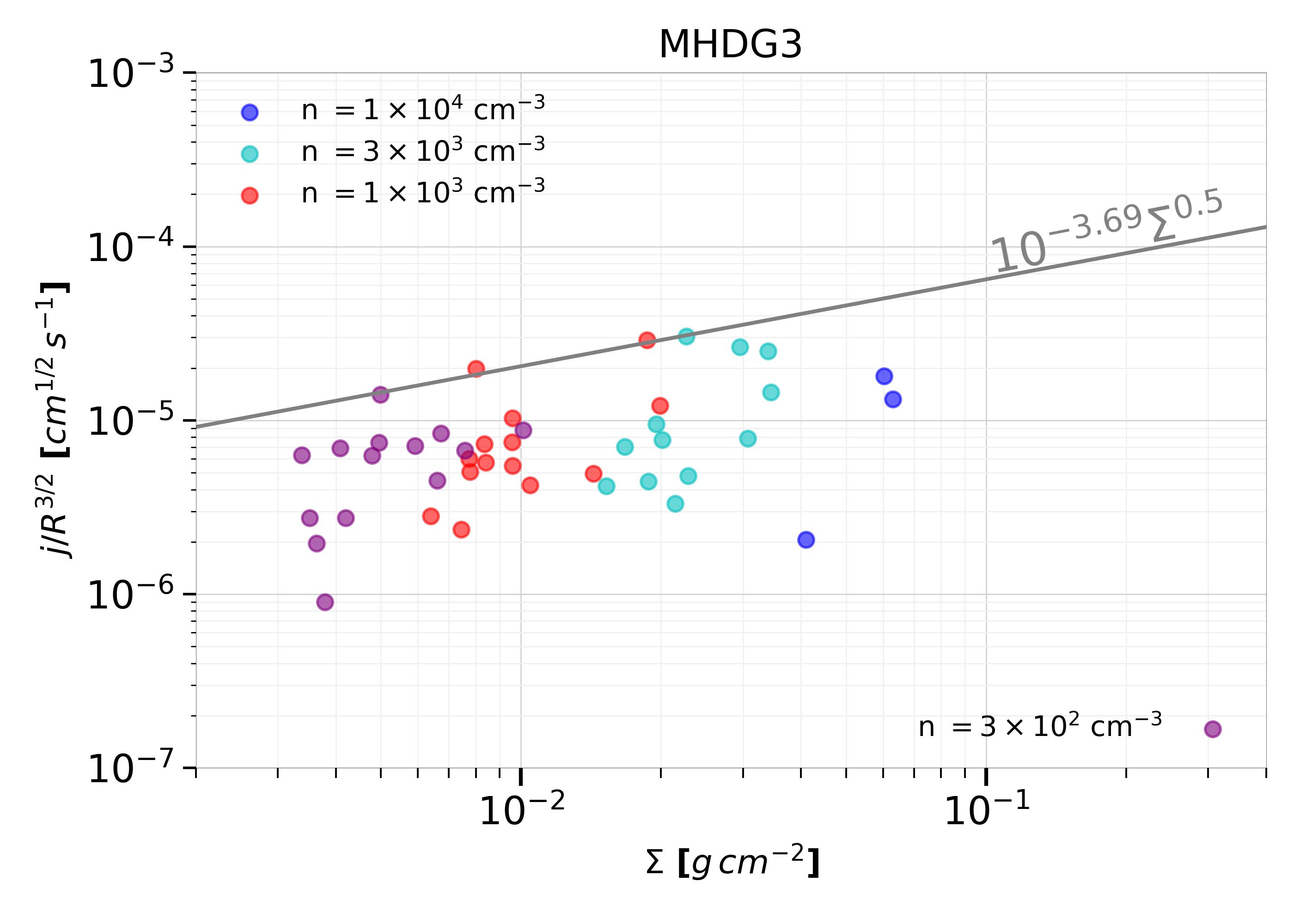}
 \caption{$j/R^{3/2}$ vs $\Sigma$ for the full (left) and reduced (right) samples in HD3 (first row), HDG3 (middle row), and MHDG3 (bottom row) simulations. Color represents the density threshold used to defined the clumps. Clumps tend to follow a trend similar to that expected, although the scatter is not significantly reduced compared to the \jR\ plots shown in Figure \ref{fig:com_clumps}.}
\label{fig:j/R-Sigma}
\end{figure*}

Interestingly, in addition to not reproducing the Larson relation like some observed examples, our filaments appear to differ from roundish clumps in the various scalings investigated in this work. This could suggest that the velocity structure required to transfer angular momentum in such a way as to reproduce the observed \jR\ relation must be close to that expected when recovering Larson's velocity dispersion-size relation, where turbulence and gravity play an important role. It then follows that the angular momentum transport mechanism proposed by \citetalias{Arroyo-Chavez.Vazquez-Semadeni2022}, driven primarily by the combined action of gravity and turbulence, might not operate in filaments.
Indeed, the mechanism proposed by \citetalias{Arroyo-Chavez.Vazquez-Semadeni2022} was based on the suggestion by \citet{BP+11} that the Larson linewidth-size relation is actually just a special case of the more general Keto-Heyer (KH) relation\footnote{This relation corresponds to approximate equpartition between the kinetic and gravitational energies.} when selection criteria restrict the column density of a sample to a narrow range of values. In consequence, \citetalias{Arroyo-Chavez.Vazquez-Semadeni2022} suggested that the fundamental scaling relation involving $j$ and $R$ should be 
\begin{equation}
j \approx \left(2 \pi \beta G \Sigma R^3\right)^{1/2}.
\label{eq:j_Sigma_R}
\end{equation} 
To test this, in Figure \ref{fig:j/R-Sigma} we plot $j/R^{3/2}$ versus $\Sigma$ . As shown in \citetalias{Arroyo-Chavez.Vazquez-Semadeni2022}, the expected dependence on the column density, $\Sigma$, is as $j/R^{3/2} \sim \Sigma^{1/2}$, which is shown by the gray line. This line was obtained assuming $\beta \sim 0.1$, which is a rough estimate of the median value of $\beta$ in Figure 14 of \citetalias{Arroyo-Chavez.Vazquez-Semadeni2022}\footnote{We note, however, that, in Figure 14 of \citetalias{Arroyo-Chavez.Vazquez-Semadeni2022}, an additional mild dependence of $\beta$ on $R$ is observed. Here we neglect this dependence, and assume that it is constant to zeroth order, as suggested by \citet{Goodman+93}. Additionally, in Appendix \ref{Appe:beta values} we present the $\beta$ values for the full and reduced samples of the simulations analyzed here.}, where values of $\beta$ were compiled for clumps and cores reported in various observational studies. Figure \ref{fig:j/R-Sigma} shows that the predicted trend is best reproduced by the full sample of run HDG3, whereas in the MHDG3 run (full and reduced samples), the values are systematically lower than those predicted. This may be indicative of lower values of $\beta$ in the magnetic case. On the other hand, the samples from run HD3 lack a large enough dynamic range in $\Sigma$ to define any significant trend.

\subsection{Measurement of the torques}
\label{subsec:measurement of the torques}

When discussing the \jR\ relation for the clump samples in each simulation in Fig. \ref{fig:com_jR}, we mentioned that even in run HD3 (without gravity and magnetic field) we recover $j$ magnitudes close to those expected for clumps of similar sizes in the observations, although both gravity and magnetic field are necessary to recover a sufficiently large dynamic range to clearly define a correlation. The fact that turbulence itself manages to produce such magnitudes suggests that indeed the angular momentum observationally measured in clumps and cores is due to the residual rotation introduced by the turbulence. But it is not entirely clear what the precise role of each type of torque (gravitational, magnetic, or hydrodynamic, the latter involving turbulent viscosity) is in the establishment of the \jR\ relation. To clarify this, in this section we perform direct measurements of the torque magnitudes respect to the center of mass from the various forces in our numerical clump sample. For this, in eq. \eqref{eq:total torque per unit mass} we write the total torque for a fluid parcel. That is
\begin{align}
 \int_{V} {\bm r} \times  \frac{\partial {(\rho \uu)}}{\partial t}\; dV =
 - {\underbrace{\textstyle  \int_{V}  {\bm r} \times \nabla \cdot (\rho \uu\uu) \; dV}_{\mathclap{\text{\textcolor{blue}{Hydrodynamic}}}}}
 - {\underbrace{\textstyle \int_{V} {\bm r} \times  \nabla P\; dV}_{\mathclap{\text{\textcolor{orange}{Pressure gradient}}}}} \nonumber & \\ 
 - {\underbrace{\textstyle  \int_{V} {\bm r} \times \rho \nabla \phi \;dV}_{\mathclap{\text{\textcolor{red}{Gravitational}}}}}
 + {\underbrace{\textstyle \int_{V} {\bm r} \times  \mu  (\nabla^{2} \uu + \nabla \nabla \cdot \uu) \;dV}_{\mathclap{\text{Viscous}}}} \nonumber & \\ 
 + {\underbrace{\textstyle \int_{} {\bm r} \times \frac{1}{4 \pi} (\nabla \times \BB) \times \BB \; dV}_{\mathclap{\text{\grn Magnetic}}}}, 
\label{eq:total torque per unit mass}
\end{align}
where the physical nature of each term is indicated below it.
A few points can be discussed regarding this equation. First, since we do not implement any contribution by molecular viscosity in our simulations, the fourth term of the right hand side cannot be measured, so we will only measure the terms corresponding to the gravitational, magnetic, hydrodynamic and pressure-gradient torques. The terms that involve the gradient and divergence have been calculated on the neighbors of each of the SPH particles in a clump using a modified version of the Python package \textsc{Plonk} \citep{Mentiplay2019}.

\begin{figure}
\centering
\includegraphics[width=0.7\linewidth]{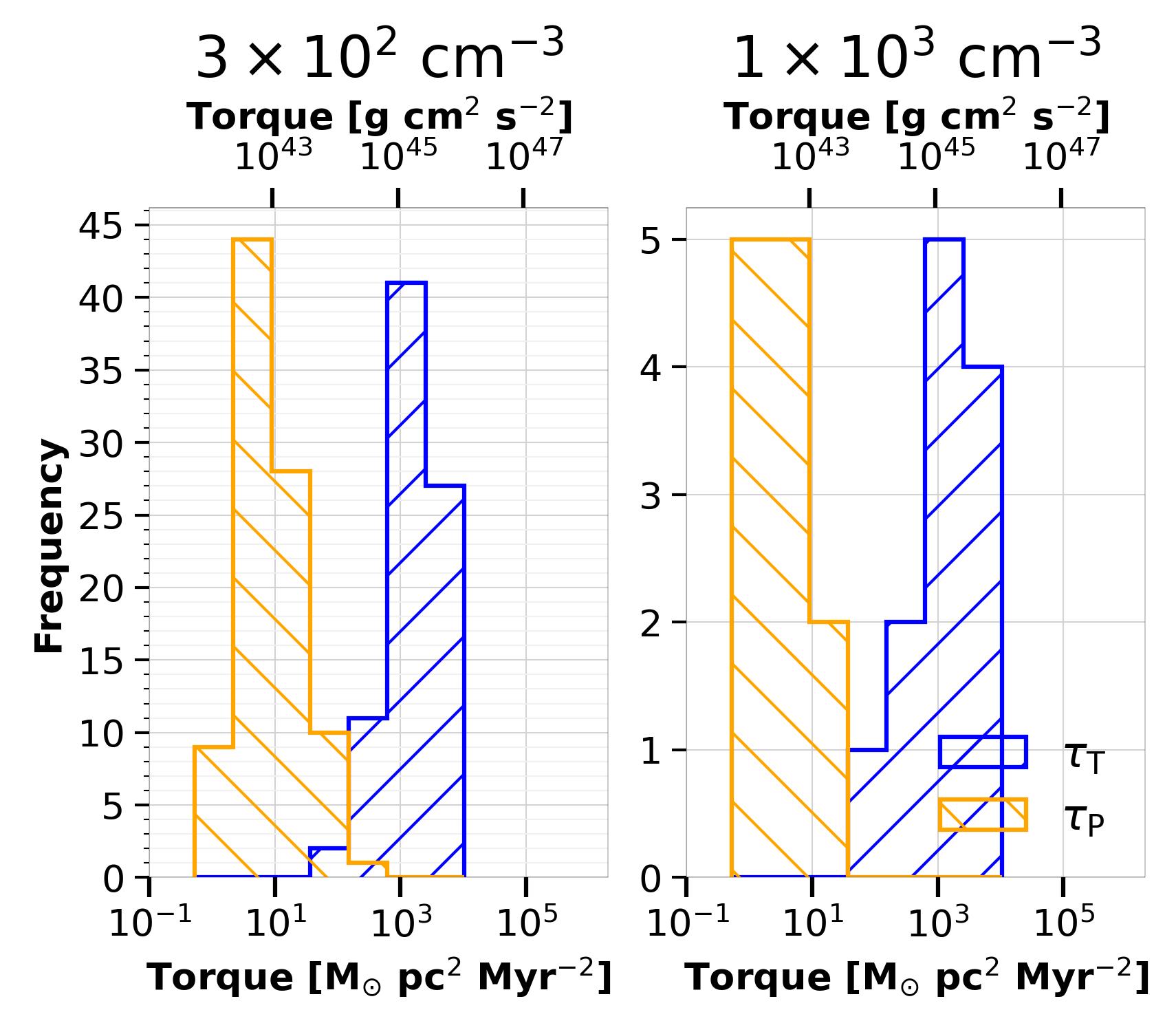}
\includegraphics[width=0.7\linewidth]{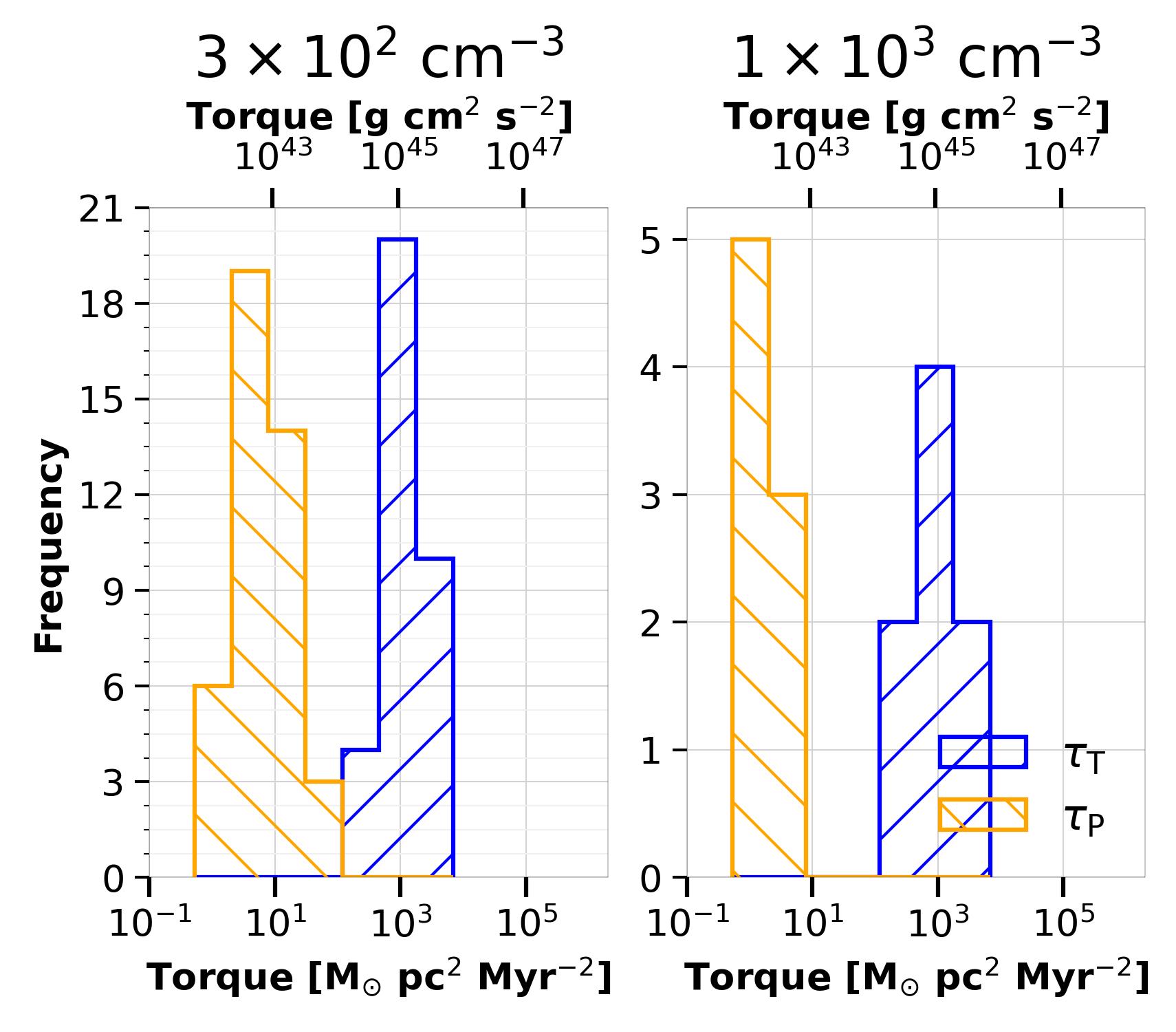}
\caption{Histograms of the value of hydrodynamic (blue) and pressure-gradient (yellow) torques for the full (first row) and reduced (second row) clump samples in the HD3 simulation, respect to the center of mass of each clump. The density threshold used to defined the clumps is shown on top of each plot. In general, the hydrodynamic torque is larger compared to the pressure-gradient torque. This difference is more noticeable for the larger density threshold; i.e., in smaller, denser clumps}.
\label{fig:hist_torques_HD3}
\end{figure}

\begin{figure*}
\centering
\includegraphics[width=\linewidth]{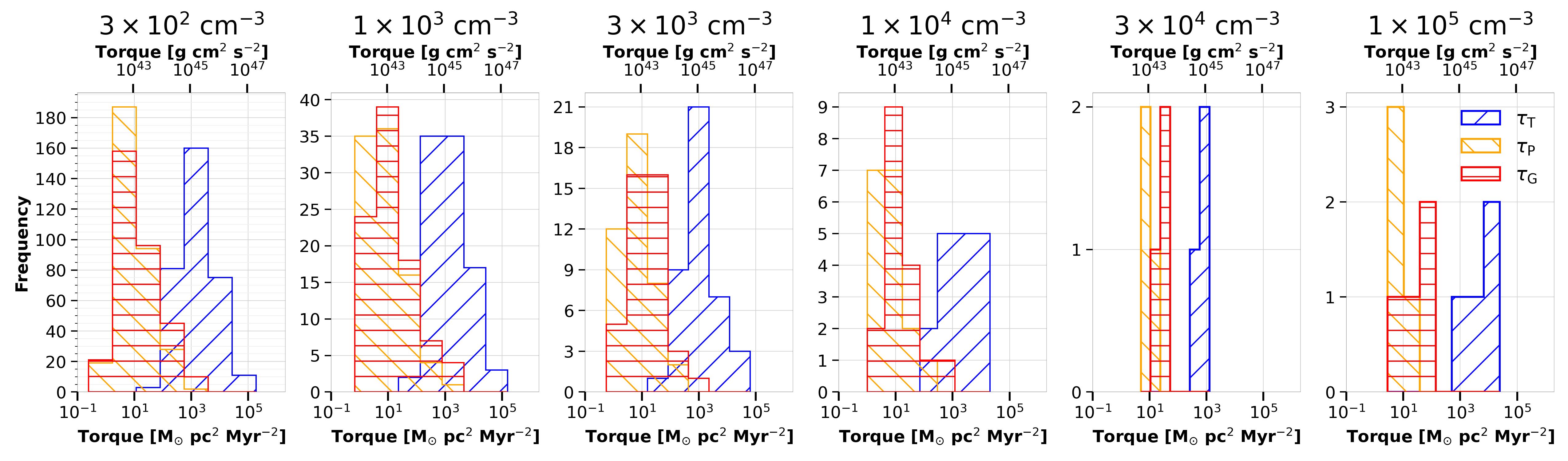}
\includegraphics[width=0.85\linewidth]{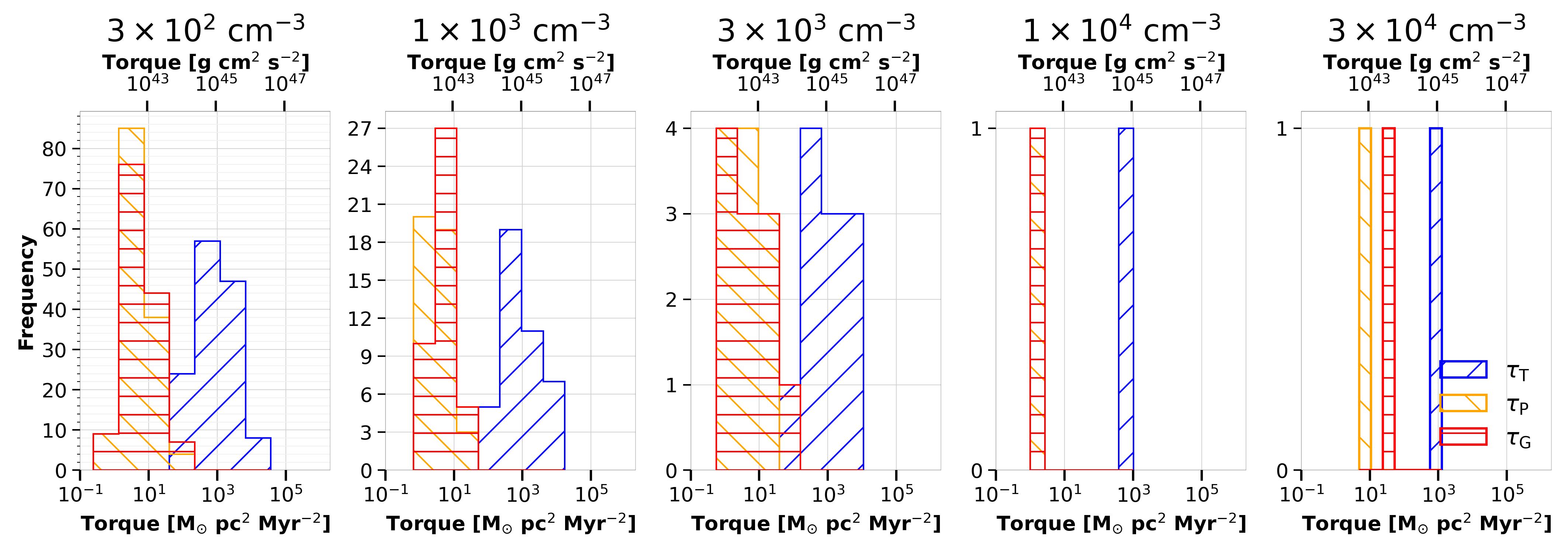}
\caption{Histograms of the value of hydrodynamic (blue), pressure-gradient (yellow) and gravitational (red) torques for the full (first row) and reduced (second row) clump samples in the HDG3 simulation, respect to the center of mass of each clump. The density threshold used to defined the clumps is shown on top of each plot. In general, the hydrodynamic torque is larger compare to the others, followed by gravitational torque This difference is more noticeable as the density threshold increases.}
\label{fig:hist_torques_HDG3}
\end{figure*}

\begin{figure*}
\centering
\includegraphics[width=\linewidth]{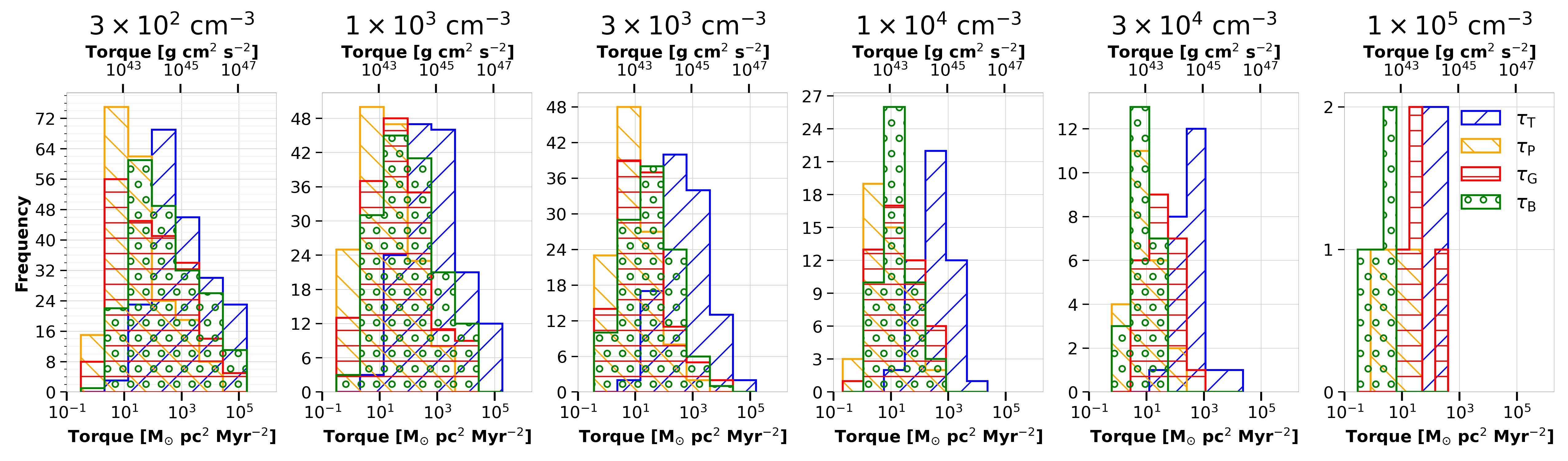}
\includegraphics[width=0.65\linewidth]{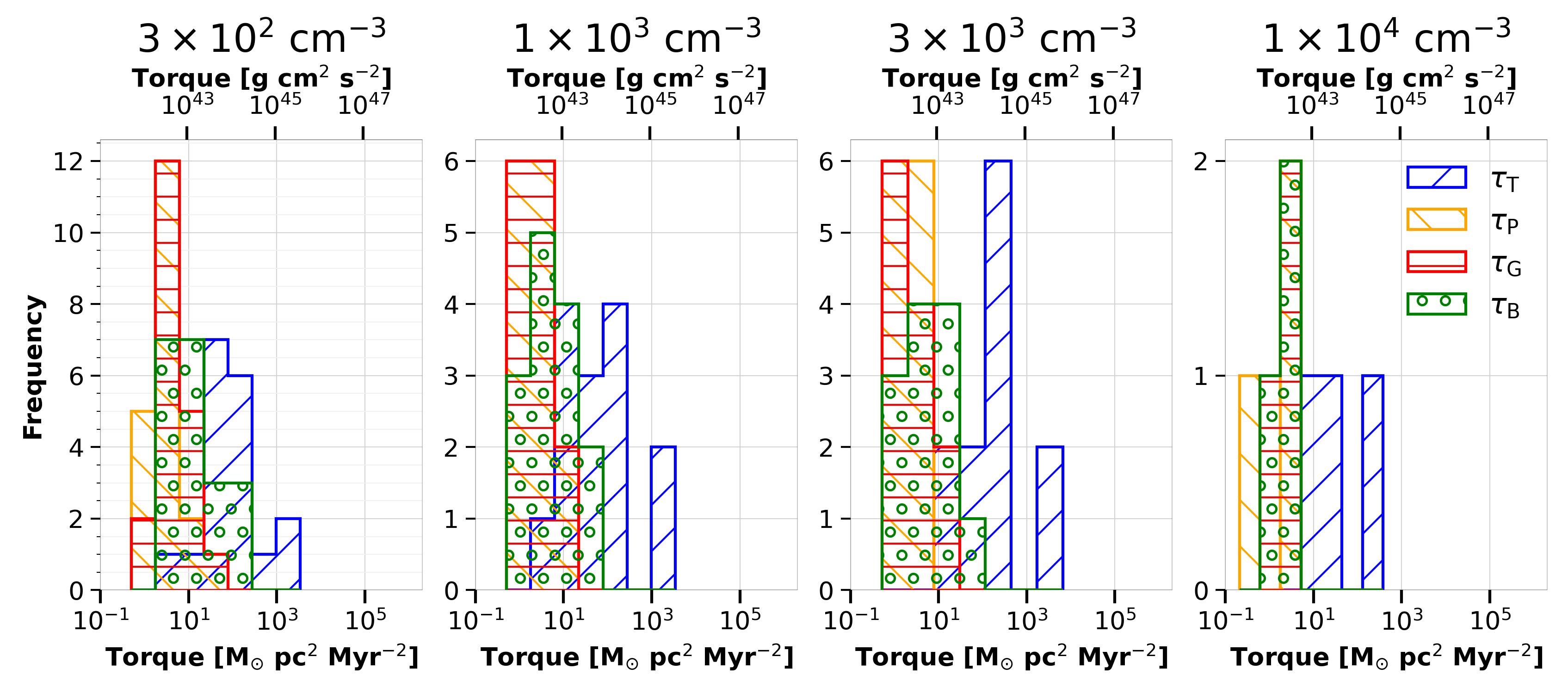}
\caption{Histograms of the value of hydrodynamic (blue), pressure-gradient (yellow), gravitational (red) and magnetic (green) torques for the full (first row) and reduced (second row) clump samples in the MHDG3 simulation, respect to the center of mass of each clump. The density threshold used to defined the clumps is shown on top of each plot. In general, the hydrodynamic torque is larger compare to the others, followed by gravitational torque, magnetic and pressure-gradient torque in that order. This difference is more noticeable as the density threshold increases.}
\label{fig:hist_torques_MHDG3}
\end{figure*}

In Figures \ref{fig:hist_torques_HD3}, \ref{fig:hist_torques_HDG3} and \ref{fig:hist_torques_MHDG3} we show histograms of the torque magnitudes for the clumps in runs HD3, HDG3 and MHDG3, respectively, with respect to the center of mass (in position and velocity) of each clump, and for the full (first row) and reduced (second row) samples. The density threshold used to define the clumps is shown on top of each plot. The colors denote each kind of torque: gravitational (red), magnetic (green), hydrodynamic (blue) and pressure-gradient (yellow). Each figure shows the torques that can be measured in each simulation according to the presence (or not) of gravity and the magnetic field.

From these plots it can be seen that, in general, the hydrodynamic torques are larger compared to the other three types. This difference is more noticeable as the density threshold increases. In the HDG3 simulation, the gravitational torque tends to be second in magnitude for some clumps, but in general the magnitude is very close to that of pressure-gradient torques. In the MHDG3 simulation, we find that the magnetic torques have a magnitude comparable to the others for the lowest density threshold. In fact, the histograms of all the torques overlap at a threshold  $n_{\rm thr}=3 \times 10^{2}\, \pcc$. As the threshold increases, it is seen that the magnitudes decrease along the sequence of hydrodynamic, gravitational, pressure-gradient, and magnetic torques. 

When comparing the full samples to the reduced ones in all simulations, it can be seen that those clumps whose gravitational, pressure-gradient, and magnetic torques are closer to the hydrodynamic torque, are precisely the most elongated ones, since by eliminating the filamentary structures, the histograms of the hydrodynamic torques tend not to overlap with the histograms of the rest of the torques. On the other hand, the largest magnitude reached by the histograms of the pressure-gradient and gravitational torques appears to increase by approximately two orders of magnitude for the clumps in the MHDG3 simulation compared to those in run HDG3. This could be due to a geometric effect introduced when measuring torques relative to the center of mass for filamentary structures, as previously discussed in section \ref{subsec:Deviations}.

These results support the suggestion in \citetalias{Arroyo-Chavez.Vazquez-Semadeni2022} that the angular momentum transfer mechanism relies mainly on the action of turbulence. The mechanism we propose also involves gravity as responsible for generating the imbalance that promotes the transfer of angular momentum by turbulence. Indeed, in \citetalias{Arroyo-Chavez.Vazquez-Semadeni2022}, we noted that, for a gravitationally contracting clump with a fixed mass, the gravitational energy per unit mass, $e_{\rm g}$, scales with radius as $e_{\rm g} \propto R^{-1}$, while, assuming angular momentum conservation, the specific rotational energy, $e_{\rm r}$, scales as $e_{\rm r} \propto R^{-2}$. Then, as the clump is compressed by gravity, its ratio of rotational to gravitational energy ($\beta$) tends to increase, increasing also their angular velocity due to the conservation of angular momentum. Then, if angular momentum were conserved during collapse, the rotational energy, $e_{\rm r}$, would eventually exceed the gravitational energy, $e_{\rm g}$, halting the collapse. In reality, however, this does not occur because angular momentum is continuously redistributed or removed. This process is precisely what allows the collapse to proceed, otherwise, the structure would become rotationally supported and further collapse would be prevented. These mechanisms must therefore operate down to the scales of disk formation. In \citetalias{Arroyo-Chavez.Vazquez-Semadeni2022}, we proposed that turbulent viscosity, or the transport of angular momentum between fluid parcels by turbulent mixing, could be one of the most efficient during the contraction and fragmentation of molecular clouds. Although we do not measure the turbulent viscosity directly, its effect is included in the term of hydrodynamic torques in this work. On the other hand, the generation of turbulence by the collapse itself \citep{Vazquez-Semadeni+98,Klessen_Hennebelle10,Robertson_Goldreich12,Murray_Chang15,Xu_Lazarian20,Guerrero.Vazquez2020} ensures that angular momentum can be redistributed through this mechanism throughout the entire process. This interplay between gravity and turbulent viscosity could explain the apparent constancy of the $\beta$ parameter.

It is worth mentioning that, so far, we have only considered the magnitude of the torques. However, since the torques are vector entities, it is also relevant to study their orientation with respect to the local flow direction, which can provide clues as to how clumps acquire (or lose) their angular momentum from the environment in which they live. We leave this study for future work.

\section{Discussion}
\label{sec:discussion}

\subsection{The crucial role of the center of mass}
\label{sec: CM discussion}

As we mentioned previously in sections \ref{subsec:sample} and \ref{subsec:measurement of the torques}, both the angular momentum and the torques have been measured with respect to the center of mass of each clump. In principle, it is possible to choose another reference point such as the density peak, which would be more in line with what is usually done in observations, using the intensity peak. By changing the reference point, it is evident that the magnitude of the angular momentum will change, since it is known that angular momentum is a non-conserved quantity when measured from a non-inertial reference frame. We now discuss the effect of this change of reference frame.

From classical mechanics, it is well known that the angular momentum of a system of particles can be expressed as that with respect to the center of mass ($\sum_{i =1}^{n} {\bm L}_{i,{\rm CM}}$, where $n$ is the number of particles), plus that of the center of mass with respect to the origin of coordinates (${\bm L}_{\rm CM}$). Therefore, the angular momentum measured with respect to the density peak satisfies
\begin{equation}
    {\bm L_{\rm DP}} = {\bm L}_{\rm CM,DP} +  \sum_{i=1}^{n} {\bm L}_{i,{\rm CM}}. \label{eq:AM_transformation}
\end{equation}
From this, it follows that the angular momentum computed with respect to any point other than the center of mass is overestimated, specifically by an amount ${\bm L}_{\rm CM,DP}$ (see Appendix \ref{Appe:AM for system of particles} for a detailed derivation).

In this sense, we expect the least elongated clumps in our present numerical sample to come closest to the values of the angular momentum obtained from the observational samples, since, in this case, the center of mass tends to coincide with the density peak, which in turn is associated with the intensity peak used in observations to define clumps. Conversely, we expect those clumps where the density peak is farthest from the center mass (as occurs with some filaments), should most poorly recover the observed \jR\ relation. Indeed, as seen in previous sections, structures with aspect ratios $>3$ deviate the most from the observed relation.

\begin{figure*}
\centering
\includegraphics[width=0.33\linewidth]{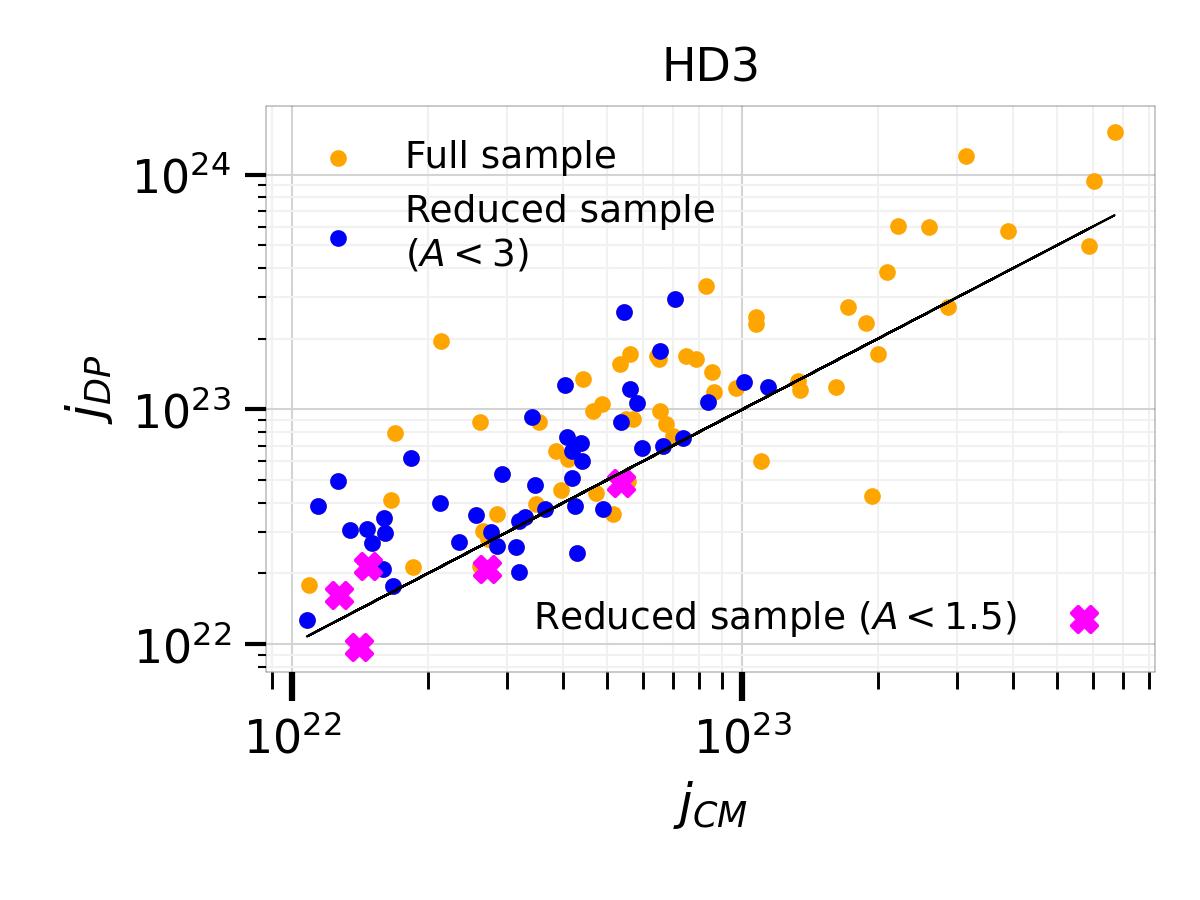}
\includegraphics[width=0.33\linewidth]{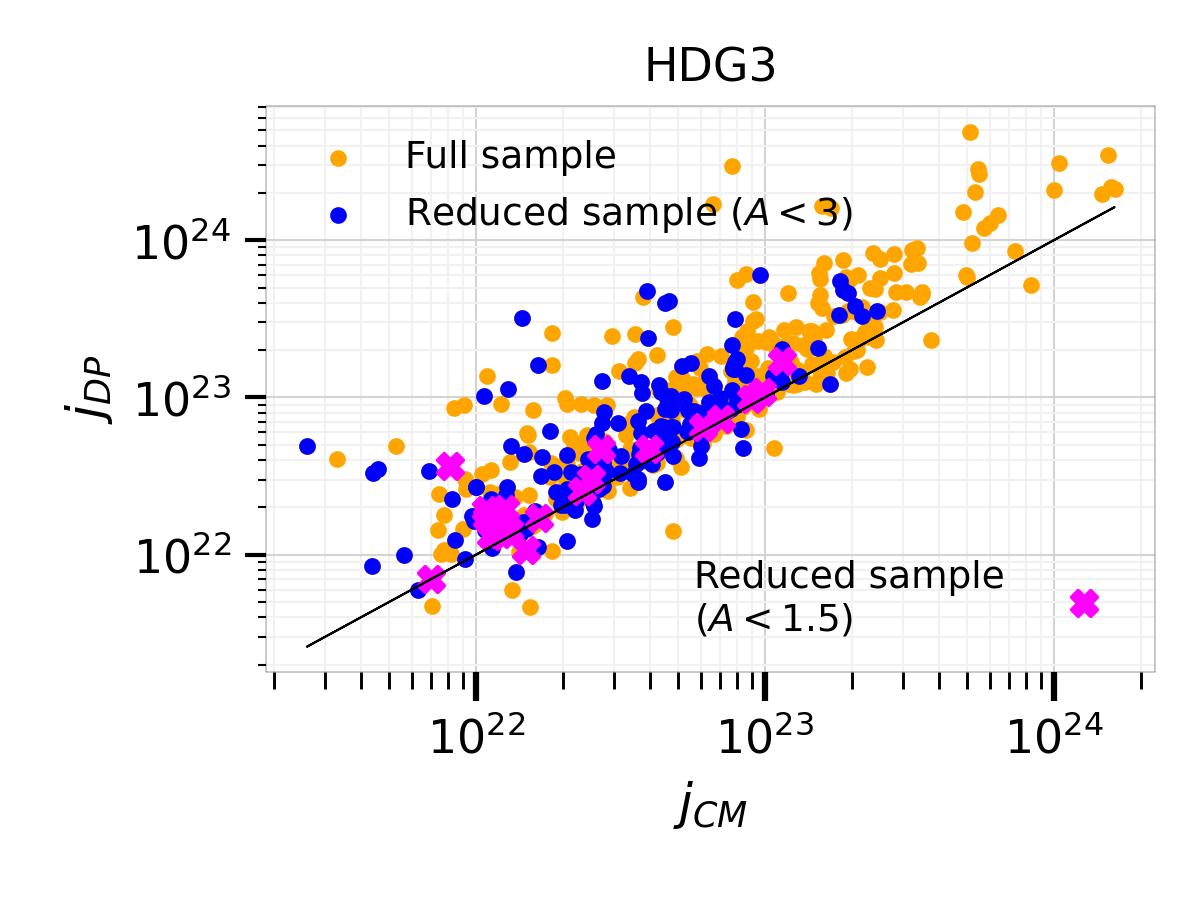}
\includegraphics[width=0.33\linewidth]{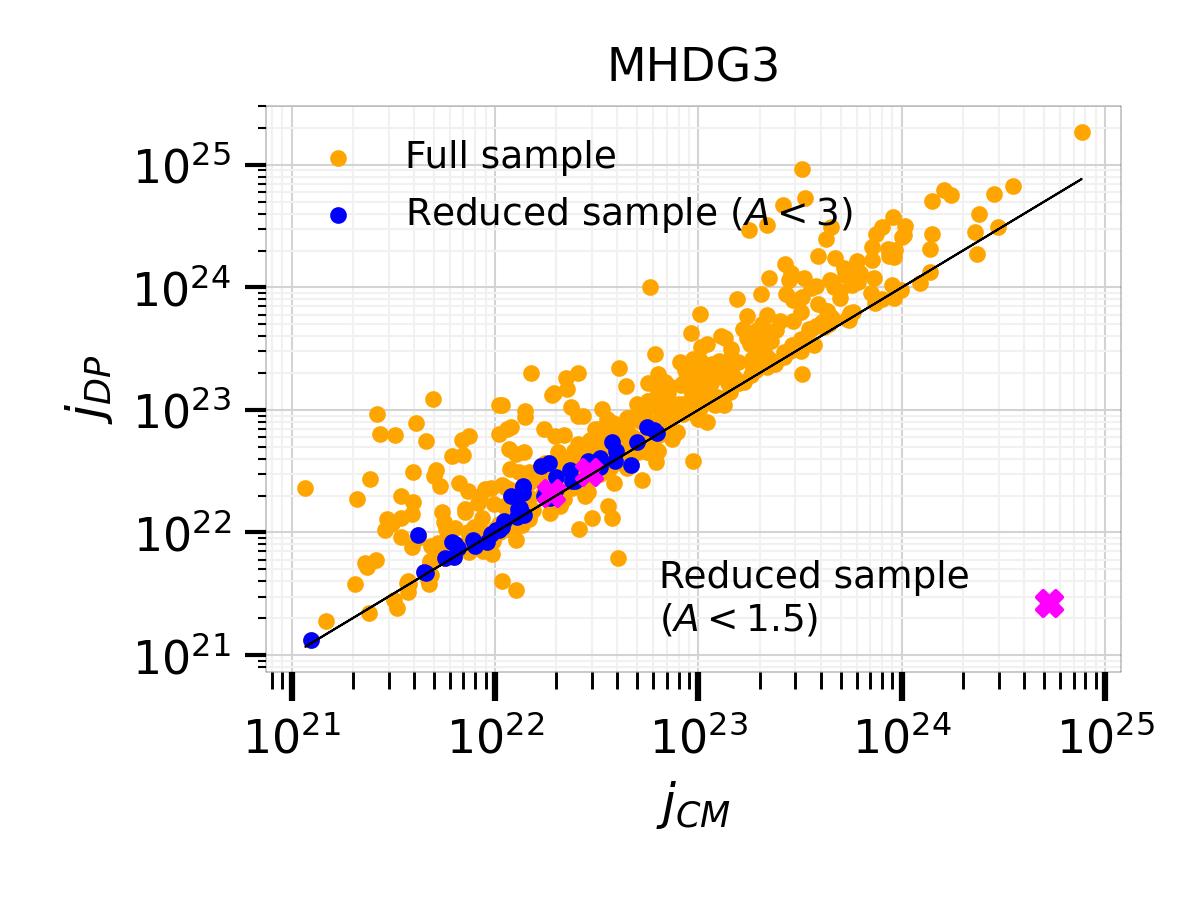}

 \caption{Specific angular momentum measured relative to the density peak ($j_{{\rm DP}}$) versus the specific angular momentum measured relative to the center of mass ($j_{{\rm CM}}$) for the simulations HD3, HDG3, and MHDG3. The full samples are shown in orange, and the reduced ones in blue ($A<3$). A second reduced sample is shown in magenta that corresponds to clumps with aspect ratios $< 1.5$. The black line in each plot represents the locus where $j_{{\rm DP}}=j_{{\rm CM}}$.}
\label{fig:jDP-jCM}
\end{figure*}

In Figure \ref{fig:jDP-jCM} we show the specific angular momentum measured relative to the density peak ($j_{{\rm DP}}$) versus the specific angular momentum measured relative to the center of mass ($j_{\rm CM}$) for the clumps in simulations HD3, HDG3, and MHDG3. The full samples are shown in orange, and the reduced ones in blue ($A<3$). A second reduced sample is shown in magenta that corresponds to clumps with aspect ratios $< 1.5$, i.e., mainly rounded clumps. The black line in each plot is the identity line.

The black line in all three cases seems to be a lower bound on the points, that is, for most of the clumps $j_{\rm DP}>j_{{\rm CM}}$. However, it can be seen that, as we restrict the sample to increasingly rounded clumps (smaller aspect ratio), the distance from the observed relation decreases together with the difference between  $j_{\rm DP}$ and $j_{\rm CM}$. In conclusion, for clumps with rounded geometry there seems to be no significant difference between using the density peak or the center of mass to measure the angular momentum. 

Our results above suggest that, when filamentary structures are considered, either in observations or in simulations, it is necessary to measure their angular momentum differently than we have done here. There are few studies in the literature on angular momentum measurements in filaments. Some velocity gradient measurements in star-forming \citep{Alvares-Gutierrez+2021} and non-star-forming \citep{Hsieh+2021} filaments show a gradient perpendicular to the principal axis of the filament that has been interpreted as rotation. However, it appears that this rotation pattern can be found only in relatively young filaments that have not undergone a period of intense star formation, where the latter causes a transition from a generalized pattern to small rotation centers at specific points. 

A possible first improvement in the measurement of angular momentum for filamentary structures could be to define it with respect to the main axis of the filament rather than with respect to a single reference point such as the center of mass. Another aspect that may affect the resulting \jR\ relation is the assumption of ideal MHD. Non-ideal effects, particularly ambipolar diffusion at these scales, can modify the redistribution of angular momentum during collapse \citep[e.g.,][]{Wurster+2018,Wurster+2022}, and may therefore alter the scaling between $j$ and $R$. Such studies will be presented elsewhere.

\subsection{The hydrodynamic torque term in the Eulerian and Lagrangian frames}

Throughout this work, we have referred to the first term on the right hand side in eq. \eqref{eq:total torque per unit mass} as the {\it hydrodynamic torque}. This term appears explicitly when writing the angular momentum evolution equation in Eulerian form---that is, for the fixed volume $(V)$ containing the clump particles at the time when the clump is defined. In this section we briefly discuss some possible physical interpretations of this term, in particular in comparison to the Lagrangian formulation. 

In the Eulerian description, the rate of change of a quantity is given by the intrinsic temporal rate of change of the quantity, plus its spatial transport by the fluid motion. In the case of the angular momentum, its net temporal rate of change includes a contribution to its transport by the velocity field itself. This process is described by the hydrodynamic torque term in eq. \eqref{eq:total torque per unit mass}. In the absence of the other torques, the angular momentum of volume $V$ can change just by pure redistribution among particles moving inertially. In this sense, particles entering and leaving the volume are partly responsible for changing the angular momentum of the volume, even without the presence of external torques. This is related to the effect of the ``intruder particles'' on angular momentum exchange discussed in \citetalias{Arroyo-Chavez.Vazquez-Semadeni2022}. 

In that paper, the  name ``intruder particles'' referred to those SPH particles that are not part of the clump at the definition time, but are able to enter and/or leave the spheroidal volume containing the original member particles after they have moved significantly from their original positions. Indeed, by Lagrangianly tracking the member particles of a few clumps, \citetalias{Arroyo-Chavez.Vazquez-Semadeni2022} found that the clumps' angular momentum ceased to be conserved when a sufficiently large number of intruder particles shared the same region of space as the particles in the Lagrangian set---the real member particles---, concluding that it is the exchange of angular momentum with the intruder particles that is responsible for the change in the angular momentum change of the Lagrangian set. In the Eulerian formulation, this behavior might be represented by the hydrodynamic torque term, which corresponds to the exchange of angular momentum across the volume's boundary due to the fluid motions. It is important to stress that this effect is apparent when the hydrodynamic torque term is written explicitly; i.e., in Eulerian form. In the Lagrangian formulation this information is lost \citep[and it is not considered in other works; e.g.,] [] {Misugi+2023, Misugi+2023_2}, since the effect of the change in angular momentum due to the simple redistribution of mass is absorbed in the total derivative. 

It is also worth noting that, in contrast to the method used in \citetalias{Arroyo-Chavez.Vazquez-Semadeni2022}, no temporal tracking of Lagrangian clumps were made in this work, and we only wrote the equation of evolution of angular momentum according to the volume defined by the clump particles at a fixed time, thus performing only an instantaneous measurement of the torques in an Eulerian frame.







\section{Conclusions}
\label{sec:conclusions}

In this paper, we have carried out simulations analogous to the one analyzed in \citetalias{Arroyo-Chavez.Vazquez-Semadeni2022} but progressively adding turbulence, gravity and the magnetic field, in order to study their effect both on the \jR\ relation and on the angular momentum transport mechanism, by investigating the \jR\ relation that develops in each case, the effect of the clump shape in the scaling, and the magnitude of the active torques for a numerical sample of clumps in each case. The clumps are defined in each simulation as sets of connected regions above a density threshold. We used six density thresholds from $n_{\rm th} =  3 \times 10^{2}$ to $10^{5}$ cm${}^{-3}$. We considered a ``full sample'' including all the clumps, and a ``reduced'' one, discarding elongated clumps with aspect ratios $A>3$. We measured the clump angular momentum by adding all individual SPH particle contributions, not by inferring rotation from large-scale velocity gradients, as is usually done, in order to obtain the true angular momentum of the clumps. Thus, the resulting vector reflects a {\it residual} angular momentum, without implying any systematic rotation around a specific axis.

Our main results can be summarized as follows:

\begin{itemize}
    
    \item The samples (full and reduced) obtained from the purely hydrodynamic simulation (HD3) are not statistically significant to safely conclude that a scaling is present, showing that it is not possible to generate this relation in the absence of gravity,   Furthermore, no elongated dense structures are generated in this simulation, indicating that hub-filament–like systems cannot arise from turbulence alone. However, we emphasize that the magnitudes of the angular momenta in the clumps are within the range of observed values, suggesting that turbulence itself may be the main driver of angular momentum redistribution at these scales. 
    
    \item The \jR\ relation is most similar (both in slope and magnitude) to the observational relation for the reduced sample in the simulation that only contains turbulence and gravity (HDG3). The elongated structures in the full sample in this run tend to deviate from the observational fit, having excessively large values of $j$. We speculate that this excess is due to the fact that, for elongated objects that may be rotatng preferentially along their principal axis, our measurement procedure for $j$, around their (point-like) center of mass, may incorrectly interpret minor turbulent motions near the tips of the structures as strongly contributing to the angular momentum. This calls for a different procedure for measuring $j$ in filamentary structures, specifically measuring it around their {\it axis} rather than around their center of mass. We defer this task for a future contribution.
    
    \item The clump sample in the simulation with gravity, turbulence, and magnetic field (MHDG3) recovers the trend of the observational \jR\ relation, despite the fact that this sample is dominated by filaments. This behavior is contrary to that observed in the HD3 and HDG3 simulations, where the filaments are precisely the ones that deviate the most from the observational trend. We speculate that this may be due to the additional effect of the magnetic field, of inhibiting turbulence generation during the assembly of structures, which may reduce the residual angular momentum magnitude, thus offsetting the excess in the measured residual angular momentum with respect to the center of mass. Note that, in the reduced sample, where the elongated clumps are discarded, 98\% of the clumps are lost, leaving a statistically non-representative sample.

    \item  From the \jR\ relation for the clump samples in the three simulations studied here (HD3, HDG3, MHDG3), we propose that turbulence is the agent determining the magnitude of the clump angular momentum, gravity is responsible for the expansion of the dynamic range of sizes (and densities), and the magnetic field acts to reduce the scatter of the relation by inhibiting the generation of turbulence.

    \item We find a clear correlation between the specific angular momentum and the 3D velocity dispersion in all simulations, whose slope increases as gravity and magnetic field are progressively added, both for the full and the reduced samples. Moreover, this relation appears to have a dependence on the density, since the samples appear to separate by density threshold, thereby increasing the scatter of the relation when all samples are considered together. This scatter is reduced by plotting the product $j\Sigma$ vs $\sigma$. Indeed, since we have chosen to measure the {\it residual} angular momentum, it is expected that a correlation between the specific angular momentum with the {\it turbulent} velocity dispersion is found, while this is not obvious if the angular momentum is calculated from large-scale velocity gradients.

    \item We measured the gravitational, magnetic, hydrodynamic, and pressure-gradient torques in the clumps of the three simulations, finding that hydrodynamic torques tend to be larger as the density threshold for defining the structures is increased. On the other hand, all four torques are indistinguishable in magnitude for elongated structures defined at lower density thresholds.
    
\end{itemize}

Our combined results from this study and its predecessor, \citetalias{Arroyo-Chavez.Vazquez-Semadeni2022} suggest that the \jR\ relation in molecular cloud cores is mainly the result of two main circumstances: first, smaller structures are {\it not} the result of a monolithic contraction of a larger structure, but rather of its {\it fragmentation} and subsequent contraction. The contracting fragment can do so freely because it transfers part of its angular momentum to the rest of the material. Second, the transfer is accomplished by the combined effect of the hydrodinamic (i.e., among neighboring flyuid parcels), gravitational, magnetic, and pressure-gradient torques at low densities, but is later dominated by the hydrodynamic torques at higher densities. In \citetalias{Arroyo-Chavez.Vazquez-Semadeni2022} we also showed that the collapsing part of a clump (i.e., the fragment) corresponds to the region that had previously lost angular momentum. Since the clumps in our simulations generally contain regions that continue to increase in density and undergo further fragmentation, it is reasonable to expect that the torques measured here remain active throughout this process. However, a more complete assessment requires following the evolution of the torques in time, in order to determine both the duration over which they act and the rate at which they remove angular momentum. Such an analysis would allow us to evaluate whether these mechanisms can quantitatively reproduce the observed \jR\ relation. We leave this investigation for future work.


\section*{Acknowledgements}

We are grateful to Professors Javier Ballesteros-Paredes, Gilberto C. Gómez, Patrick Hennebelle, Shu-ichiro Inutsuka, Aina Palau, and Manuel Zamora-Avilés for their valuable comments and insightful feedback during the review of the doctoral dissertation work that led to this article.

G.A.-C.\ and E.V.-S.\ thankfully acknowledge financial support from UNAM-DGAPA PAPIIT grant IG100223.

\software{Phantom, \citep{Price+2018}, Numpy \citep{numpy}, \citep{Robitaille12}, Matplotlib \citep{matplotlib}, Sarracen \citep{Harris.Tricco2023}, Plonk \citep{Mentiplay2019}, Scipy \citep{SciPy-NMeth}, scikit-learn \citep{scikit-learn}}

\facility{IRyA HPC, UA HPC}




\appendix

\section{Transformation of the angular momentum between reference frames}
\label{Appe:AM for system of particles}

\begin{figure}
\centering
\includegraphics[width=0.5\linewidth]{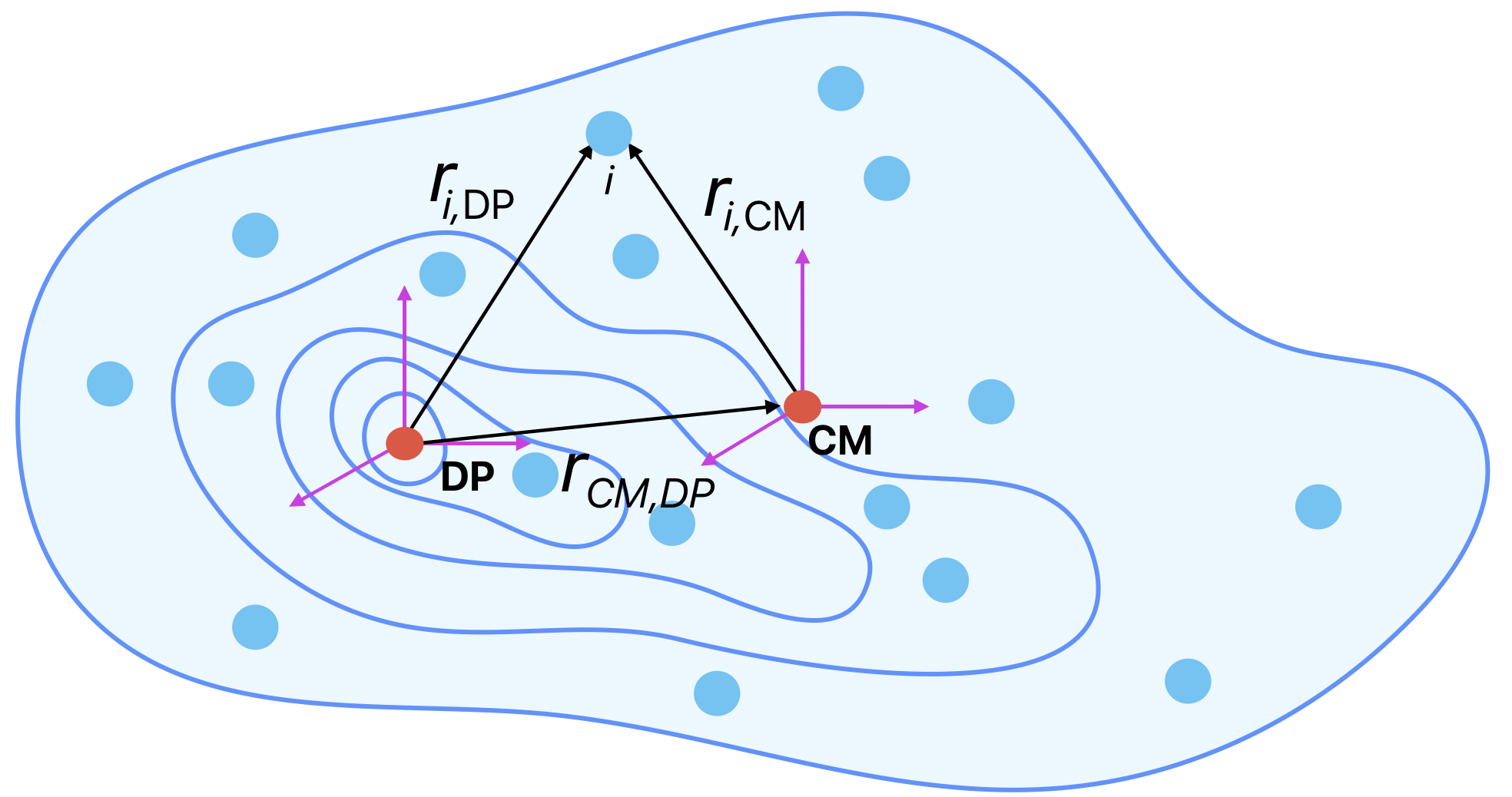}
\caption{Decomposition of particle position vectors in a cloud, represented by blue density contours, relative to two reference frames: the density peak (DP) and the center of mass (CM).}
\label{fig:CM_moving}
\end{figure}

This appendix presents a comparison between the angular momentum calculated with respect to the center of mass (subscript CM) and that obtained relative to another reference point in the cloud, for example, the density peak (DP). The comparison is illustrated using a classical particle system composed of the SPH particles (filled blue circles) that form the clump, represented by the blue density contours in Fig. \ref{fig:CM_moving}.. 

In general, we can write the angular momentum of a system of particles with respect to a reference frame (taken as the CM) moving with velocity ${\bm v}$ with respect to another (taken as the DP). Thus, the position and velocity vectors of the particles with respect to the DP can be written as:
\begin{align}
    {\bm r}_{i, {\rm DP}} = {\bm r}_{{\rm CM,DP}} + {\bm r}_{i,{\rm CM}},\\
    {\bm v}_{i,{\rm DP}} = {\bm v}_{{\rm CM,DP}} + {\bm v}_{i,{\rm CM}},\nonumber
\end{align}
 where ${\bm r}_{{\rm CM,DP}} = {\bm r}_{{\rm CM,DP}}(t)$, and ${\bm v}_{{\rm CM,DP}} = d{\bm r}_{{\rm CM,DP}}/dt$.

Then, the total angular momentum of the system in the system of the density peak (DP) is given by:
\begin{align}
\label{eq:L from CM}
    {\bm L_{\rm DP}} =& \sum_{i}^{n} {\bm L}_{i,{\rm DP}} 
    = \sum_{i}^{n} {\bm r}_{i,{\rm DP}} \times {\bm p}_{i,{\rm DP}} 
    = \sum_{i}^{n}  {\bm r}_{i,{\rm DP}} \times m{\bm v}_{i,{\rm DP}} \nonumber \\
    =& \sum_{i}^{n} ( {\bm r}_{{\rm CM,DP}} + {\bm r}_{i,{\rm CM}}) \times m ({\bm v}_{{\rm CM,DP}} + {\bm v}_{i,{\rm CM}}) \nonumber \\
    =& \sum_{i}^{n} m [ {\bm r}_{{\rm CM,DP}} \times {\bm v}_{{\rm CM,DP}}  + {\bm r}_{{\rm CM,DP}} \times {\bm v}_{i,{\rm CM}} \nonumber \\
    & + {\bm r}_{i,{\rm CM}} \times {\bm v}_{{\rm CM,DP}}  + {\bm r}_{i,{\rm CM}} \times {\bm v}_{i,{\rm CM}}]. 
\end{align}

In principle, the previous equation is valid for any two reference systems that behave like those in Figure \ref{fig:CM_moving}, but having taken one of them as the center of mass, from the definition of center of mass we have that
\begin{align}
    \sum_{i}^{n}  m({\bm r}_{i,{\rm DP}} - {\bm r}_{{\rm CM,DP}}) =  \sum_{i}^{n} m{\bm r}_{i,{\rm CM}} = 0; \\ \sum_{i}^{n} {\bm p}_{i,{\rm CM}} =  \sum_{i}^{n} m {\bm v}_{i,{\rm CM}} = 0.
\end{align}
Thus, it is seen that the middle terms in Equation \eqref{eq:L from CM} vanishes, and we recover the well-known result that the angular momentum of a system of particles can be expressed as the angular momentum of the particles with respect to the center of mass, plus the angular momentum of the center of mass with respect with respect to any other point, which in this case is the density peak. That is:
\begin{align}
    {\bm L_{\rm DP}} &= {\bm r}_{{\rm CM,DP}} \times {\bm v}_{{\rm CM,DP}} +  \sum_{i}^{n}  {\bm r}_{i,{\rm CM}} \times {\bm v}_{i,{\rm CM}}  \nonumber \\
    & = {\bm L}_{\rm CM,DP} +  \sum_{i}^{n} {\bm L}_{i,{\rm CM}}.
\end{align}
Therefore, it can be seen that for a point other than the center of mass, specifically by the term ${\bm L}_{\rm CM,DP}$.

\section{$\beta$ values}
\label{Appe:beta values}

\begin{figure*}
\centering
\includegraphics[width=0.33\linewidth]{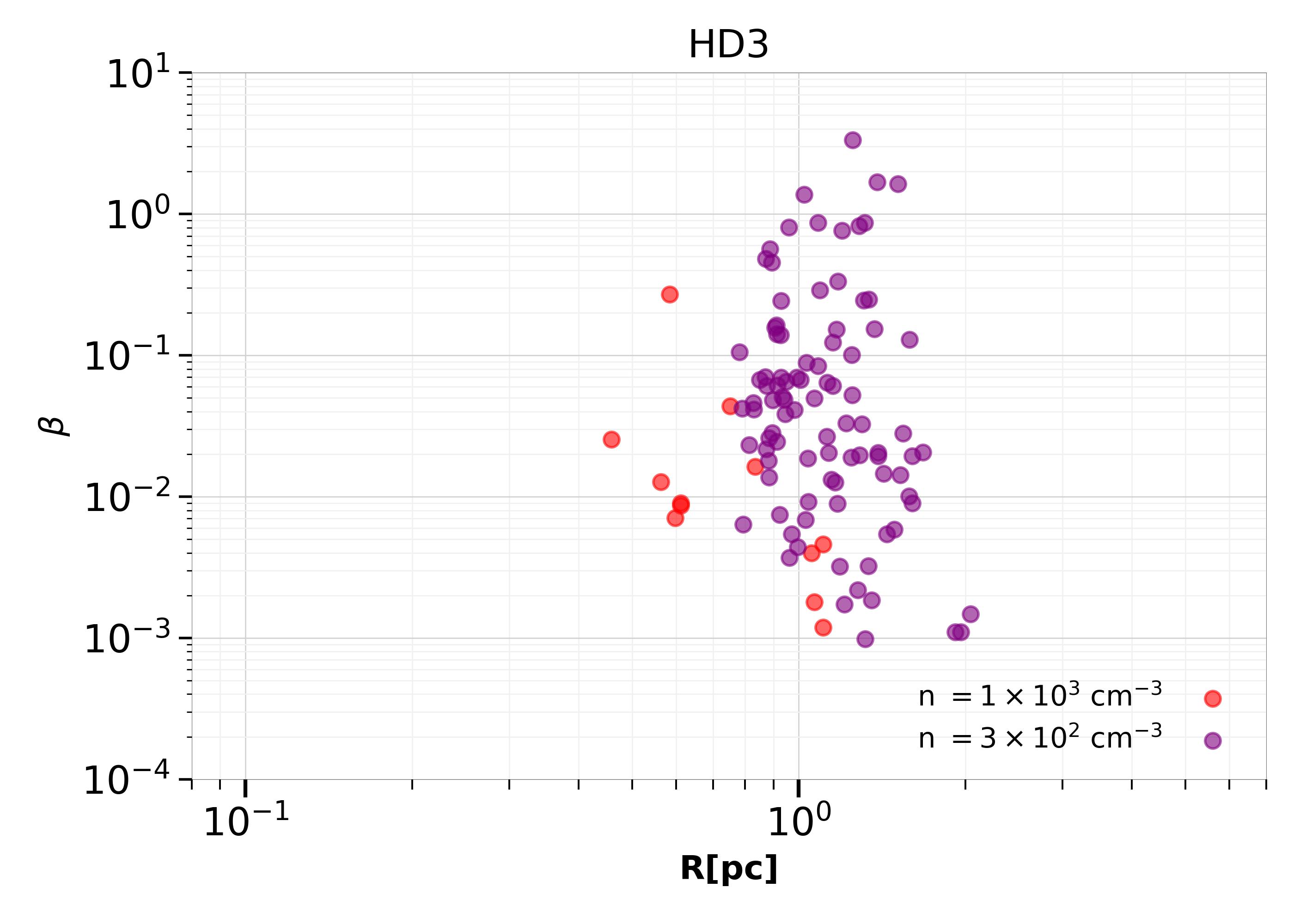}
\includegraphics[width=0.33\linewidth]{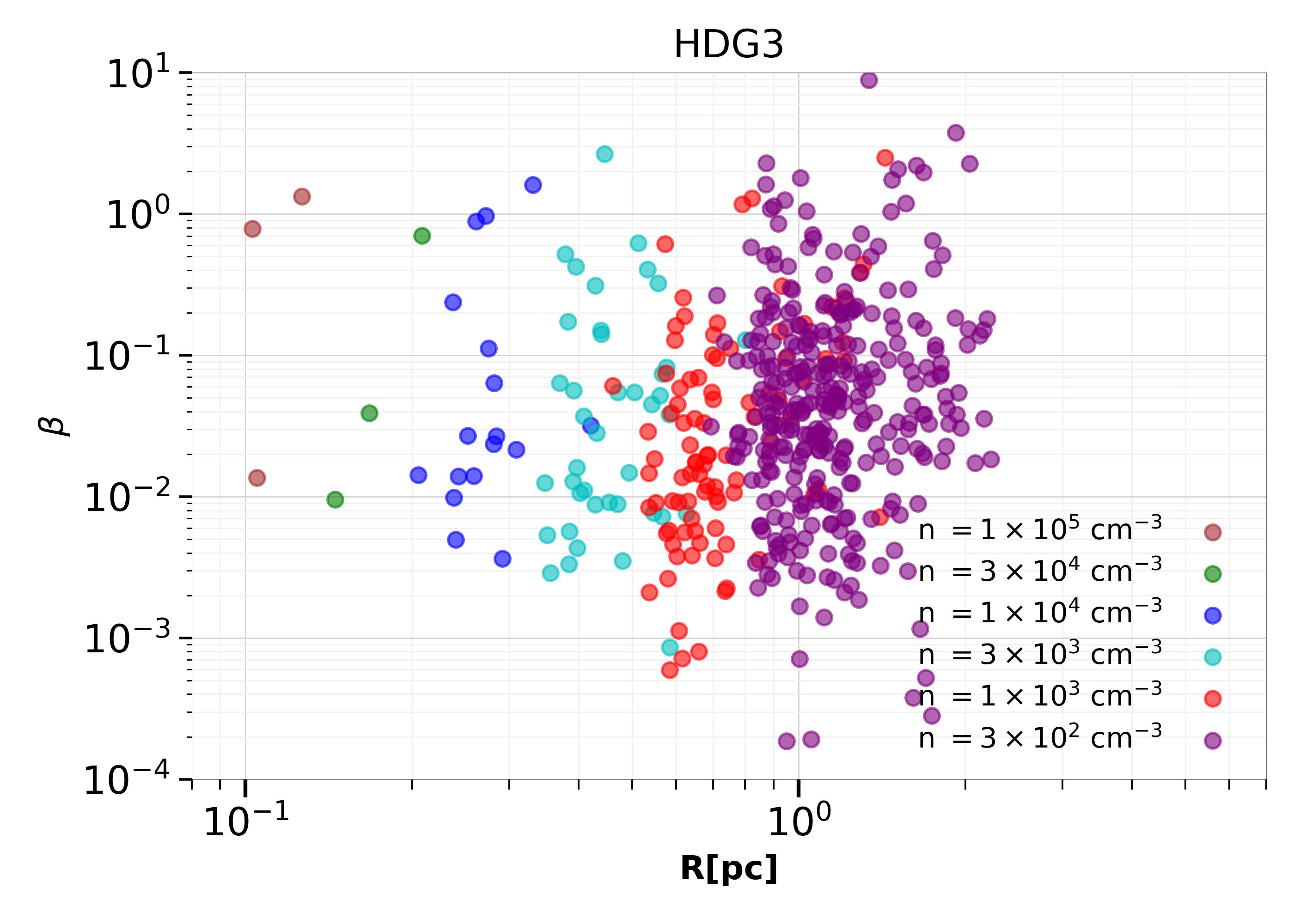}
\includegraphics[width=0.33\linewidth]{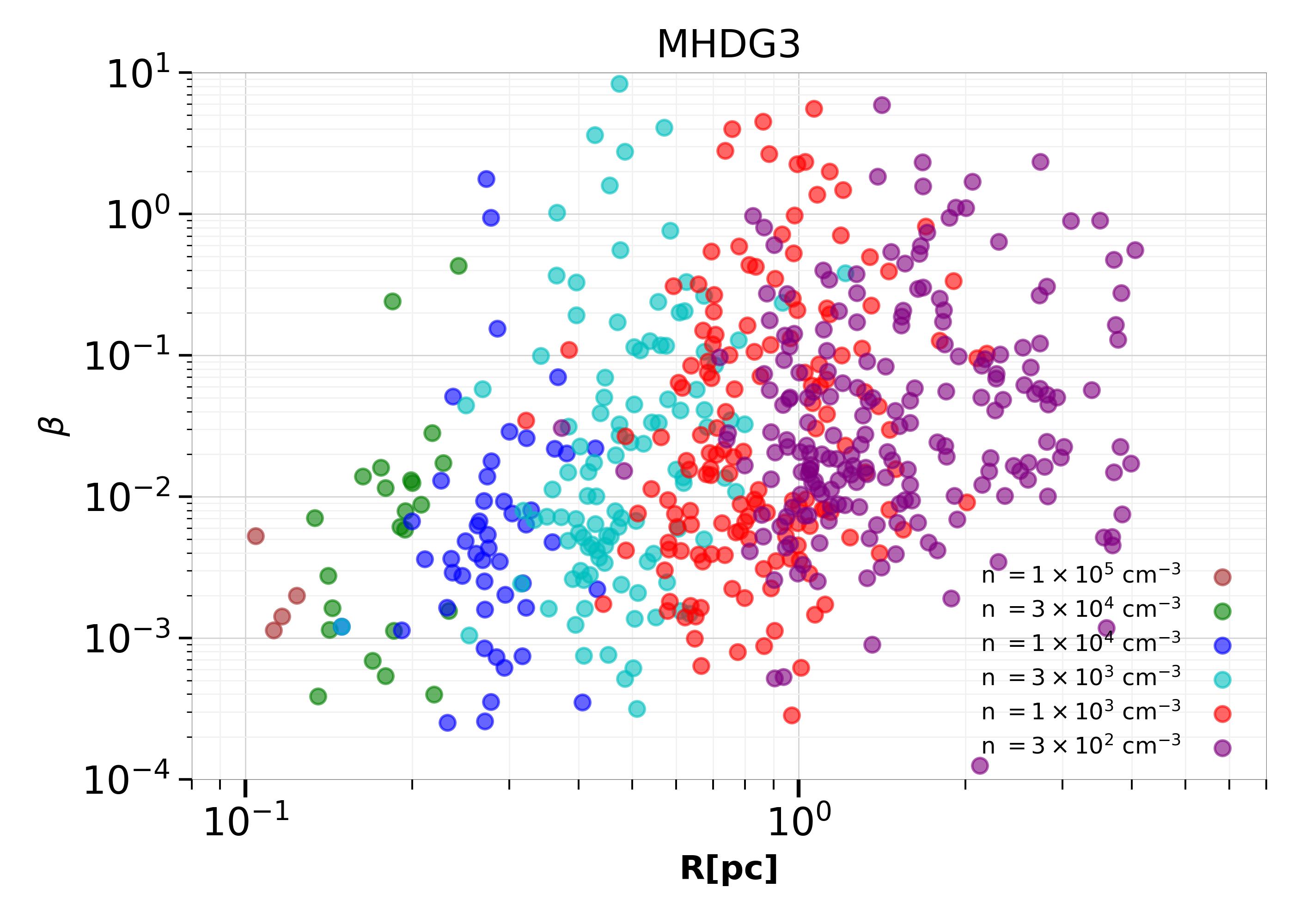}
\includegraphics[width=0.33\linewidth]{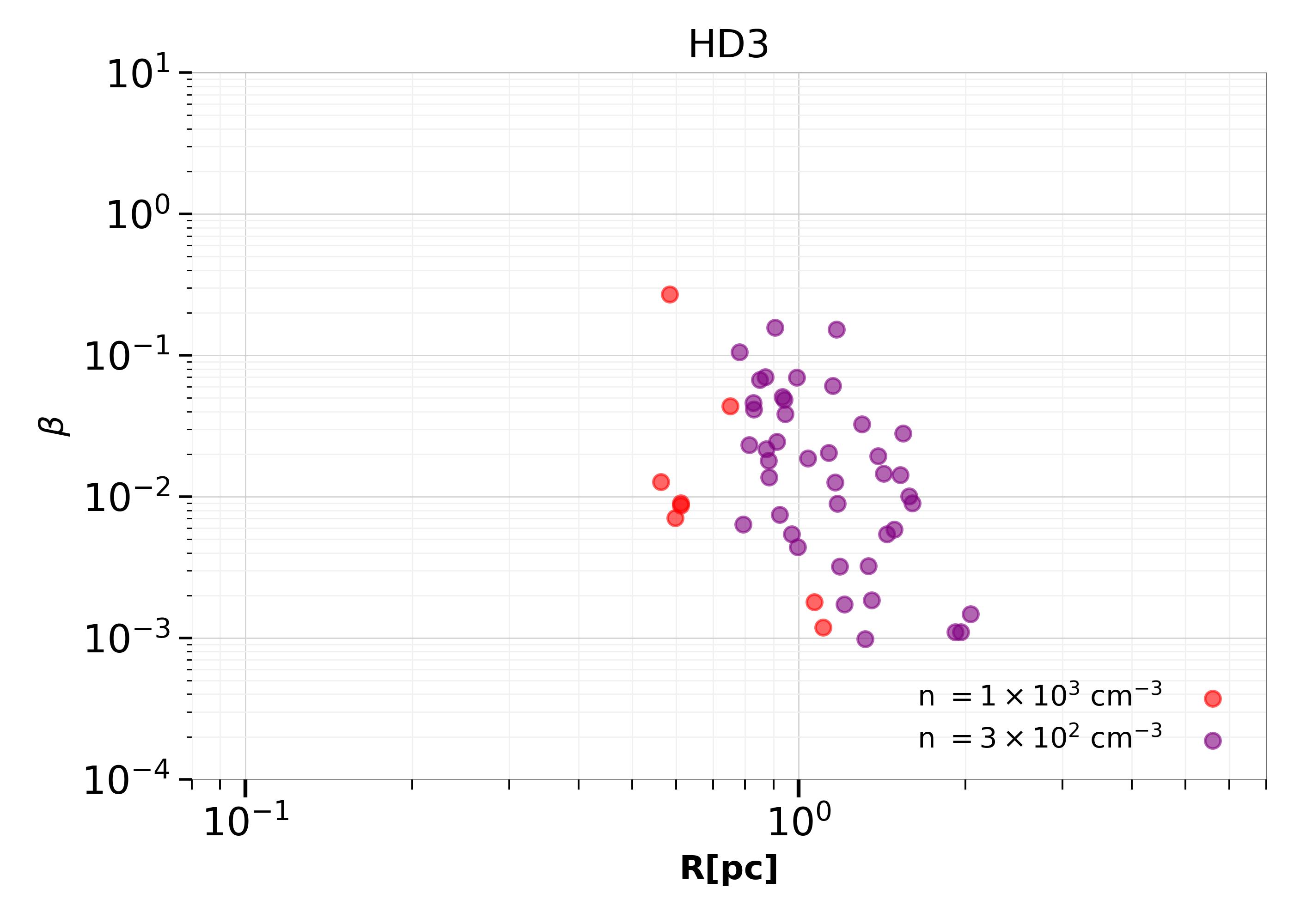}
\includegraphics[width=0.33\linewidth]{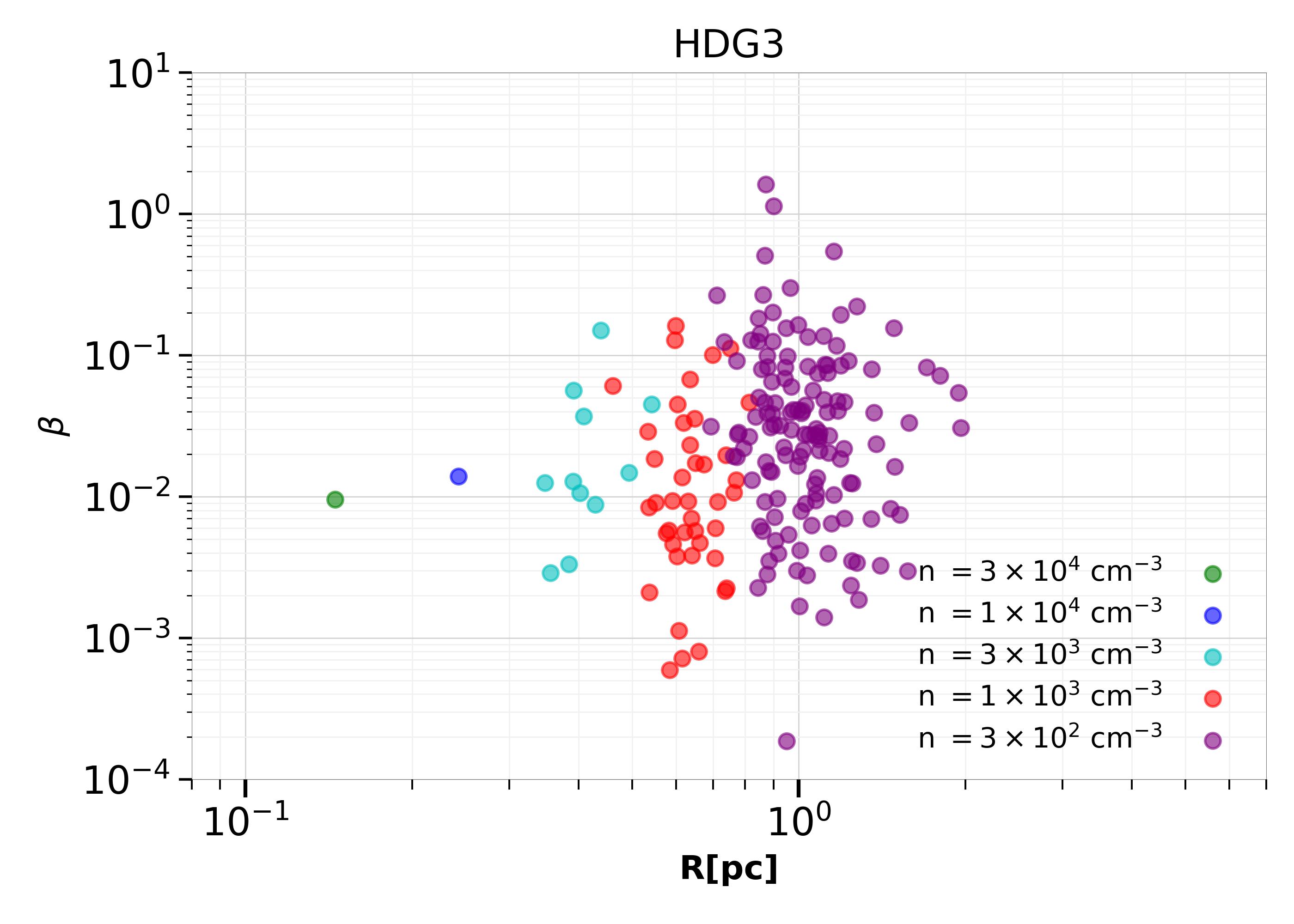}
\includegraphics[width=0.33\linewidth]{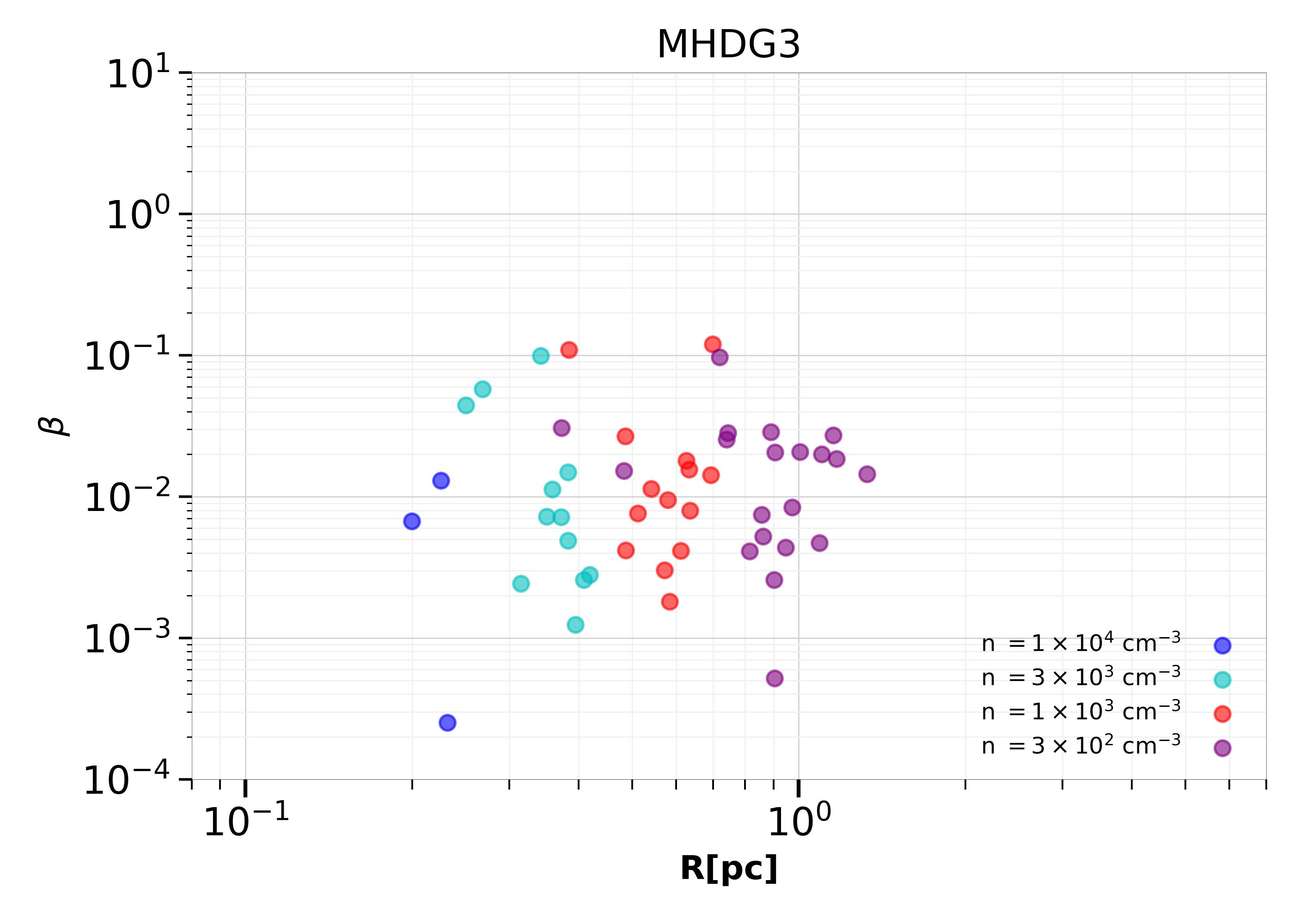}
 \caption{Values of $\beta$, the ratio of rotational energy to gravitational energy, calculated as shown in eq. \ref{eq:beta}, for the full (first row) and reduced (second row) samples in each simulation.}
\label{fig:betas}
\end{figure*}

In Figure \ref{fig:betas} we show the values of $\beta$, the ratio of rotational energy, $e_{\rm r}$, to gravitational energy, $e_{\rm g}$, estimated as
\begin{equation}
    \beta = \frac{e_{\rm r}}{e_{\rm g}} \approx \frac{j^{2}}{2GMR},
    \label{eq:beta}
\end{equation}
where $j$ is the specific angular momentum, $M$ is the mass, and $R$ is the size of the clump for the full (first row) and reduced (second row) samples in each simulation. Despite the scatter in $\beta$, no clear dependence on the size is evident, as reported in \citetalias{Arroyo-Chavez.Vazquez-Semadeni2022}. However, a possible dependence can be seen in the full sample of the MHDG3 simulation. A more detailed analysis is required to assess whether this trend is related to magnetic field effects or instead results from geometric biases introduced in the estimation of $j$ used to compute $\beta$. The latter possibility appears plausible, since the reduced sample does not show this behavior.

\section{$B$-density relation}
\label{Appe:B vs n}
\begin{figure}
\centering
\includegraphics[width=0.5\textwidth]{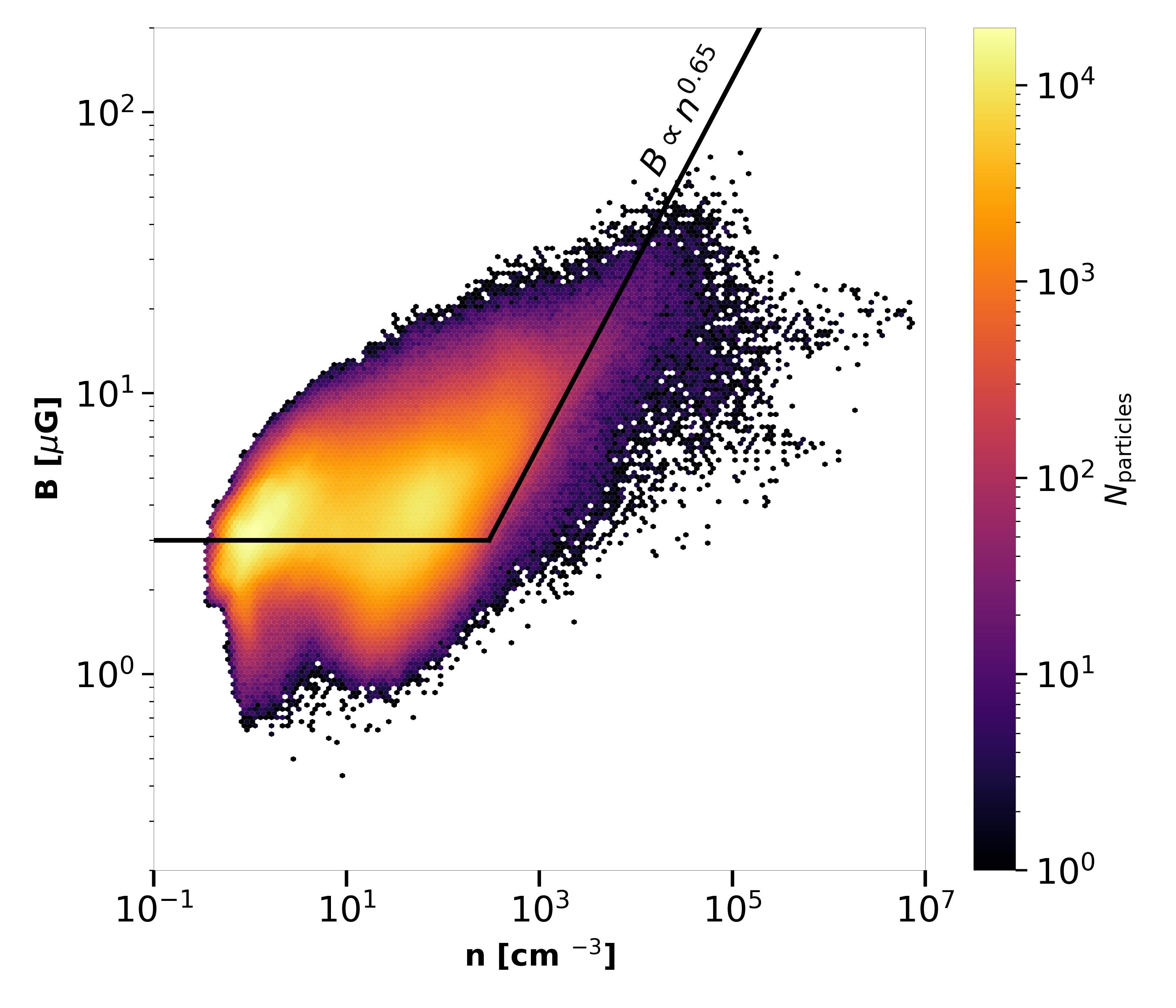}
\caption{Magnetic field strength, $B$, versus number density, $n$, for SPH particles in the MHDG3 simulation, using the three snapshots considered for clump identification. The color bar shows the particle frequency. The black lines correspond to the slopes reported by \citet{Crutcher+2010} for a background magnetic field $B_{0} = 3,\mu\mathrm{G}$, consistent with the initial conditions in the simulation.}
\label{fig:B vs n}
\end{figure}

In order to provide an idea of the typical magnetic field strength, $B$, for run MHDG3, in Figure \ref{fig:B vs n} we show the relation between $B$, and number density, $n$, for all SPH particles in the three snapshot considered. The figure is presented as a hexbin map, where the color scale indicates the particle frequency. The black lines correspond to the slopes reported by \citet{Crutcher+2010} for a background magnetic field $B_{0} = 3,\mu\mathrm{G}$, consistent with the value adopted as initial condition. Overall, the distribution follows the expected trend, with the magnetic field remaining approximately constant at low densities and increasing with density at higher values.

\section{Aspect ratio - size relation}
\label{Appe:A vs R}

\begin{figure*}
\centering
\includegraphics[width=0.33\linewidth]{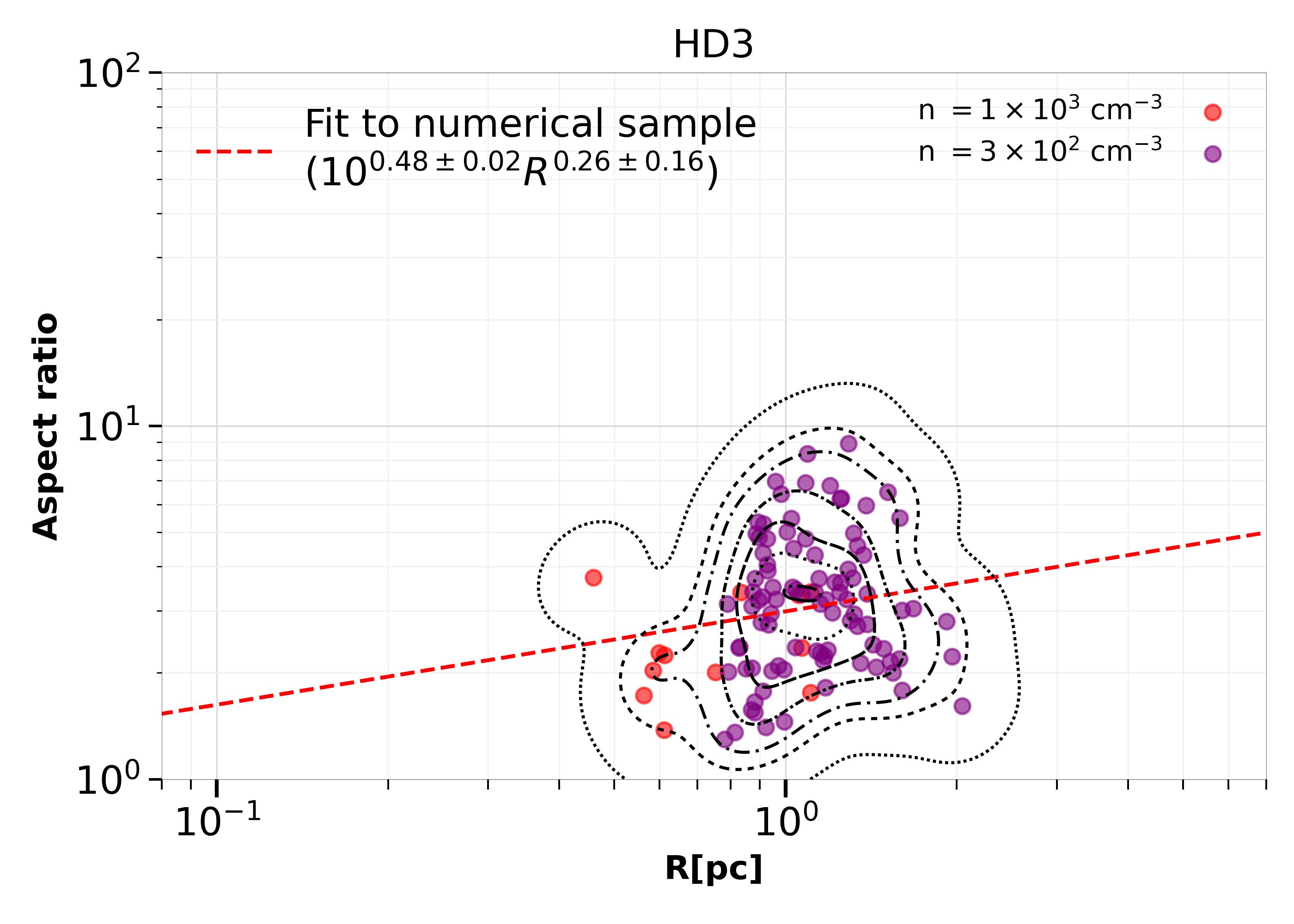}
\includegraphics[width=0.33\linewidth]{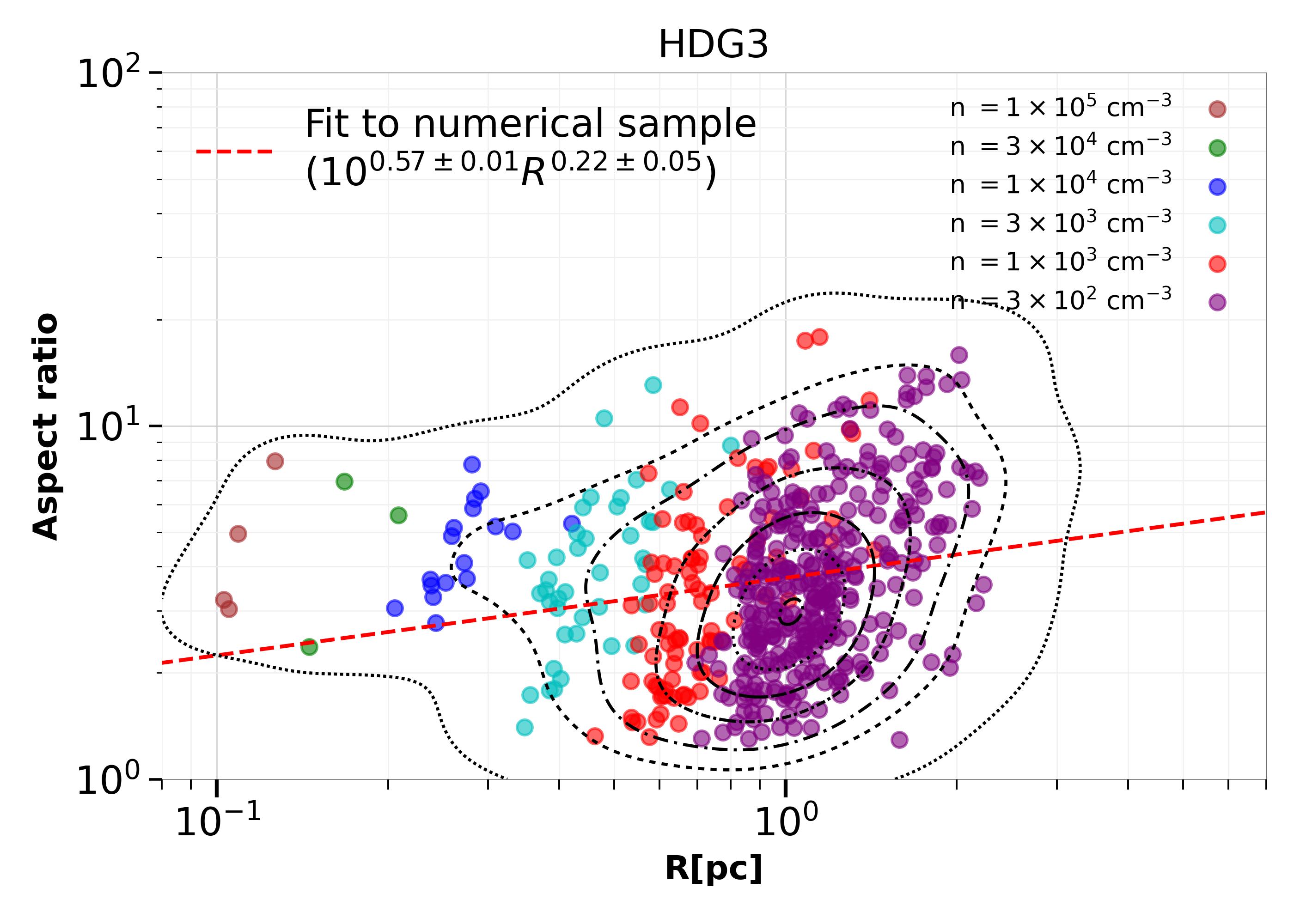}
\includegraphics[width=0.33\linewidth]{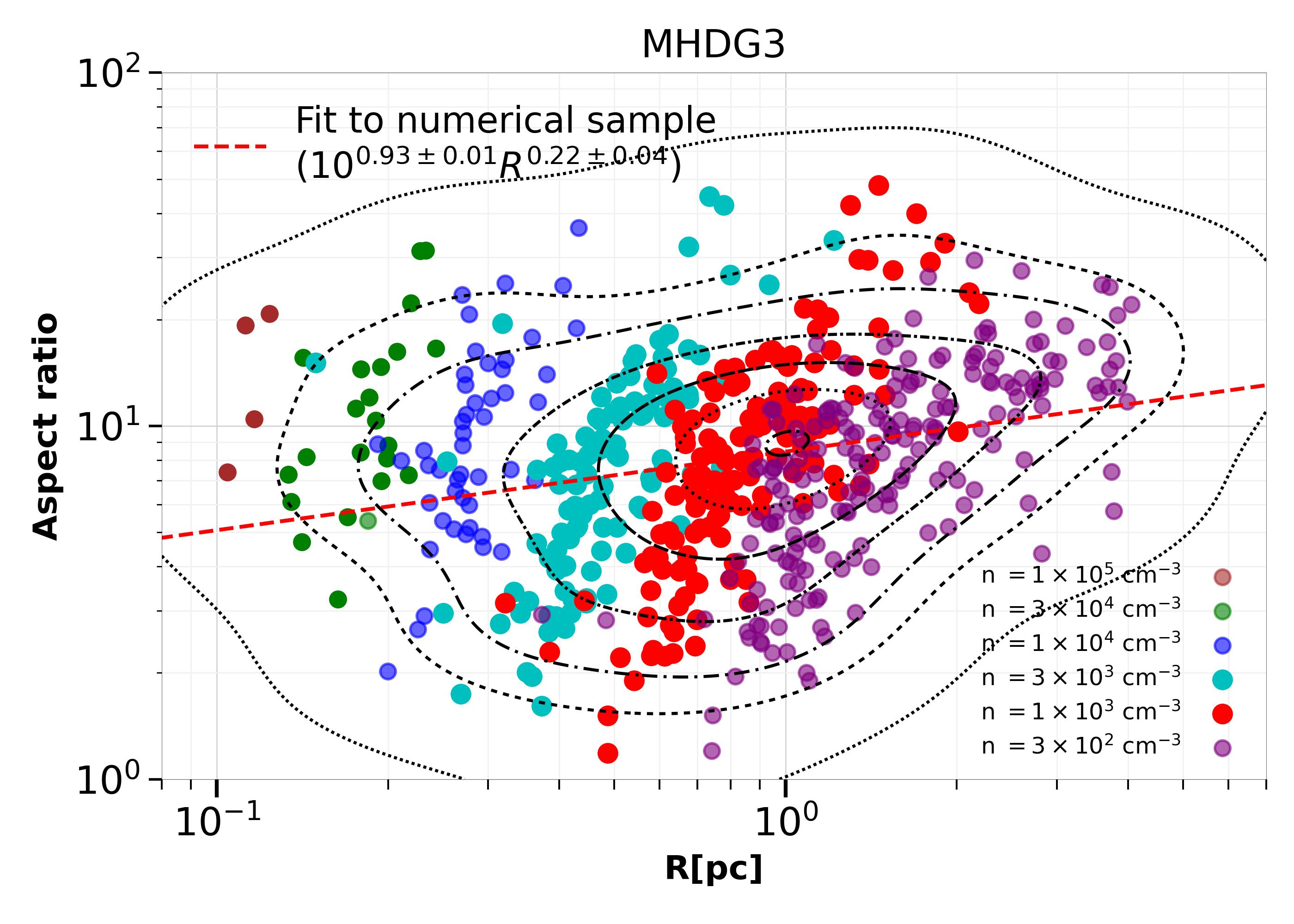}
 \caption{Aspect ratio, $A$, as a function of size, $R$, for the full samples in each simulation. The black lines show isocontours of the 2D density distribution of the points, with levels normalized to the peak in point density and set at $0.01$, $0.1$, $0.2$, $0.4$, $0.6$ , $0.8$, and $0.99$ shown with different line styles.}
\label{fig:AR}
\end{figure*}

In Fig. \ref{fig:AR}, we show the aspect ratio, $A$, as a function of size, $R$, for the full samples in each simulation. It can be seen that, in addition to producing a larger number of filamentary structures, the MHDG3 run also forms the largest ones. This effect contributes to steepening the slope of the \jR\ for the sample shown in the bottom left panel of Fig. \ref{fig:com_jR}, as the ends of elongated filaments provide a dominant contribution to the angular momentum when it is computed with respect to the center of mass.

\bibliography{biblio}
\bibliographystyle{aasjournalv7}


\end{document}